%% file: amarino.tex
\documentclass[useAMS,usenatbib,usegraphicx,usepsfig]{mn2e}
\usepackage{natbib} 
\begin{document}
 \topmargin=-1.3cm
\title[Galaxy evolution in nearby groups. II.]{Galaxy evolution in nearby loose groups. II.\\{  
  Photometric and kinematic characterization of USGC U268  and USGC U376 group members in the Leo cloud\thanks{Based on
   {\it GALEX}  (GI6-6017, PI A. Marino)  and PUMA (PI M. Rosado)
   observations.}}}  
\author[A. Marino et al.]{A. Marino$^{1}$\thanks{E-mail:antonina.marino@unipd.it}, H. Plana$^{2}$, R. Rampazzo$^{3}$, L. Bianchi$^{4}$, M. Rosado$^{5}$, D. Bettoni$^{3}$, 
\newauthor{G. Galletta$^{1}$, P. Mazzei$^{3}$, L. Buson$^{3}$, P. Ambrocio-Cruz$^{6}$}, R. F. Gabbasov$^{5}$ \\    
$^{1}$Dipartimento di Fisica e Astronomia G. Galilei di Padova,  vicolo dell'Osservatorio~3, I-35122  Padova, Italy \\
$^{2}$Laboratorio de Astrof\'isica Te\'orica e Observacional, Universidade Estadual de Santa Cruz, Rodovia Ilh\'eus-Itabuna Km 16 \\
 45650-000 Ilh\'eus - BA, Brazil \\
$^{3}$INAF-Osservatorio Astronomico di Padova, vicolo dell'Osservatorio~5, I-35122  Padova, Italy\\
$^{4}$Dept. of Physics and Astronomy, Johns Hopkins University, 3400 North Charles
           Street, Baltimore, MD 21218  USA\\ 
$^{5}$Instituto de Astronom\'ia, Universidad Nacional Aut\'onoma de M\'exico,
Av. Universidad 3000, Ciudad Universitaria, \\ C.P. 04510
M\'exico D.F., M\'exico\\
$^6$Universidad Aut\'onoma del Estado de Hidalgo,
\'Area Acad\'emica de Ciencias de la Tierra y Materiales
Ciudad Universitaria,\\ Km 4.5 Carretera Pachuca - Tulancingo; Col. Carboneras
C.P. 42184, Mineral de la Reforma, Hidalgo, M\'exico
            }

\date{Accepted. Received}

\pagerange{\pageref{firstpage}--\pageref{lastpage}} \pubyear{2012}

\maketitle

\label{firstpage}

\begin{abstract}
The paper is the second of a series in which we are exploring the co-evolution of  galaxies and groups
in the Local Universe,  adopting a multi-wavelength approach. 
Here we present  the photometric and kinematic characterization of
two groups, USGC U268 and USGC U376 (U268 and U376 hereafter) located in different regions of the Leo cloud. 
 We revisit the group membership, using  results coming from recent red-shift surveys, and  we
investigate their substructures.    
U268, composed of 10 catalogued members and 11 new added members,  has a small fraction ($\approx$24\%) of early-type galaxies (ETGs).  
U376 has 16  plus  8 new added members,  with  $\approx$38\% of ETGs.  We  find the presence of significant 
 substructures in both groups suggesting that they are likely accreting galaxies.  U268 is located in  a more loose  environment
 than  U376. 
For each member galaxy, broad band integrated and surface photometry have been obtained in
far-UV (FUV) and near-UV (NUV)  with {\it GALEX}, and in $u,g, r, i, z$ (SDSS) bands.  
H$_{\alpha}$ imaging and 2D high resolution kinematical data have been obtained 
using PUMA Scanning Fabry-Perot interferometer at the 2.12 m telescope in San Pedro M\'artir,  (Baja California, M\'exico).  
We improved the galaxy classification and we detected morphological and kinematical distortions
that may be connected to either on-going and/or past interaction/accretion events or
environmental induced secular evolution.
 U268 appears more active than U376, with a large fraction of galaxies showing interaction
signatures (60\% vs. 13\%). The presence of bars among late-type galaxies is $\approx$10\% in   U268
and $\approx$29\% in U376. The cumulative distribution of (FUV - NUV) colours  of galaxies in U268 is significantly
different than that in   U376 with galaxies in   U268  bluer than those in   U376.
In the $(FUV-r$ vs. $M_r)$  and  $(NUV-r$ vs. $M_r)$  planes no members of   U268 are found in the `red sequence', even early-type
galaxies lie  in the `blue sequence' or in the `green valley'.  
Most (80\%) of the early-type members in   U376
inhabit the `red sequence',  a  large fraction of galaxies, of different morphological types, 
are located in the `green valley', while the `blue sequence' is under-populated
with respect to   U268.  

\end{abstract}


\section{Introduction}
In the Local Universe, the distribution of galaxies is bimodal in the color space 
and relates to galaxy morphology \citep[e.g.][]{Strateva01, Balogh04}. 
In the  ($u$ - $r$) vs. M$_r$ color Ð magnitude diagram (CMD), early-type quiescent galaxies 
populate the `red sequence' and  late-types, with active star formation, the blue one
\citep[e.g.][]{Baldry04}.  
This bimodality is ubiquitous, extending from the general field, to groups 
and to clusters \citep[e.g.][]{Lewis02}.
The physical origin of this  bimodality is still under debate. However,
there are strong evidences that the two distinct
populations are the result of transformations driven by the  environment. 
The galaxy evolution from the blue to the `red sequence', i.e. from star forming to quiescent galaxies, occurs via
transition that leads galaxies in an intermediate zone of the CMD,  
the `green valley' \citep{Martin07}.  Since the color bimodality is mainly driven by the {\it on-off} of
star formation activity and  the UV bands are an excellent tracer
of recent  star formation, it is not surprising that the galaxy
color sequences are especially well separated in the UV vs. optical
CMD \citep[e.g.][]{Wyder07}. 
Investigating the mechanisms governing the {\it on-off}
of the star formation in different environments is one of the main topic of
the observational cosmology.
In this respect, galaxy groups are important mainly from two facts. 
First, they contain a large fraction ($\sim$ 50-60\%) of the galaxies 
in the Local Universe \citep[e.g.][]{Geller83, Tully88, Ramella02, Eke04, Tago08}.   
Second, since the galaxy velocity dispersion in groups is comparable to
the inner velocity dispersion of individual galaxies, 
severe galaxy-galaxy interaction, accretions and merging are favoured  in groups 
with respect to clusters  \citep[e.g.][]{Mamon92, Schweizer96, Zablu98}.  
Interactions, accretions and mergers {\bf induce star formation \citep[e.g.][]{Kennicutt96, Ellison08, Ellison10, Scudder12}}.
The fraction of star-forming galaxies in groups lies 
 between  clusters ($<$ 30\%) and  field ($>$ 30\%) \citep{Calvi12}.  
This fact suggests the existence of pre-processing mechanisms 
acting on galaxies during the formation/virialization of groups,  
partly quenching star formation by ejecting the interstellar medium  via starburst, 
AGN or shock-driven winds \citep[e.g.][]{Dimatteo05} well before a group 
eventually fall into  a cluster \citep[see e.g.][and references therein]{Zablu98, Bai10}.
Ram-pressure stripping  phenomena may also have a role e.g. in removing the hot gas halos 
\citep{Rasmussen06, Kawata08, McCarthy08} i.e. the fuel for possible future star formation. 
Galaxy-galaxy interaction, accretions, and merging, ultimately control the members' morphology 
\citep{Toomre72, Barnes96, Barnes02} typically turning late-type galaxies into ETGs. 
When and how the disks are transformed  in spheroids, the star formation is quenched
and stellar mass is accreted in the massive galaxies  
need to be further investigated.

In spite of the fact that groups are the most common environmental phase experienced by
galaxies, and able to affect galaxy populations until transform their morphology,  
our understanding of the groups is still scarce. We intend to map the different
forms of interactions within groups with the aim of 
investigating the co-evolution of groups and their members. This  requires a 
considerable observational effort, starting from the group definition, considering 
the presence of  possible substructures, and a correct galaxy morphological (and kinematical)
classification.  Multi-wavelength surface photometric  
and 2D kinematical studies are necessary to identify the effect of interaction
and accretion events, i.e.  the main galaxy transforming mechanisms, as well as the secular evolution 
and to map their effect on the UV-optical CMDs.
 
In this context,  we use Galaxy Evolution Explorer ({\it GALEX})  and
Digital Sky Survey (SDSS) imaging  to analyze  a set of groups likely spanning 
a wide range  of  evolutionary phases, from analog to the Local Group 
to systems totally virialized (Marino et al. in prep.).   
We already presented a first study of three groups dominated by 
late-type galaxies in \citet[][hereafter Paper I]{Marino10}.

  Here we present the UV and optical observations and the data  analysis of two galaxy groups in the Leo cloud
\citep{Tully88}.   
We focus in particular  on   U268 and  U376.  
Starting from their definition given in \citet{Ramella02} which identify 
bright group members, we revisit the two groups including fainter members,
provided by recent redshift surveys, sharing the same spatial and velocity extent.
For each galaxy group we investigate morphology, measure surface photometry in  the UV, and optical bands and obtain
UV - optical CMDs.
 We also  achivied   2D H$\alpha$  data cubes  
 for a fraction\footnote{Since these observations are very time consuming.} of their bright late-type members.

The paper is arranged as follows. Section~2   describes  
the characteristics of the sample. 
  Section~3 presents the photometric and kinematics observations 
and the data  reduction. The photometric   
and   kinematic results  are presented in Section~4.  
The results  are discussed and summarised in Sections~5 and 6 respectively. 
H$_0$=75 km~s$^{-1}$Mpc$^{-1}$ is used throughout the paper.

\input{tab1.tex}

\section{Sample}
 
The group sample has been selected  from the catalog of \citet{Ramella02}
which lists 1168 groups of galaxies  covering  4.69 steradians to a limiting magnitude 
of  m$_{B}\approx$15.5.
In order to make a detailed morphological and photometric 
analysis, we have chosen only groups within 40 Mpc   (V$_{hel} <$ 3000 kms$^{-1}$), 
and composed of at least 8 galaxies, {\bf to map from intermediate to
rich groups},  minimizing contamination by interlopers and spurious galaxies. \\
The cross-match of the member galaxies of the catalog with the {\it GALEX} and 
Sloan Digital Sky Survey  (SDSS, \citet{York00}) archives led to a sample of 13 nearby groups having almost of the 
members covered by both surveys. 
 We further obtained new  NUV imaging of most of the remaining 
galaxies in the {\it GALEX} GI6 (program 017).  
The sample contains groups  having between 
8 and 47 members,  
 and  a fraction of ETGs, according to the   
Hyper-Lyon-Meudon Extragalactic DAtabase ({\tt HYPERLEDA} hereafter) 
\citep[][]{Paturel97}
 classification, varying from those found in the field   to $\approx$70\%, typical 
of dense environments.
The analysis of the entire sample
will be presented in a forthcoming paper. 

Here, we focus on 2 of the above groups, namely  U268 and U376,
 since they map different regions within the Leo cloud \citep{Tully88}, in
 particular the associations 21-12+12 and 21-1+1, respectively.
 In addition to the  UV and optical images, bright spirals  of the two groups,
  were selected for  a 2D kinematic study in order to
map the galaxy velocity field and possible kinematical distortions.    

In the catalog of \citet{Ramella02}, U268 is composed of 10 galaxies with 
$\langle V_h \rangle = $1454$\pm$67 kms$^{-1}$, apparent 
B magnitude of $\langle B_T \rangle $ = 14.26$\pm$0.98 and 
  $\sim$30\% of ETGs. 
 U376 is composed of 16 galaxies with  
 $\langle V_h \rangle$ =  1110$\pm$240 km~s$^{-1}$, apparent B magnitude of  
$\langle B_T \rangle =  $12.81$\pm$1.45 and $\sim$50\% of ETGs. 

 We also include as new members of the two groups all galaxies found in the {\tt HYPERLEDA} database
within the spatial and velocity extent defined  by the catalog of \citet{Ramella02}. 
All galaxies in the {\tt HYPERLEDA} database with velocity $\pm$ 3$\sigma$ within the group average velocity given
above and within a diameter of $\sim$ 1.5 Mpc about the group center identified by \citet{Ramella02} were included. 
 All members thus chosen and their main characteristics  are listed in Table 1.
The table provides for each group, the  name, the J2000 coordinates, the morphological
type, the foreground galactic extinction, the inclination, the  major axis diameter D$_{25}$,  the axis ratio, 
the position angle (P.A.), the  heliocentric systemic velocity and the B total 
apparent magnitude of the galaxy members. In Figure 1 (top panels), we show the 
projected spatial distributions of the group members. Galaxies are separated in B magnitudes bins and
for morphological types. ETGs,  Spirals and Irregulars, with absolute magnitudes M$_B >$ -14, -18 $<$M$_B <$ -14 and, M$_B <$ -18 are 
indicated with  squares, triangles and circles of increasing size, respectively.
In U376, the ETGs appear concentrate  while the brightest Spirals form a substructure (see next section). 
On the contrary, U268 appears dominated by late type galaxies with  only three S0s in the outskirts of the group. 
\citet{Dressler88} `bubble-plots' are also overlaid (see next subsection).

\subsection{Substrucure}
The presence of substructure in a galaxy group is believed to be a signature of recent accretion and can be used as a probe
of evolution of their members \citep[e.g.][]{Lacey93, Zablu98, Firth06, Hou12}. Substructure manifests itself as a deviation in the spatial and/or velocity arrangement of
the system.

 If a galaxy group is a dynamically relaxed system, then the spatial distribution of galaxies should be approximately spherical and the velocity distribution a Gaussian. The presence of substructure indicates a 
 departure from this quasi-equilibrium state. Substructure is indicated by at least one of the following
 characteristics: (1) significant multiple peaks in the galaxy position distribution, (2) significant departures 
 from a single-Gaussian velocity distribution, and (3) correlated deviations from the global velocity 
 and position distributions.  The 1D (radial velocity histogram) and 2D (projected sky positions) tests 
 could be questionable for galaxy groups, due to the relatively small number of galaxies members.
The Anderson-Darling  test, that \citet{Hou09} report as a reliable tool to detect departures 
from Gaussian velocity distribution in small data sets, applied to the  heliocentric radial velocity 
distributions of  our two groups, does not significantly depart from normality.
 
The 3D tests use both velocity and spatial information to find substructures. Pinkey et al. 96 determined
 that the \citet{Dressler88} test (DS test hereafter)  is the most sensitive test  for systems 
 with as few as 30 members.  \citet{Zabludoff98, Firth06, Hou12}  show that the DS test is suitable
  to be applied to groups with a minor number of members. 
The DS test identifies a fixed number of nearest neighbors on the sky around each galaxy, 
computes the local mean velocity and velocity dispersion of the subsample, and compares 
these values with the average velocity and velocity dispersion of the entire group. The deviations
 of the local average velocity and the dispersion from the global ones are summed. 
 In particular, for  the galaxy $i$, the deviation of its projected neighbors is defined as 
$\delta_i$ = (nn +1)/$\sigma_{gr}$ [ (v$_{loc}$ - $\bar{v})^2$ + ($\sigma_{loc} - \sigma)^2$]
where  $\bar{v}$ and  $\sigma^2$ are the average velocity and the velocity dispersion of the entire group,
v$_{loc}$ and $\sigma_{loc}$ are the local ones, and nn is the numbers of the nearest neighbour.
The total deviation for the group is defined as the sum of the local deviations 
$\delta_i$, $\Delta$ = $\sum_i^N\delta_i$,
where N is the number of the group members. 
If the group velocity distribution is close to  Gaussian and the local variations are only random 
fluctuations, $\Delta \simeq$ N, whereas $\Delta >$ N is an indication of probable substructure. 
To compute $\delta$, we set the number of neighbors at  4 $\approx$  N$^{1/2}$ \citep[see e.g.][]{Silverman86}.  
Since $\delta_i$ are not statistically independent and velocity distribution could not be Gaussian 
even if there are no subgroups, it is necessary to test the reliability of the $\Delta$ statistic by 
comparing it with a Monte Carlo analysis.
The velocities were randomly shuffled among the positions and the $\Delta$ was 
recomputed 10000 times producing the probability that the measured  $\Delta$ is a random result. 
The significance of having substructure (the p-value) was quantified by the ratio of the 
number of the simulations in  which the value of $\Delta$ is larger than the observed value, 
and the total number of simulations (N($\Delta_{sim} > \Delta_{obs})/N_{sim}$). 
A small  p-value, corresponds to  a high significance of  substructure detection.
In the top panels of Figure 1, we also over plot the Dressler-Shectman `bubble-plots' 
 on each galaxy in the group as a circle whose radius scales with e$^\delta_i$. 
 Larger circles indicate larger deviations in the local kinematics compared to the global one. 
 A local grouping of galaxies with similarly large circles may indicate a kinematically distinct system,
 i.e. a substructure.
Both  the groups present substructures, $\Delta$ is greater than N at a confidence 
level  $>$ 95\% and $>$99\% respectively (see Tab~\ref{ds}).

\begin{table}
\centering
\caption{Results of the Dressler-Shectman substructure test.}
\label{sub}
\scriptsize
\begin{tabular}{lllll}
\hline\hline
Group & N$_{gal}$  & $\Delta/N_{gal}$ &P-value \\
\hline
U268 & 21 & 1.37  & 0.037 \\
U376 & 24 & 1.63 &0.001\\
\hline
 \end{tabular}
 \label{ds}
\end{table}

\begin{figure*}
\vspace{-0.5cm}
  \begin{tabular}{cc}
  \vspace{-0.7cm}
 \includegraphics[width=7cm]{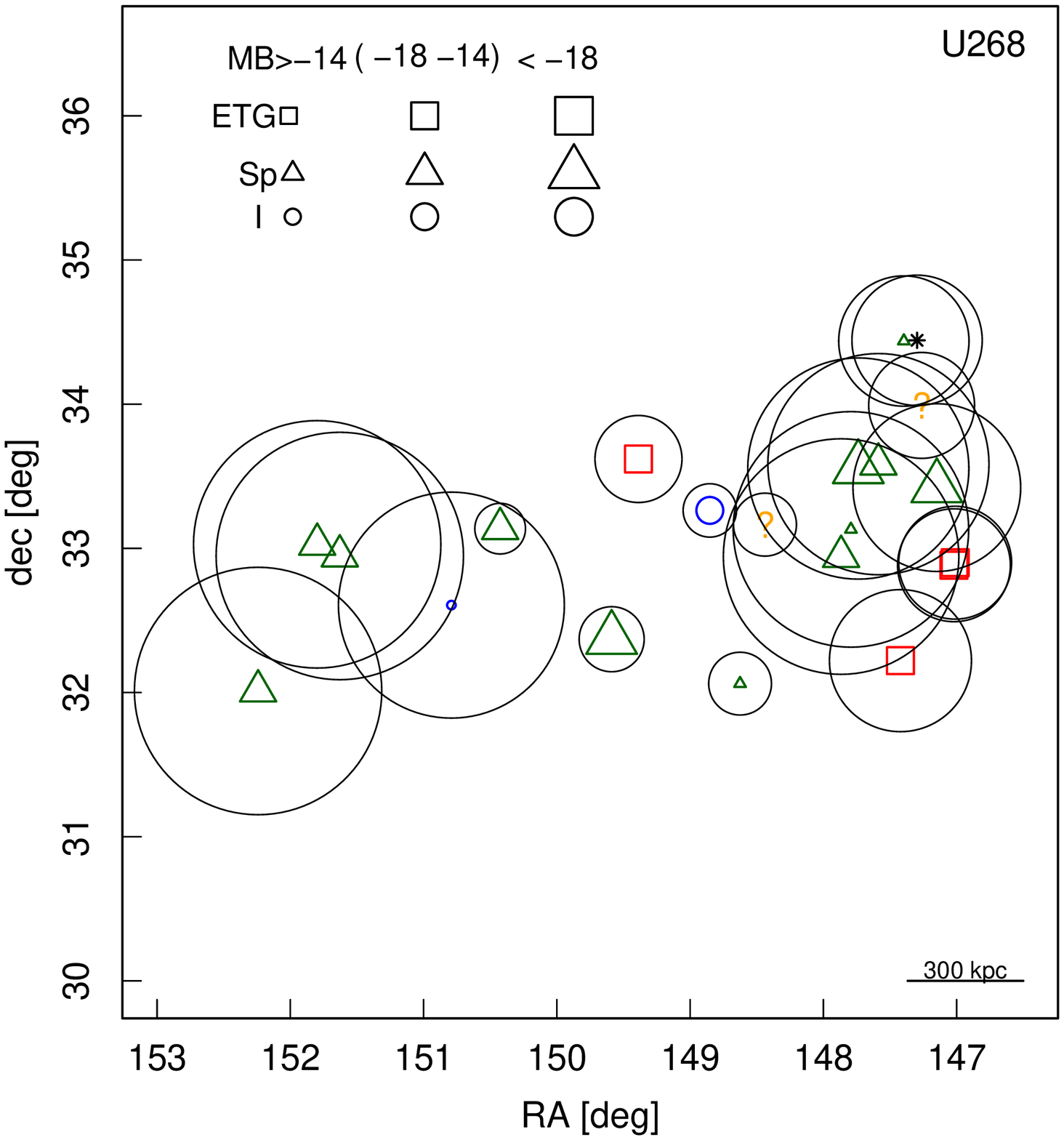} &
  \includegraphics[width=7cm]{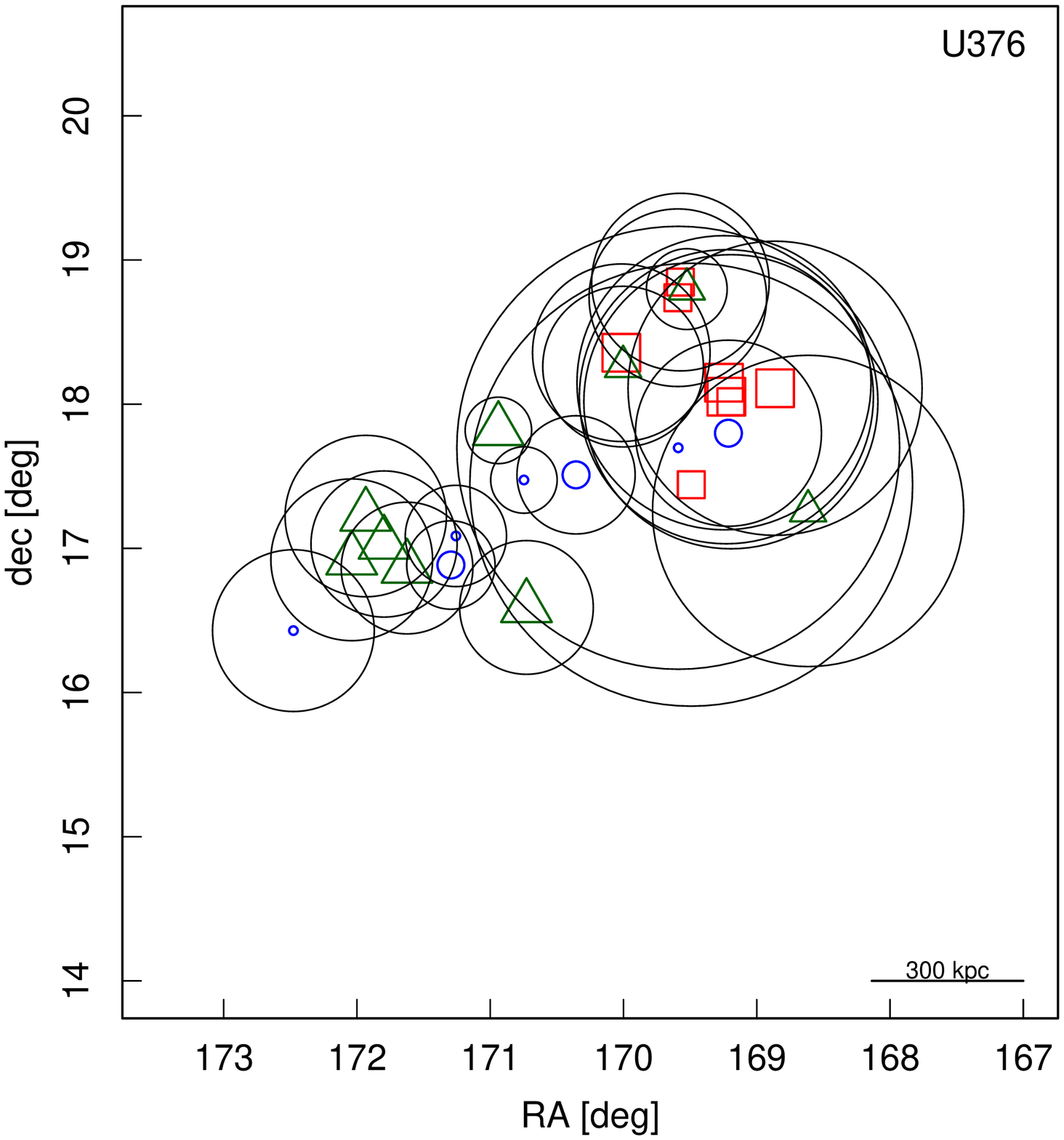}\\
  \vspace{-0.5cm}
  \includegraphics[width=6.cm,angle=-90]{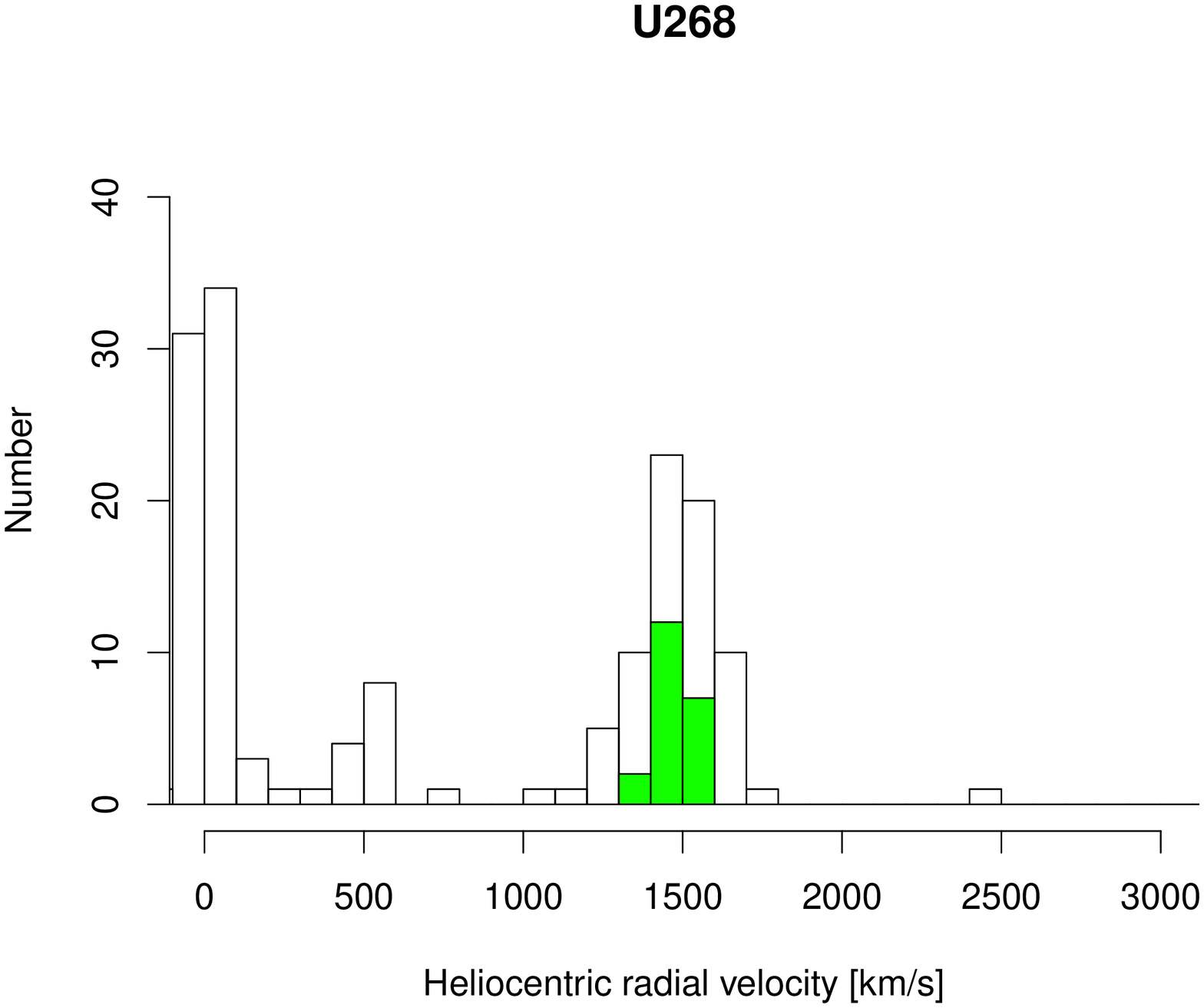} &
  \includegraphics[width=6.cm,angle=-90]{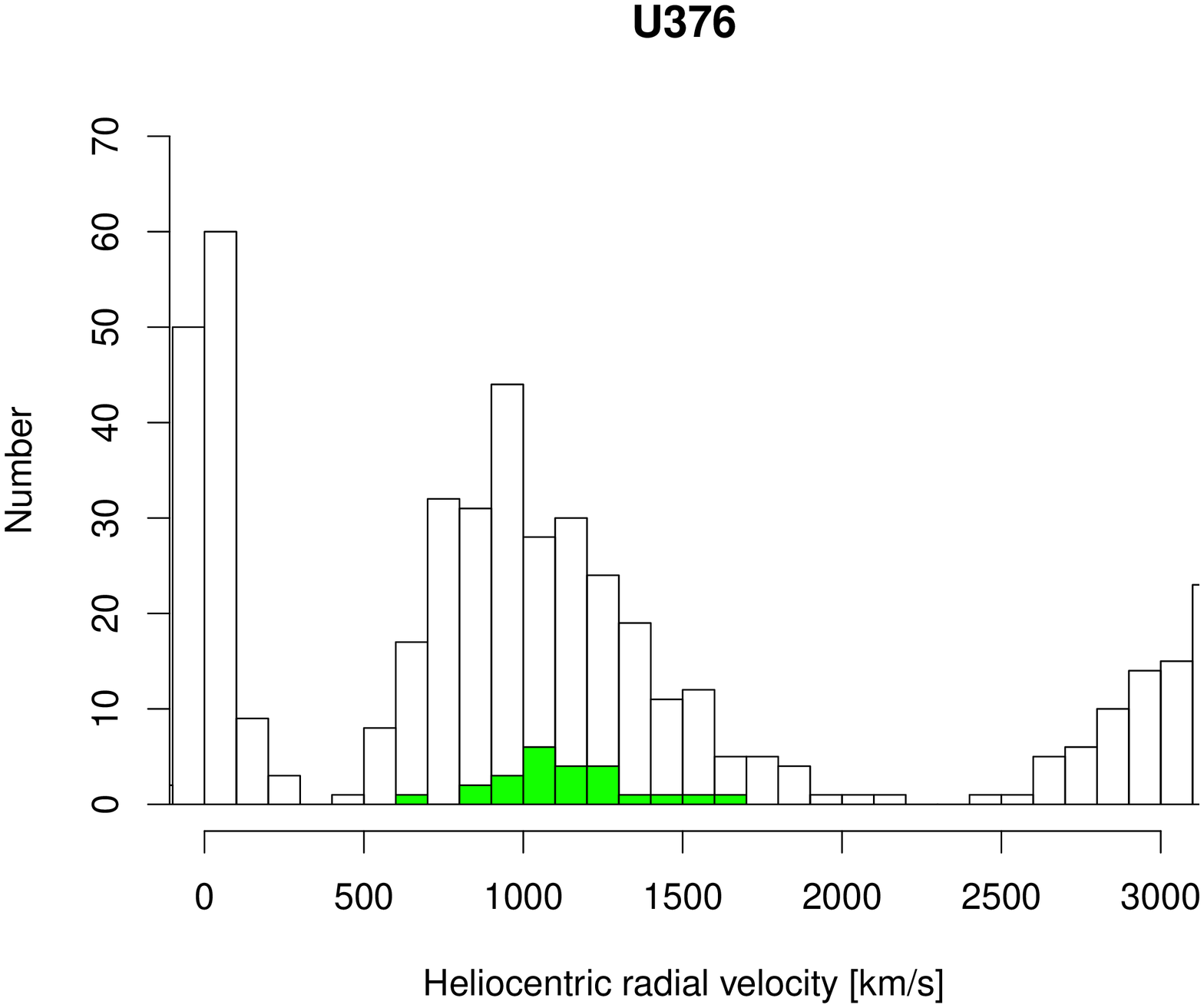} \\
\vspace{-0.3cm}
   \includegraphics[width=7.5cm,angle=-90]{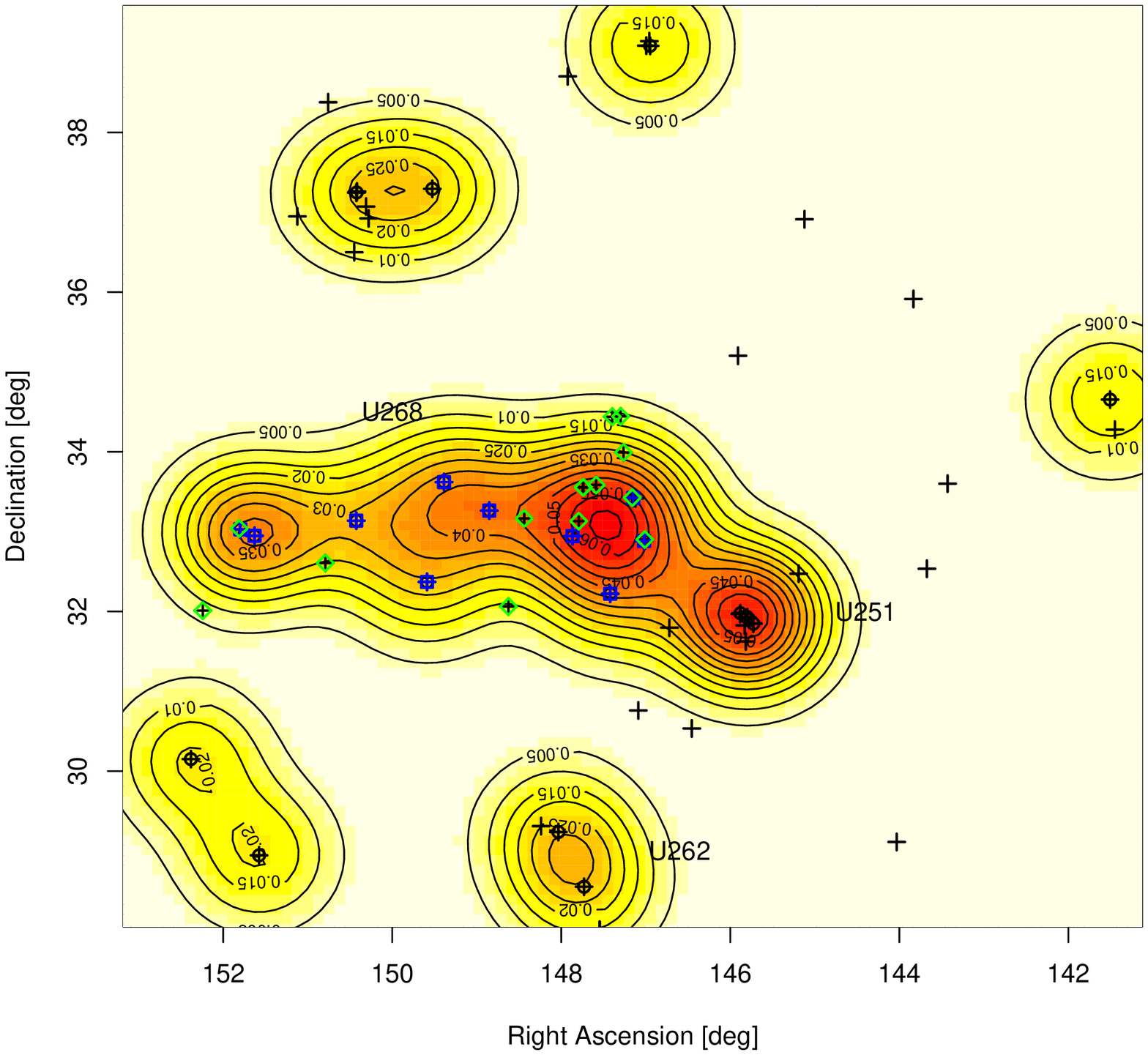} &
    \includegraphics[width=7.5cm,angle=-90]{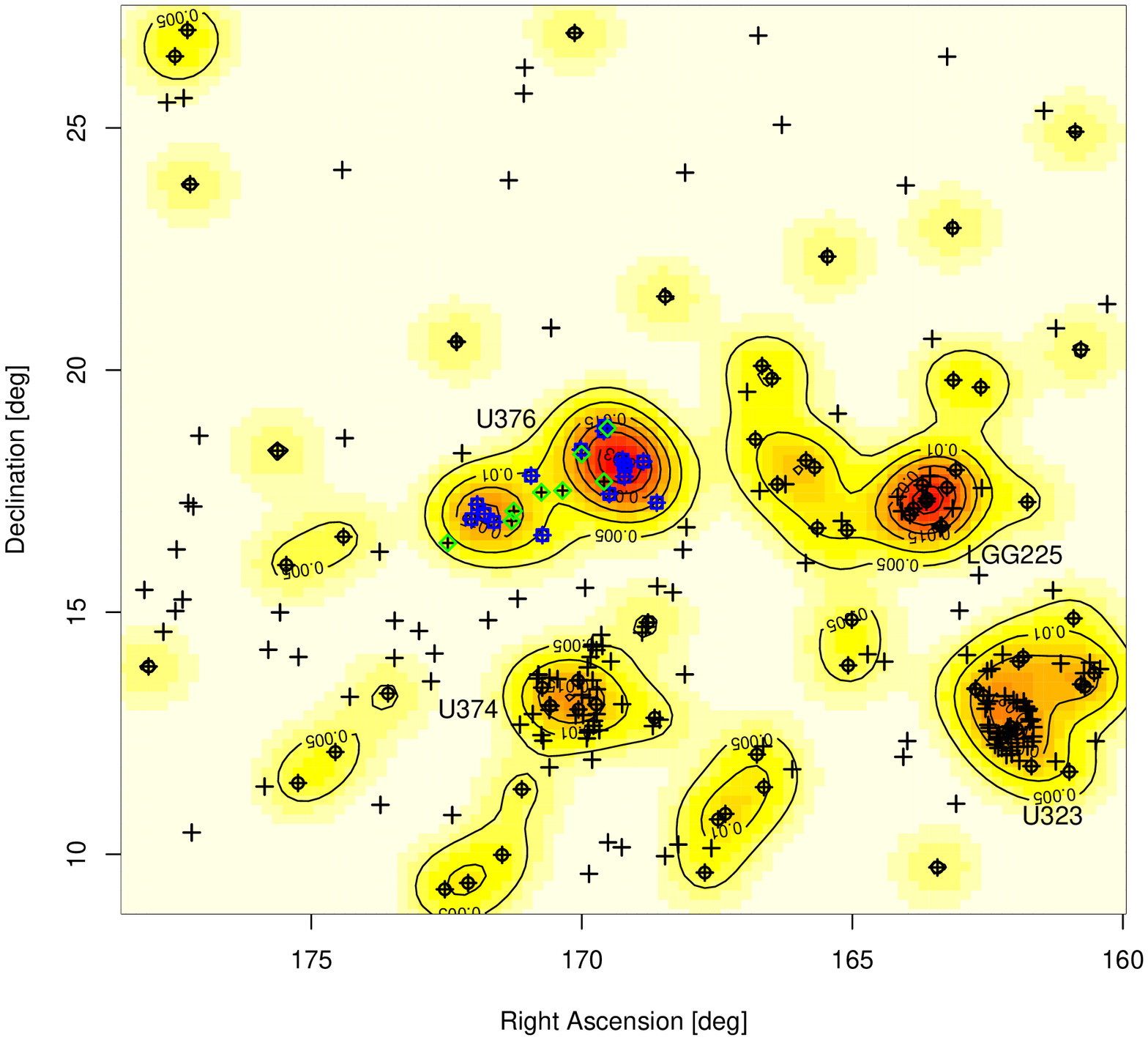}
    \end{tabular}
 \caption{({\it Top panels})   
Spatial distribution of galaxies we consider as  group members. 
Galaxies are separated in B magnitude bins and for morphological type.  Unknown morphological types with known B magnitudes are 
labelled with "?". The smallest symbols are galaxies with no B magnitudes. Black asterisks  refer to
galaxies with no B magnitudes and no morphological types. \citet{Dressler88} `bubble-plots' are also overlaid.
 ({\it Middle panels}) 
 Histogram of heliocentric radial velocity in the range 10-3000 kms$^{-1}$ 
of galaxies within a box of 4 Mpc centred on the respective B brightest  member of the two groups. The width of the velocity bins is 100 kms$^{-1}$.
Green bins show the velocity extent  of the members of the two groups. 
({\it Bottom panels}) Spatial distribution of galaxies {\bf (black plus)} within the velocity extent defined by the members in the catalog of \citet{Ramella02}.   
The two squares are  
centred on NGC 3003 and NGC 3067 in U268 and U376 respectively.  Blue square symbols show  the members of the two groups listed in the catalog of \citet{Ramella02}, green diamond ones indicate the added   members.
 The 2D binned kernel-smoothed number density contours  for the galaxies with m$_B$ $\leq$ 15.5 (circle + cross) are also shown.  } 
 \label{histVh}
  \end{figure*}
  
  \subsection{U268 and U376 environment} 
  
 Are the large scale environments of the two groups in the Leo cloud
similar? To investigate  the environment of each  group we considered
the  galaxy distribution within a box of 4 Mpc, about two times the size of a
typical group, centred on the  brightest galaxy {\bf in B band} of each group 
(NGC 3003 and NGC 3067 in U268 and U376, respectively).   
From the {\tt HYPERLEDA} database we select galaxies with a heliocentric radial velocity within  
$\pm$ 3$\sigma$ of  the group mean velocity given in the catalog of
\citet{Ramella02} finding  64  and 268  galaxies in U268 and U376 box, respectively 
(+ symbols in  Figure \ref{histVh}, bottom panels, see also Appendix B).   
In the middle panels of Figure \ref{histVh} we highlights (filled bins) 
the velocity distribution of group members  given in Table~1
 superposed to the distribution of the heliocentric radial velocity, 
in the range 10 and  3000 kms$^{-1}$, of galaxies in the box of 4 Mpc. 
 
Among these galaxies, to mark the `bone' of the galaxy groups, we selected 
 only galaxies  more luminous than 15.5 B-mag, i.e the magnitude limit 
of the galaxies in \citet{Ramella02}. On this sample we performed a density analysis. 
The 2D binned kernel-smoothed number  contours maps are shown 
in the bottom panels of  Figure \ref{histVh}). Density in the box is
colour coded. The highest densities correspond to the red regions; 
density levels above 0.005 are in yellow.  
Blue squares show the members of the two groups in the catalog of 
\citet{Ramella02}, green diamonds indicate
the new members we added as explained in  section 2.1.

Density contour maps show that galaxy members of U268 and U376  
are, in projection, part  of elongated structures, revealing sub-structures, 
as found in the previous section. The quite large   separation   of $\sim$ 1.6 Mpc
among the members, suggests that U268 is a very loose configuration, 
at the borderline with the field. 
In U268 the two substructure, identified in the previous section,
 appear also likely connected to another structure which includes
USGC U251 detect by \citet{Ramella02} as a single group.   
 
Galaxy members of U376 have a more {\bf patchy} configuration than U268. 
The environment appears  rich of galaxies and crowded by tight associations. 
Several other galaxy groups (USGC U323 and USGC U374),   
including LGG 225, already analysed in Paper I,  
have been identified in the nearby surroundings  of U376
(Figure~\ref{histVh}, bottom right panel).

\section[]{Observations and data reduction}

 \subsection{UV and optical data}	   	    
The UV imaging was obtained from {\it {\it GALEX}} \citep{Martin05, Morrissey07} GI6-6017 (PI A. Marino) and  archival data   
 in two ultraviolet bands, far-UV (FUV, 1344 -- 1786\AA) 
 and near-UV   (NUV, 1771 -- 2831\AA). The instrument has a very wide field of view (1\degr.25 
 diameter) and a spatial resolution of $\approx$ 4\farcs2  and 5\farcs3 FWHM in FUV and NUV
 respectively, sampled with 1\farcs 5$\times$1\farcs 5 pixels \citep{Morrissey07}.
			   	    
The exposure times (see Table \ref{tab2}) for most of our  sample are $\sim$100 sec (limiting AB magnitude in FUV/NUV of $\sim$ 19.9/20.8 \citep{bianchi09}). NGC 3011 in U268 and  a few galaxies in U376  have an exposure time $\sim$10 times
longer ($\sim$ 2.5 AB mag fainter limit).    We used FUV and NUV   intensity images   to compute integrated photometry of the galaxies and light profiles, as described in Sect. 4.1.\\
 In addition, we used optical Sloan Digital Sky Survey (SDSS) archival data  in
  the u [2980-4130 \AA], g [3630-5830 \AA], r [5380-7230 \AA], i
[6430-8630 \AA], z [7730-11230 \AA] filters \citep{Ade08}.\\
Figure \ref{A1} and Figure  \ref{A2} display  colour composite UV (FUV blue, NUV yellow, 
left panels) and optical (SDSS, g blue, r green, i red)  images  of the galaxy 
members of   U268 and  U376 respectively as listed in the catalog of \citet{Ramella02}.
All UV and optical colour composite images of the added galaxy members but PGC 2016633 are shown in
Appendix A.

\input{tab2.tex}

\input{tab2a.tex}

 \begin{figure*}
      \includegraphics[width=15cm]{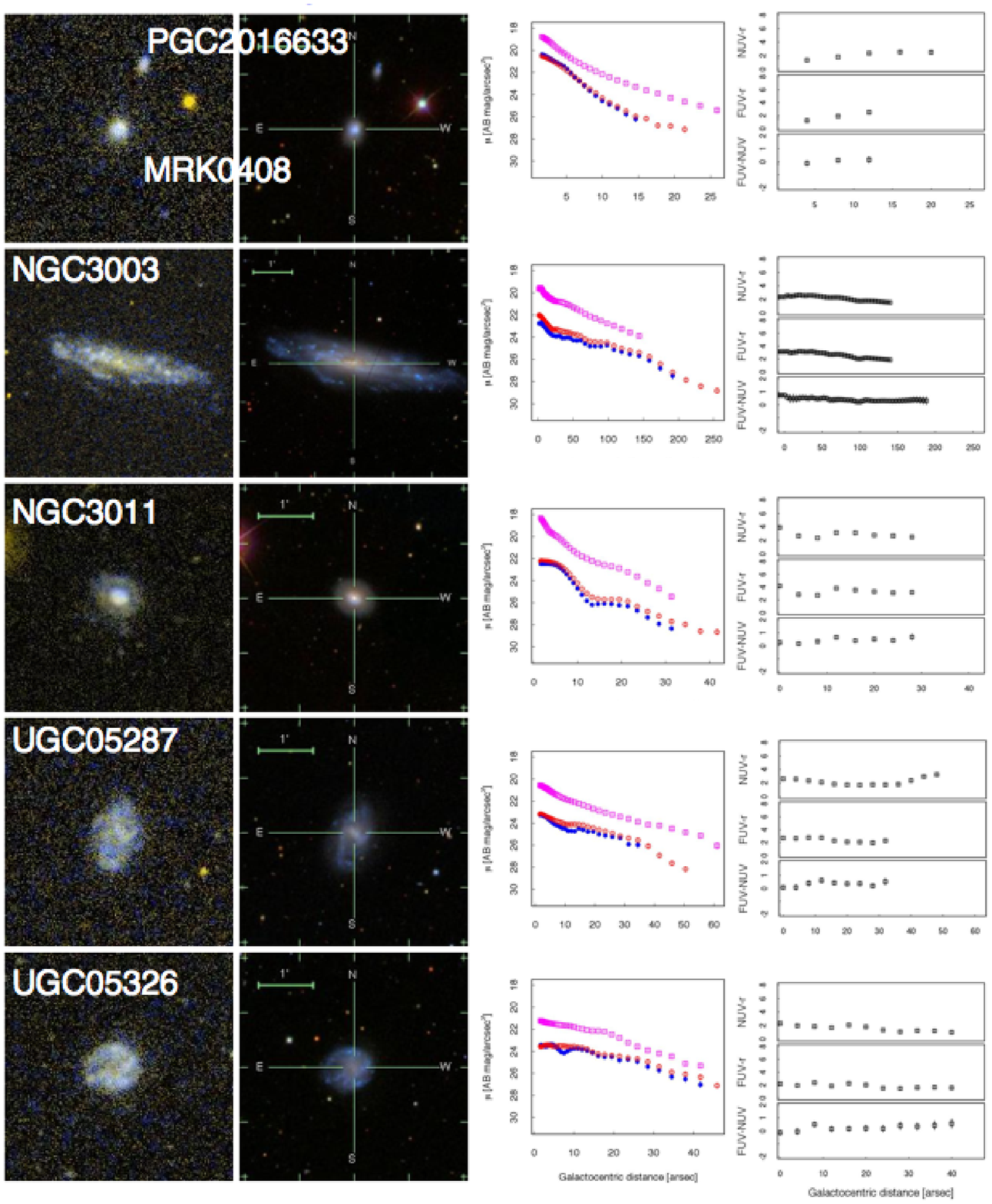}
      \caption{From left to right: colour composite UV (FUV blue, NUV yellow) and optical
     (SDSS, $g$ blue, $r$ green, $i$ red)  images of 5 members of U268; 
      UV (FUV blue, NUV red) and optical (SDSS-$r$ magenta) surface
      luminosity, and colour profiles, corrected by galactic extinction.}
     \label{A1}
    \end{figure*}
    
\begin{figure*}
    \includegraphics[width=15cm]{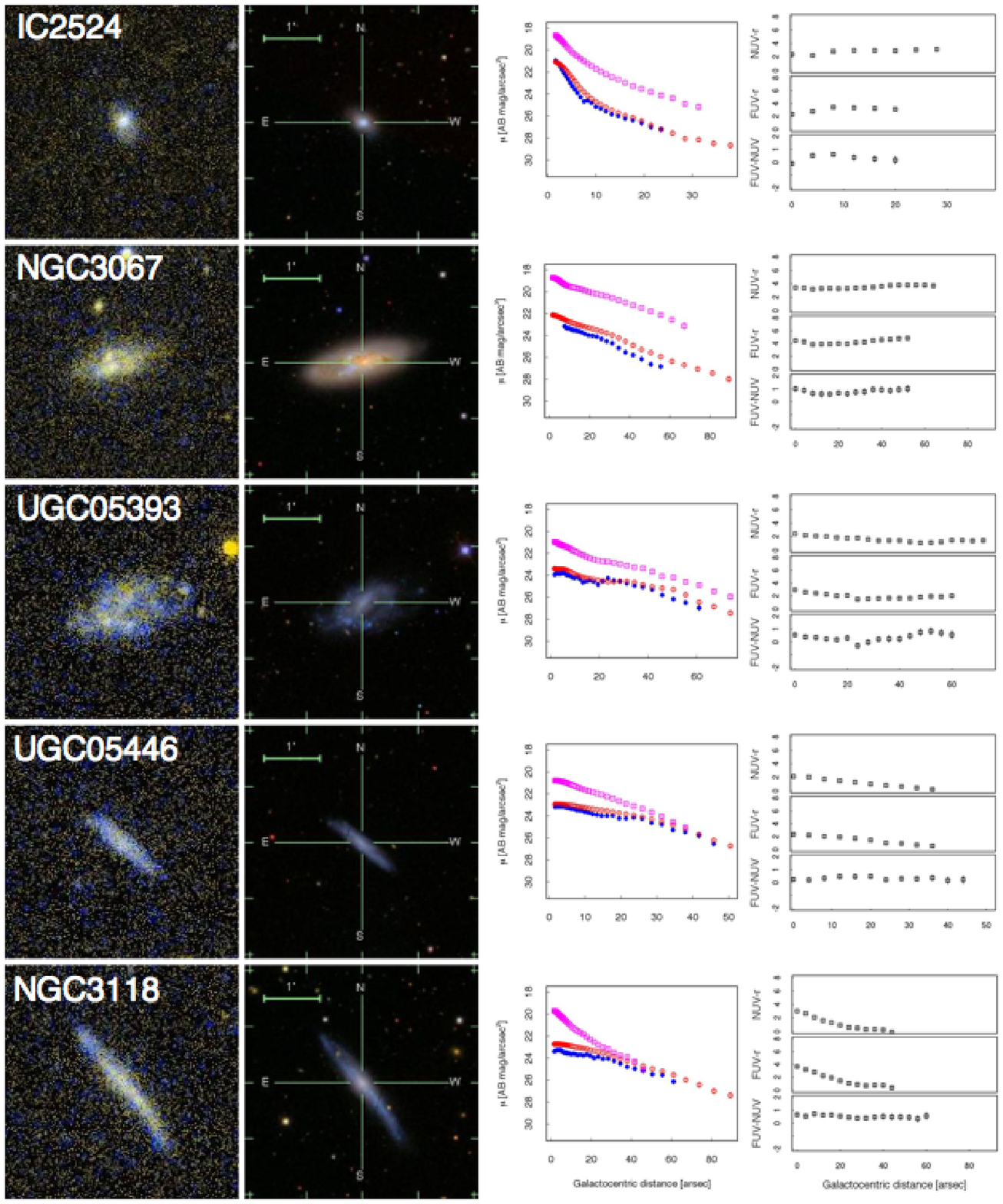}
  \addtocounter{figure}{-1}
   \caption{{Continued.}}  
      \end{figure*}	
	
\begin{figure*}
     \includegraphics[height=18cm]{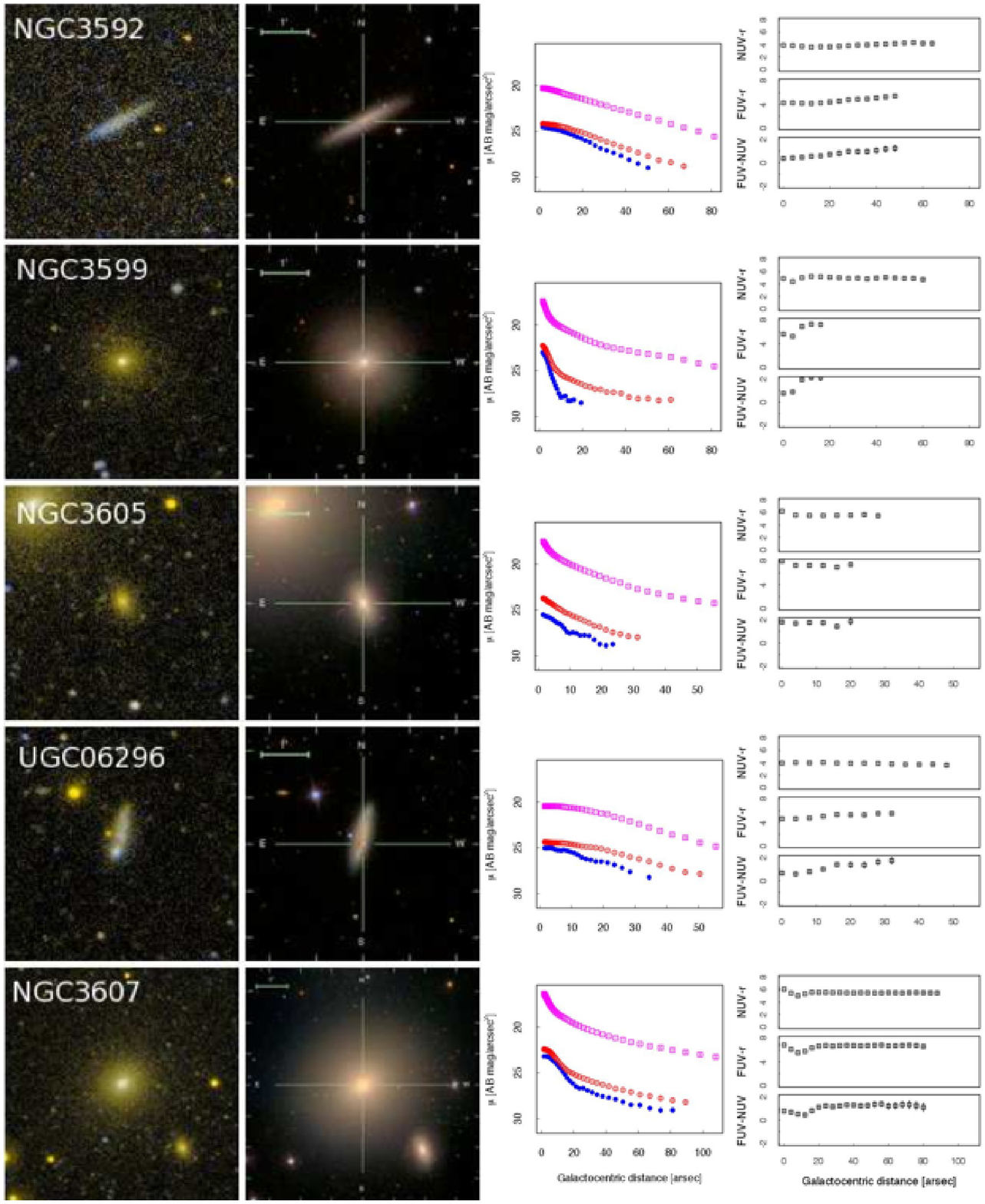}
   \caption{As in Figure 2 for U376 galaxy members.} 
   \label{A2}
   \end{figure*}	

\begin{figure*}
   \includegraphics[height=18cm]{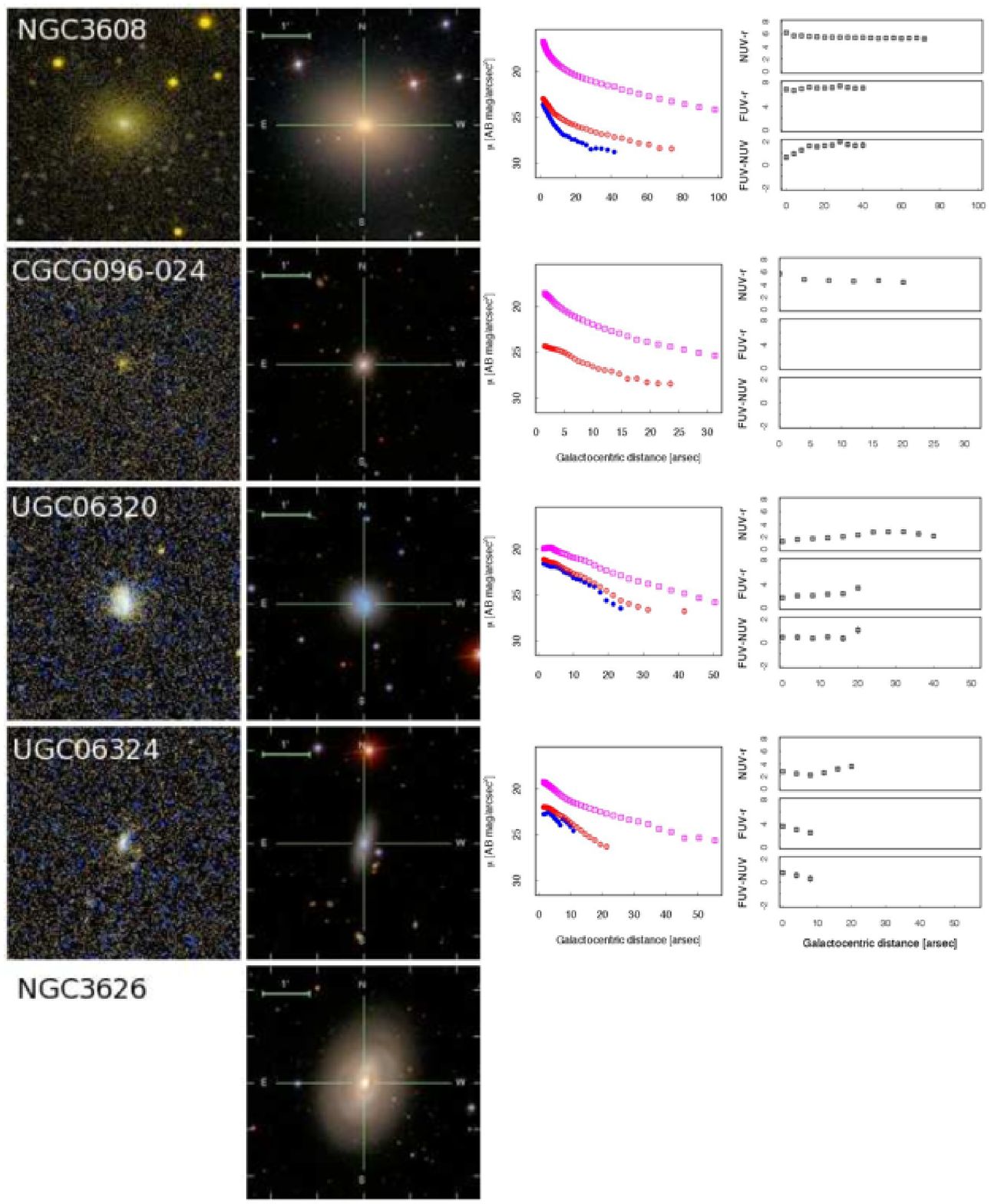}
   \addtocounter{figure}{-1}
   \caption{{Continued.}}  
   \end{figure*}	
   
\begin{figure*}
   \includegraphics[height=18cm]{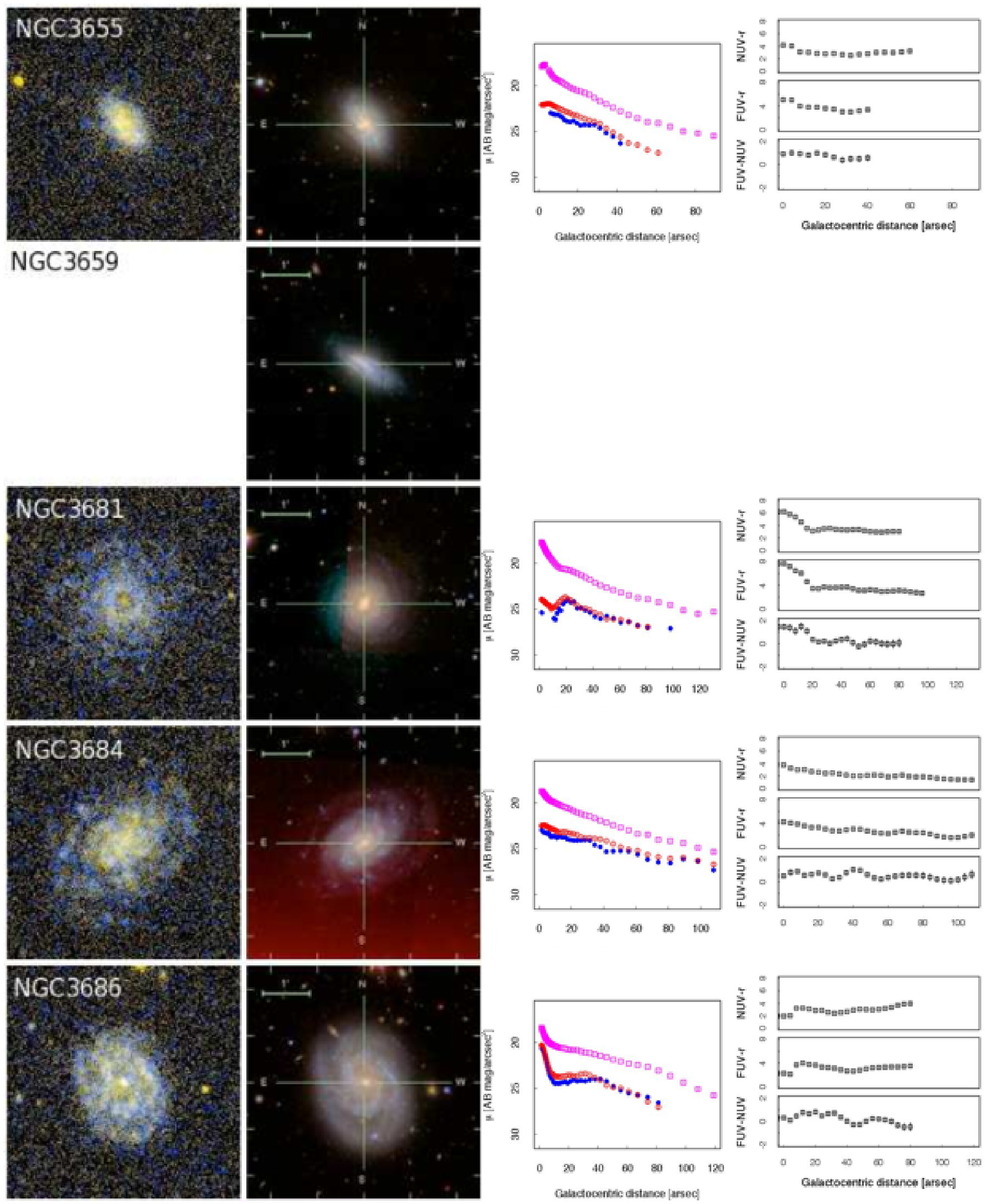}
   \addtocounter{figure}{-1}
   \caption{{Continued.}}  
    \end{figure*}	

\begin{figure*}
  \includegraphics[height=4cm]{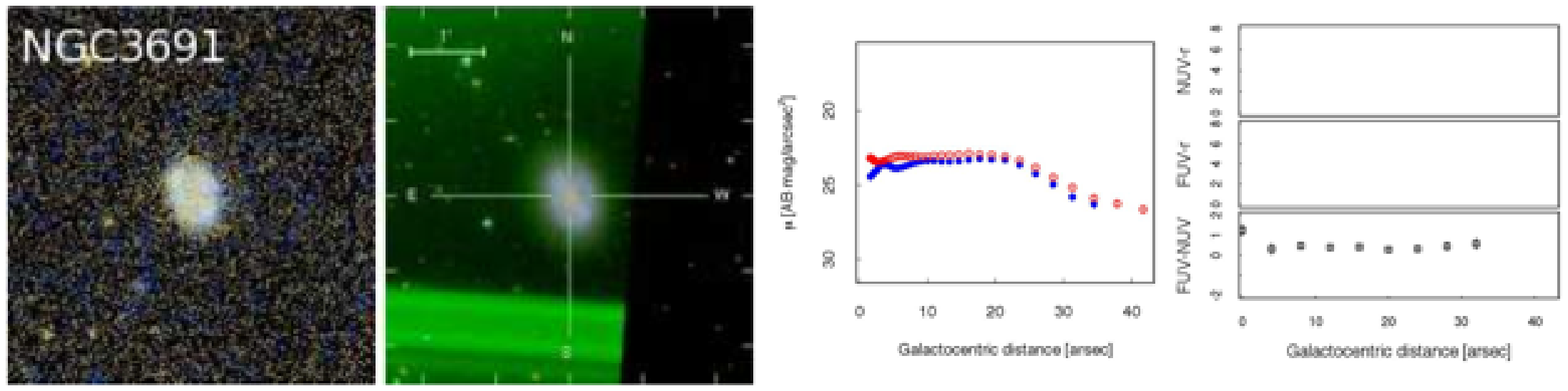}
   \addtocounter{figure}{-1}
   \caption{{Continued.}}  
     \end{figure*}	

 \subsection{2D kinematical observations and data reduction of a subset of spirals} 
 
Observations were done in 2011 February, at the Cassegrain focus of the  2.12m telescope  at the Observatorio Astron\'omico Nacional in San 
Pedro M\'artir (M\'exico), using the scanning           
 Fabry - Perot  interferometer\footnote{All Fabry - Perot data 
 (cubes, moment maps and rotation curves) are available at FabryPerot.oamp.fr, in the loose group survey}  PUMA \citep{Rosado95}. PUMA instrument uses 
a Thomson2k 2048x2048 pixels CCD detector with a pixel size of 15$\times$15$\mu$.  
The readout noise was 5.7e$^{-}$.  
 A 1500$\times$1500 px window has been used, and a 3$\times$3 binning performed, 
 giving an pixel equivalent to 1.07\arcsec  on the sky. 
Table {\bf 4} provides the journal observations and the characteristics 
of the interferometer we used.

\input{tab3.tex}

\subsubsection{Moment maps}

PUMA data have been reduced using the ADHOCw software  procedures
of Boulesteix\footnote{Available  at http://www.oamp.fr/adhoc/distribution/}, and \citet{Daigle06}\footnote{Available at http://www.astro.umontreal.ca/~odaigle/reduction/}. 
The first step, before  the  correction phase, was to perform 
the standard CCD data reduction by applying bias and flat field corrections. 
The data reduction  procedure has been extensively described in \citet{Amram96,  
  Daigle06, Epinat08a}, \citet{Epinat08b}. 
Wavelength calibration was obtained by scanning the 
narrow NeI 6599\AA\ line under the same observing conditions. 
Velocities measured relative to the systemic velocity 
are very accurate, with an error of a fraction of a channel 
width ($< 3$ km~s$^{-1}$) over the whole field. 
The signal measured along the scanning sequence was 
separated into two parts: (1) an almost constant level produced 
by the continuum light in a 15 \AA\ passband around H$\alpha$ 
(continuum map), and (2) a varying part produced by the 
H$\alpha$ line (H$\alpha$ integrated flux map). The continuum 
is computed by taking the mean signal outside the emission line.
The H$\alpha$ integrated flux map was 
obtained by integrating the monochromatic profile in each 
pixel. The velocity sampling was 19 km~s$^{-1}$. Strong OH 
night sky lines passing through the filters were subtracted 
by determining the level of emission away from the galaxies \citep{Laval87}. 
In order to improve the S/N ratio, we used  the adaptive smoothing 
(Voronoi {\bf tessellation}) as described in \citet[][and references within]{Daigle06}.   
The different maps presented in this paper has been made
adopting a final SNR of 5 or 6, depending of the galaxies. 
In any case, a rectangular smoothing of 3 channels have 
been also applied.

\begin{figure*} 
 \begin{tabular}{ll}
 \includegraphics[width=6.5cm]{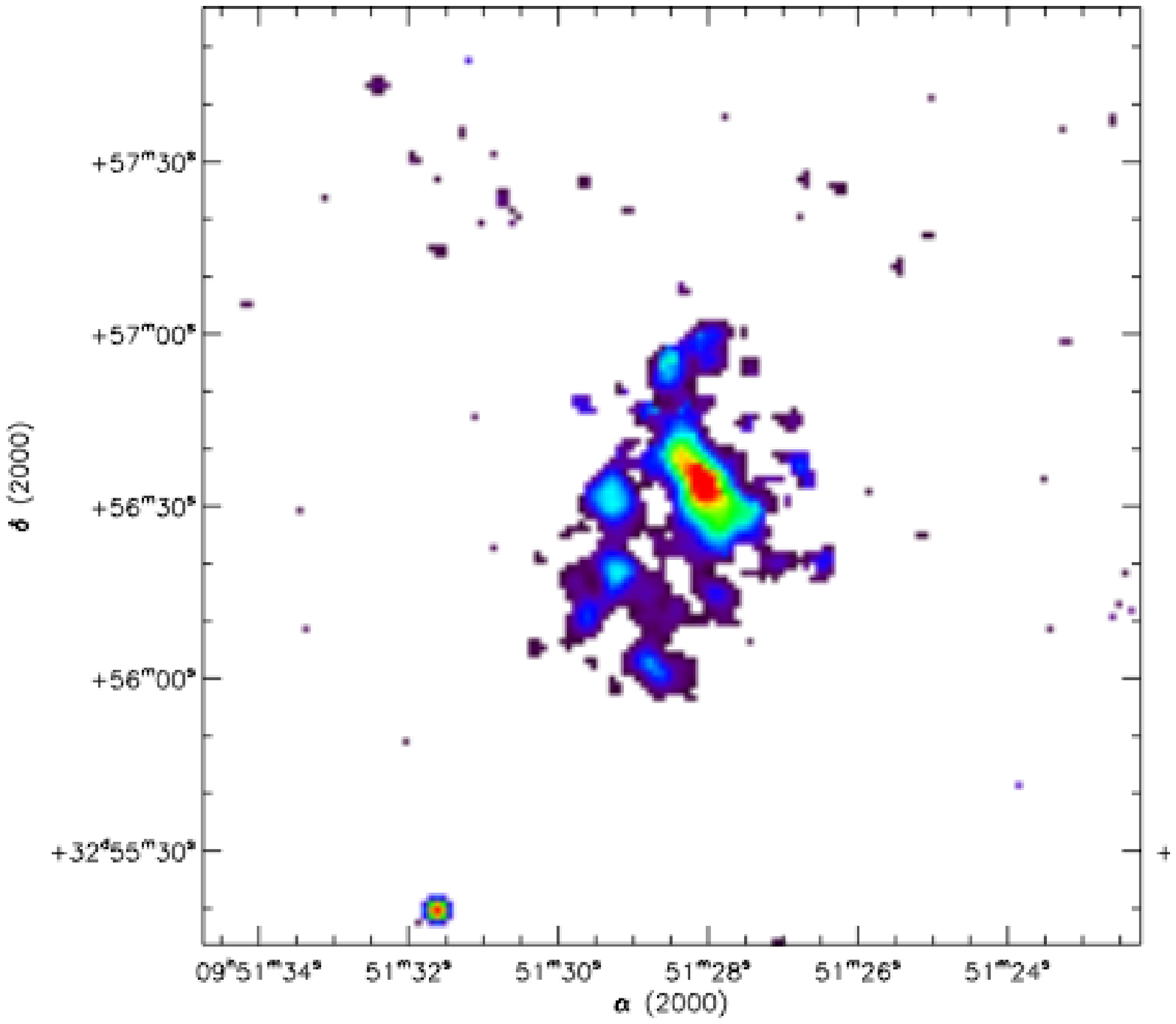}  &   \includegraphics[width=6.5cm]{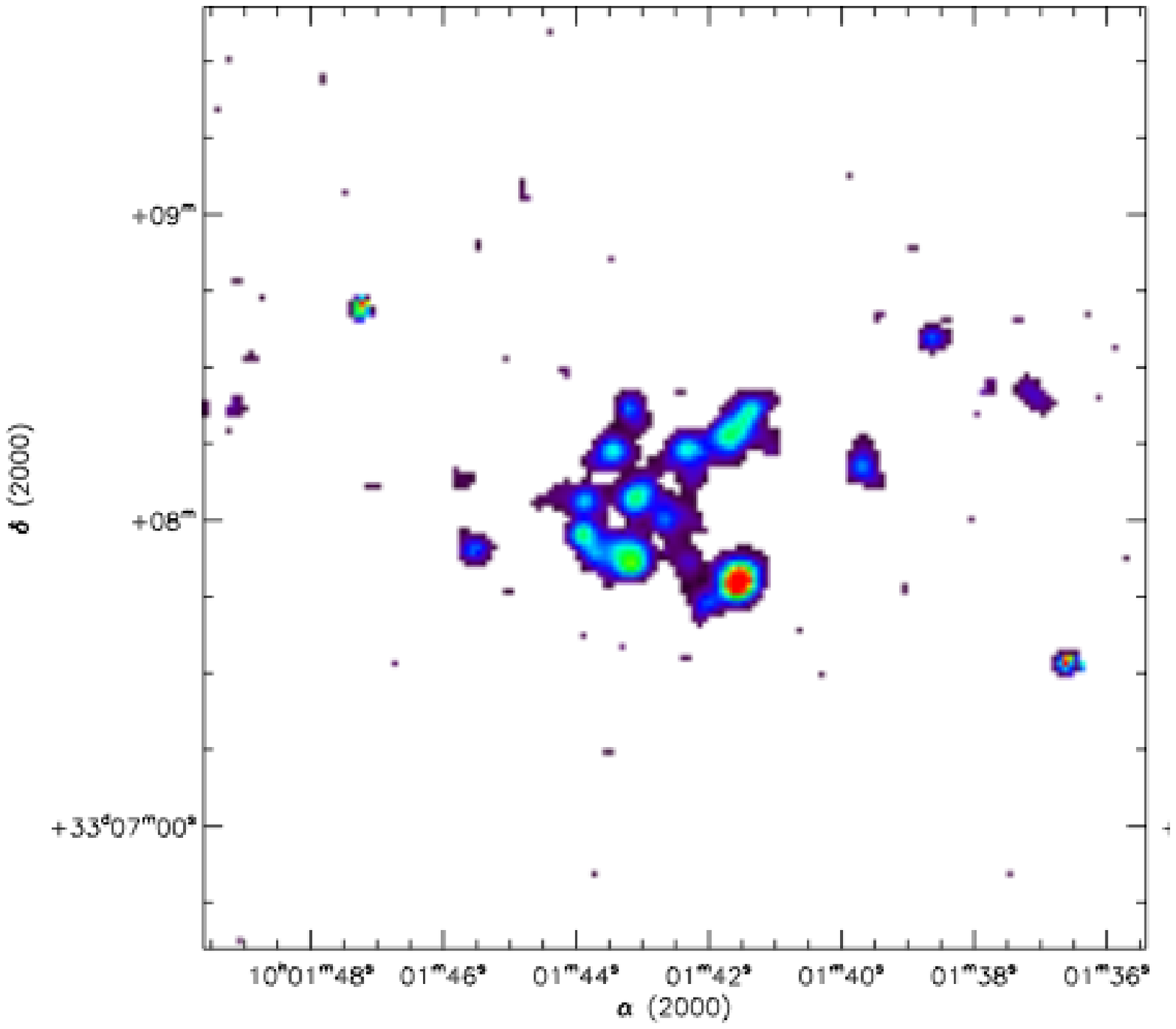} \\
 \includegraphics[width=6.8cm]{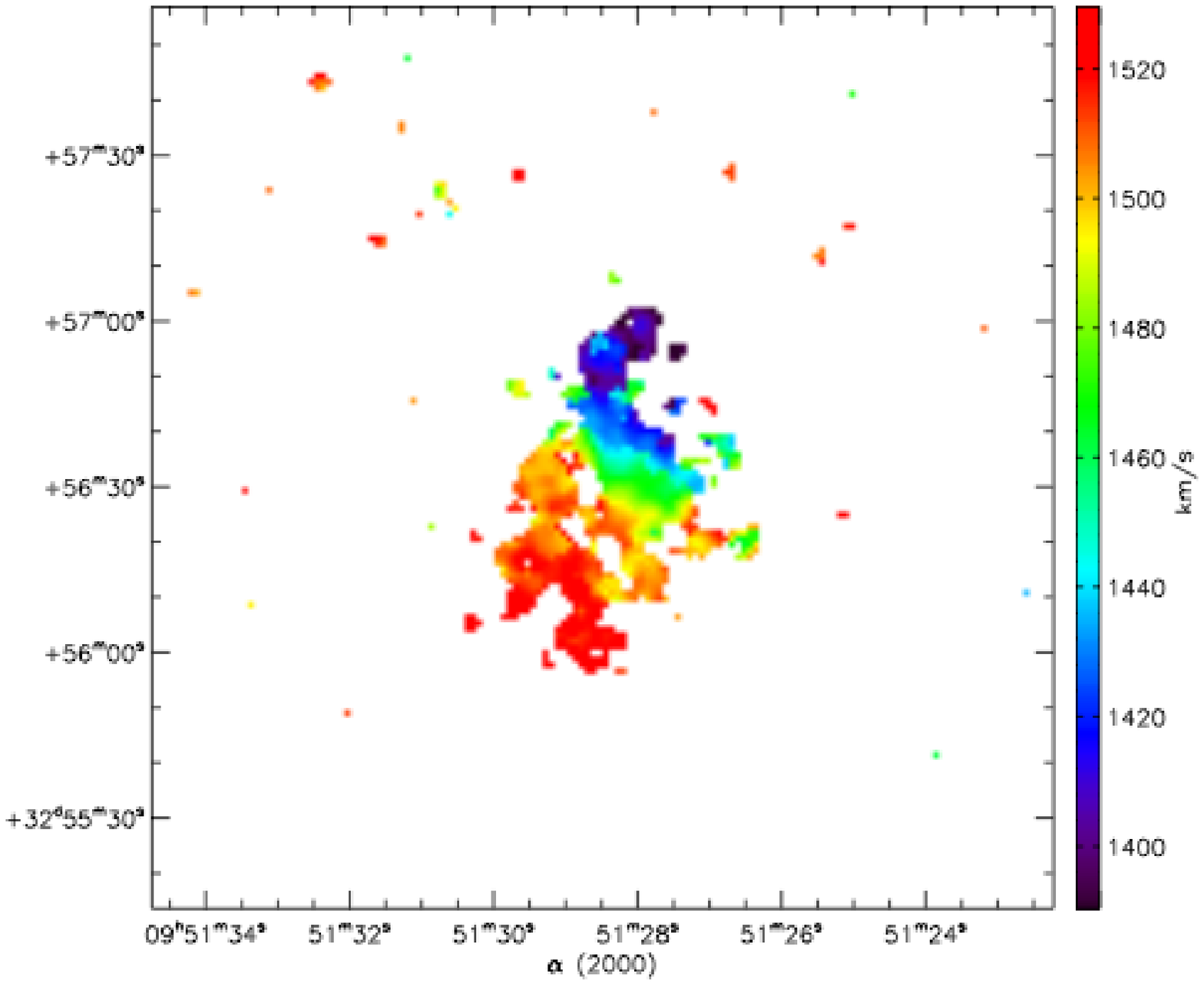}&  \includegraphics[width=6.8cm]{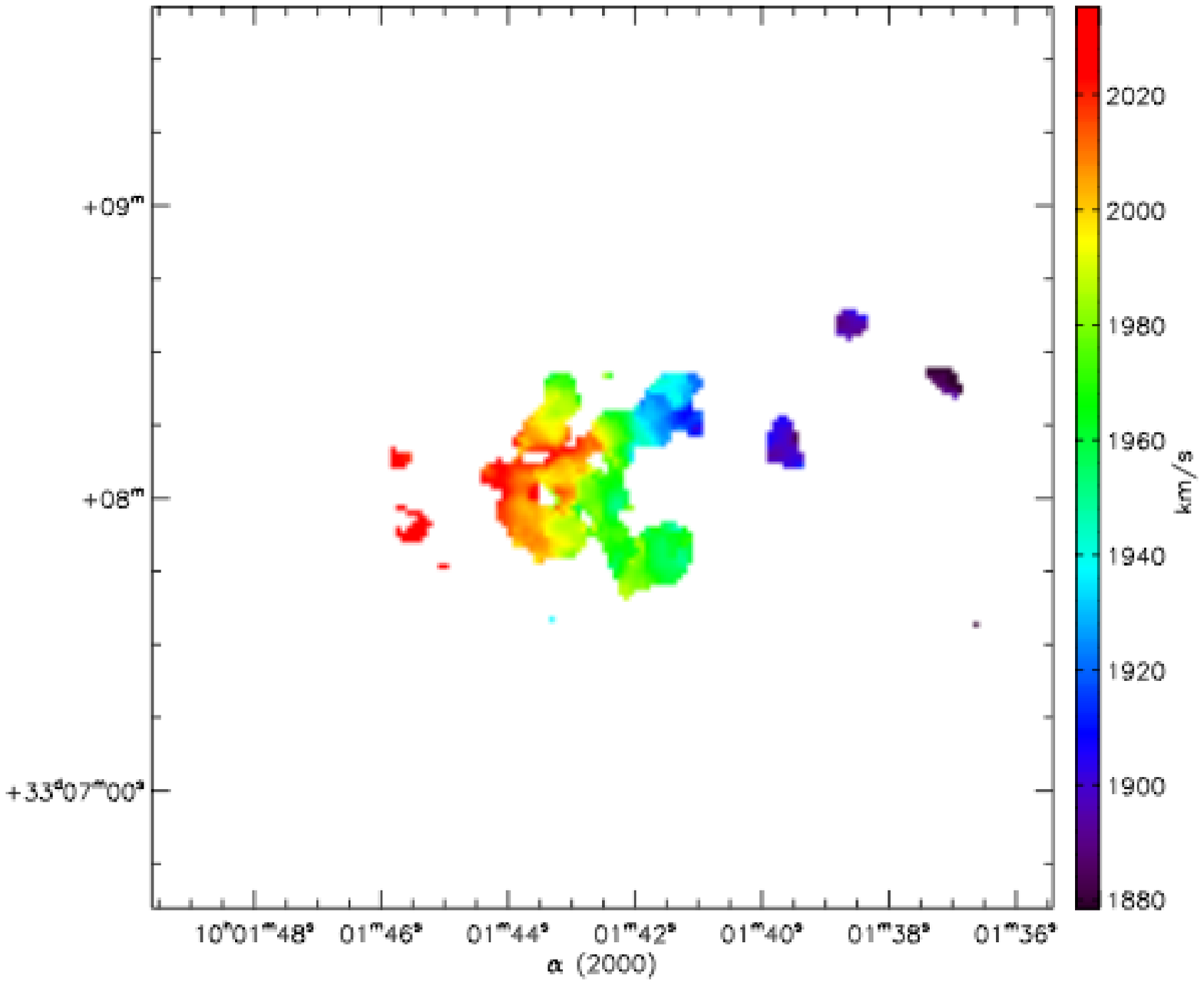}\\
  \includegraphics[width=6.8cm]{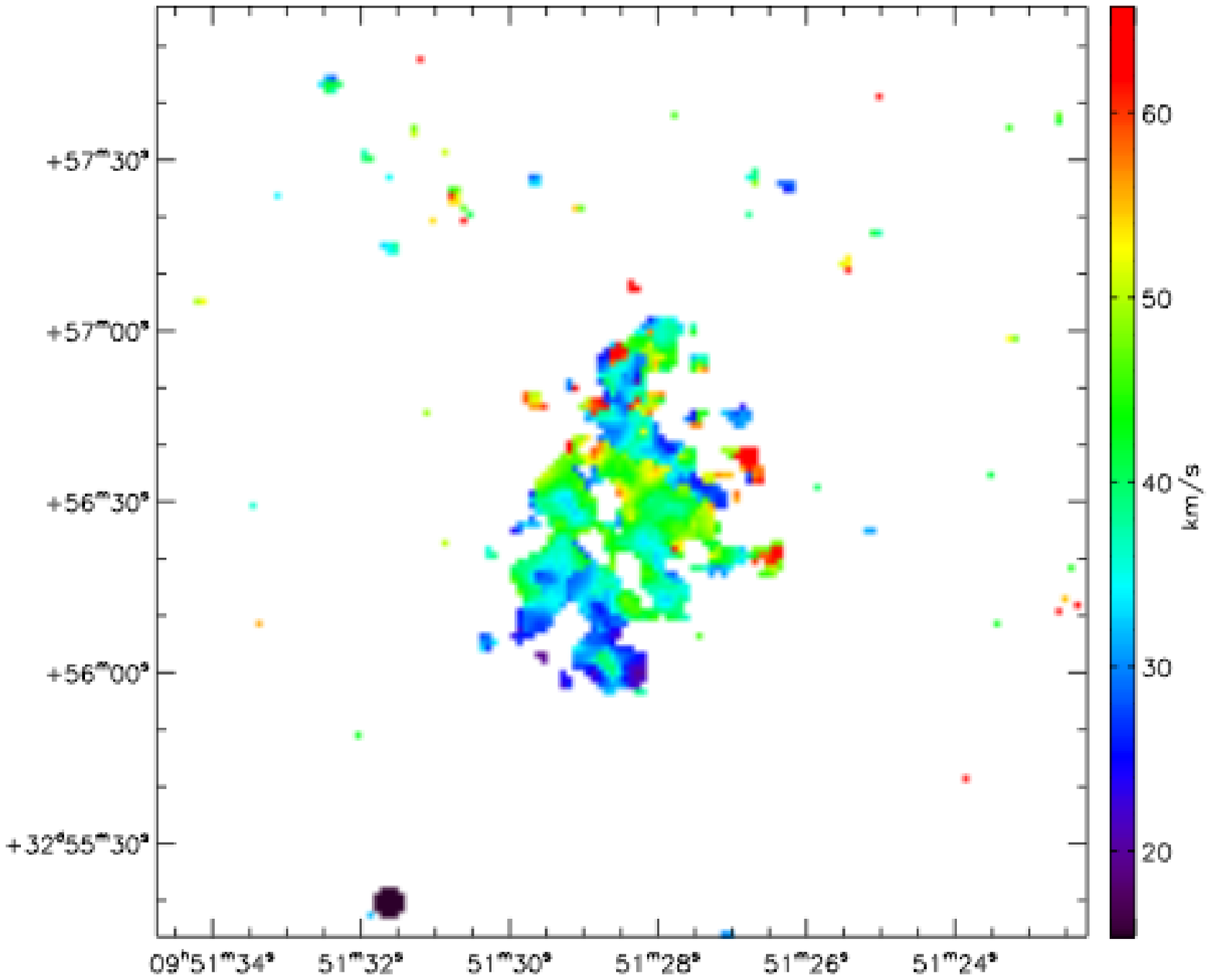} &\includegraphics[width=6.8cm]{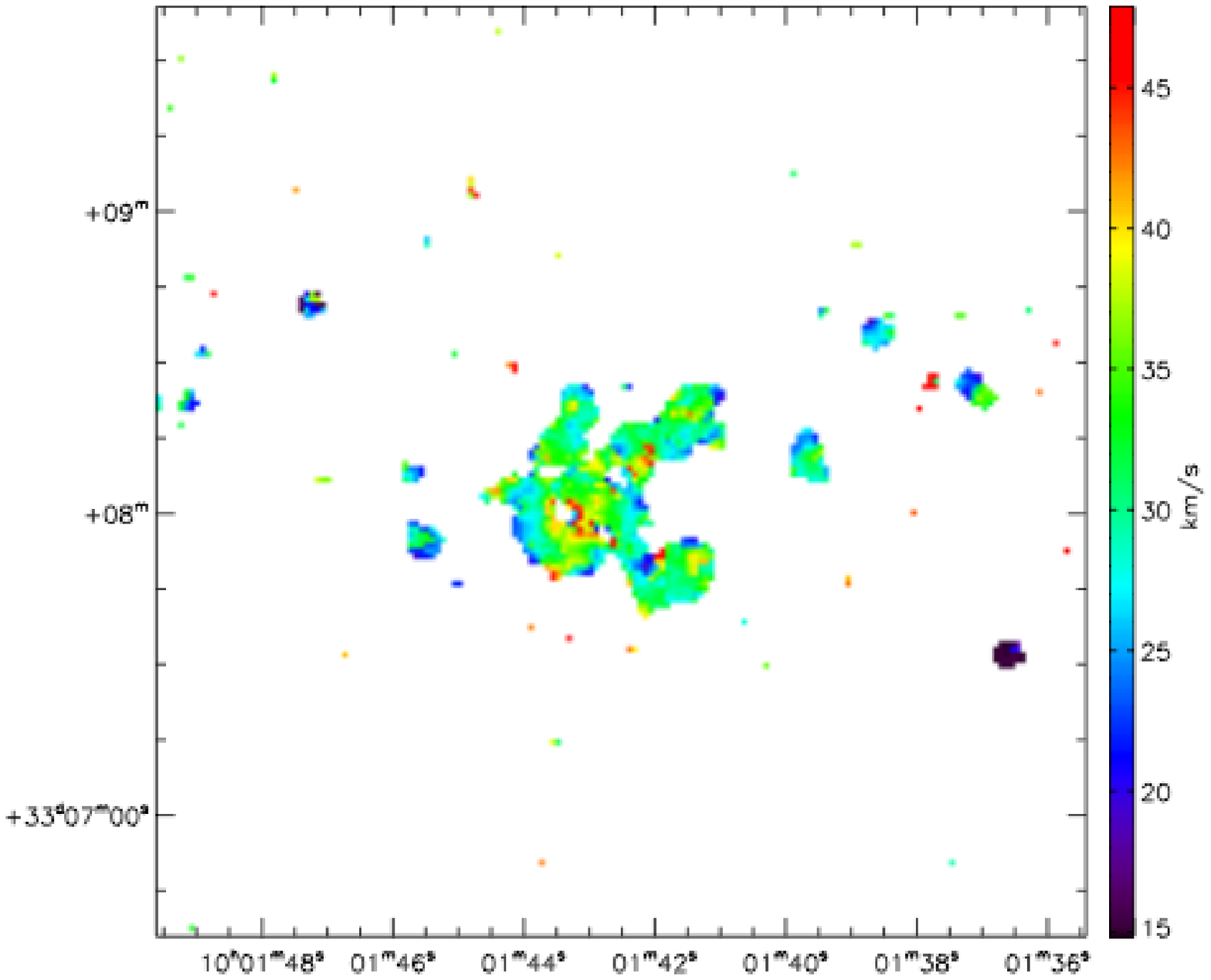} \\
 \end{tabular}
 \caption{From top to bottom: monochromatic H$\alpha$ image, velocity field and ionized gas velocity dispersion 
 of  UGC 05287  (left panels) and  UGC 05393 (right panels) in U268.
  }
  \label{maps1}
 \end{figure*}

Figures \ref{maps1} and \ref{maps2}  show the monochromatic H$\alpha$ maps,   the 2D velocity maps   
and the dispersion velocity maps of spirals, in U268 and U376, respectively.

\subsubsection{Rotation Curves}

Figure \ref{rc}  shows the {\bf H$\alpha$} rotation curves (RCs) of UGC 05287 in U268 and NGC 3655, NGC 3659, NGC 3684, 
NGC 3686, NGC 3691 in U376.  

\citet{Epinat08b} provide the details about the method used to derive the RCs. 
The fitting method is similar to that adopted by \citet{Barnes03}, the main difference 
being in the determination of the velocity uncertainties. The adopted  
technique takes into account errors due to non circular motions (e.g. presence of a bar, 
spiral arms, etc.). The code fits different parameters of the 
modeled velocities in polar coordinates for different bins or crowns
\citep[][their Appendix A]{Epinat08b}. The input parameters, coordinates of the  centre (x$_c$, y$_c$), 
systemic velocity (v$_{syst}$), position angle (PA) and galaxy inclination ($i$) are taken from 
the literature in order to constrain the fit. 
Table \ref{kp}  summarises the kinematical results. Inclination  and position angles are derived
from both velocity and continuum maps.
 
\begin{figure*}
 \begin{tabular}{ll}
 \includegraphics[width=6.5cm]{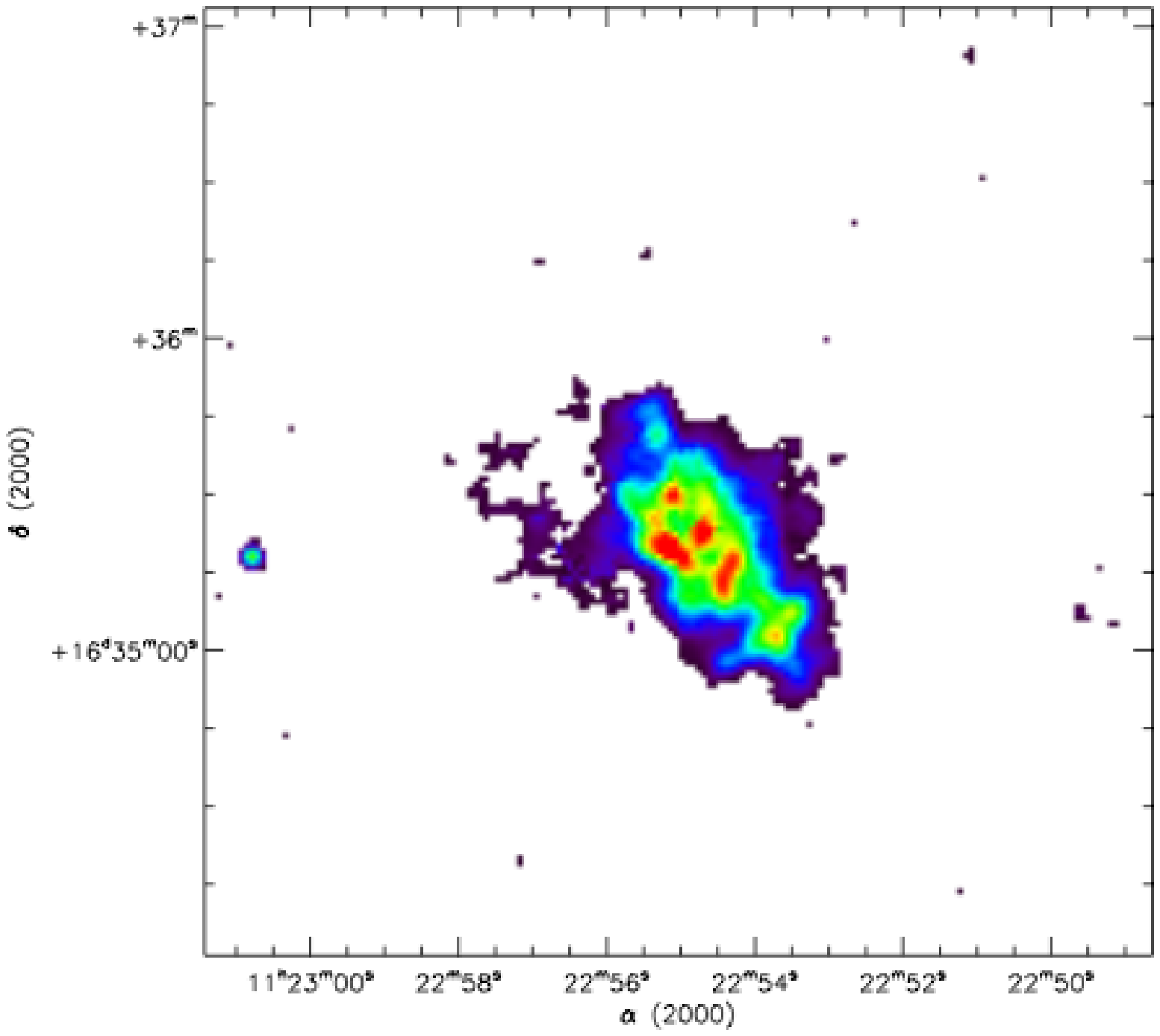}  & \includegraphics[width=6.5cm]{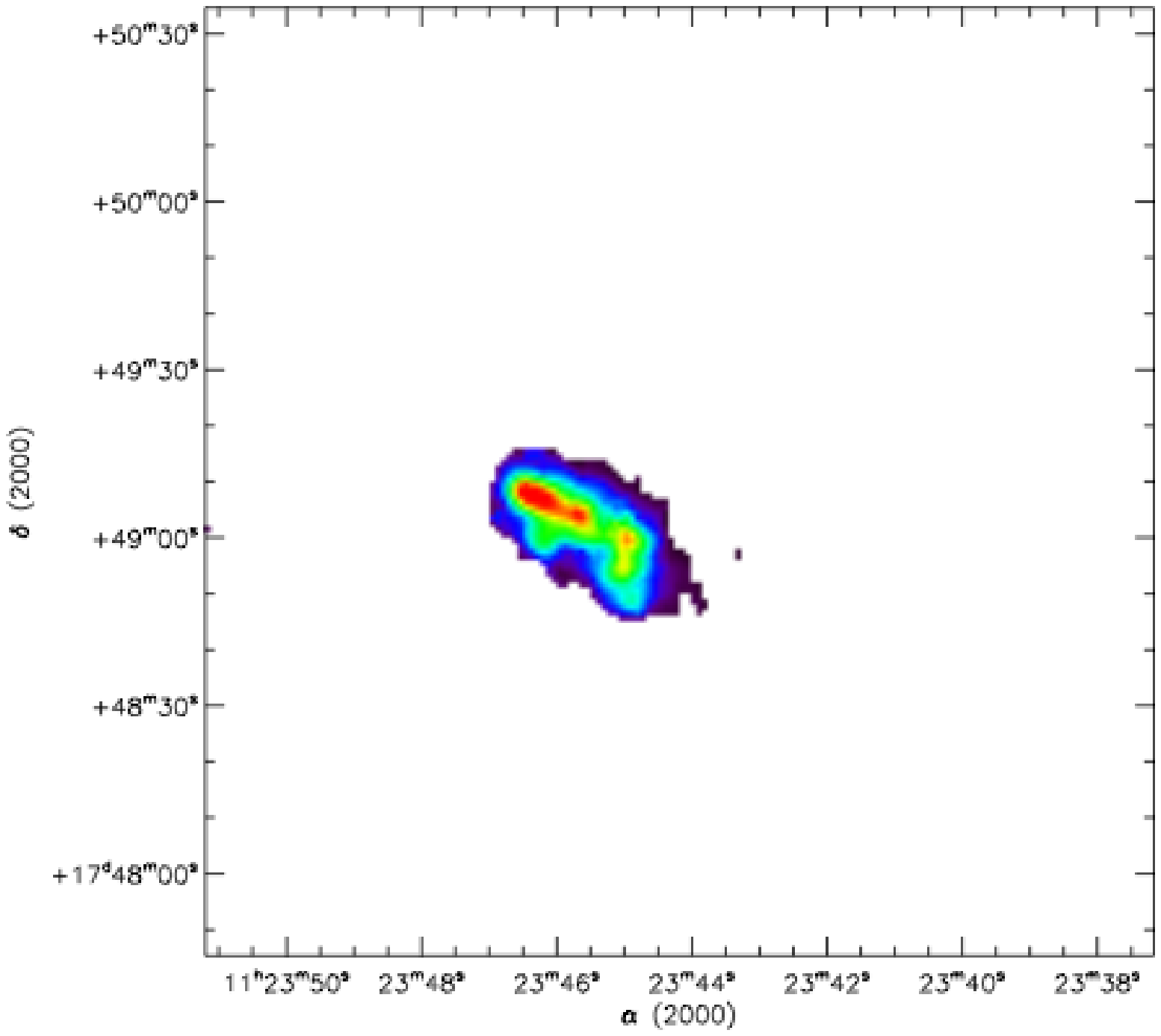} \\
 \includegraphics[width=6.8cm]{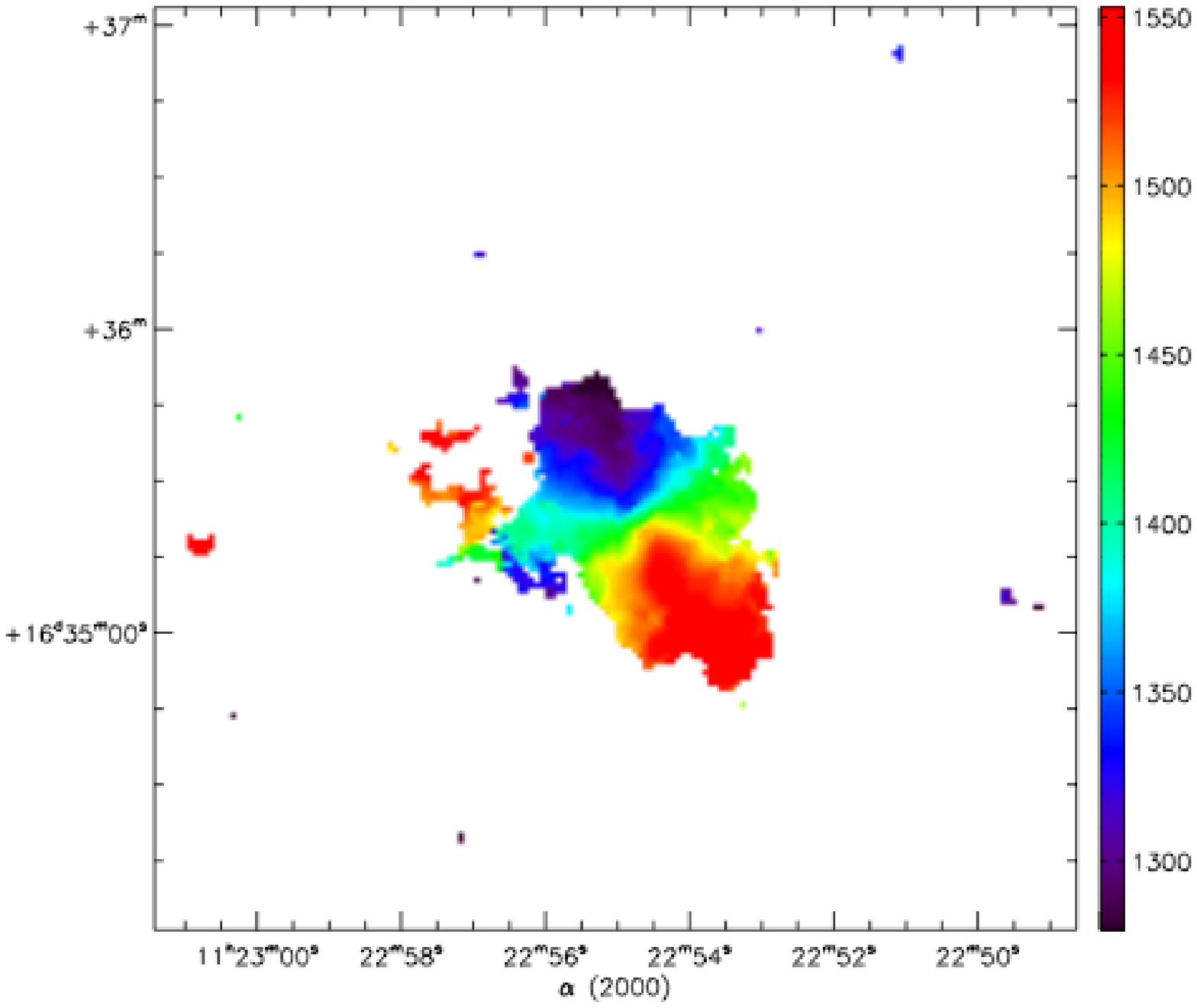}& \includegraphics[width=6.8cm]{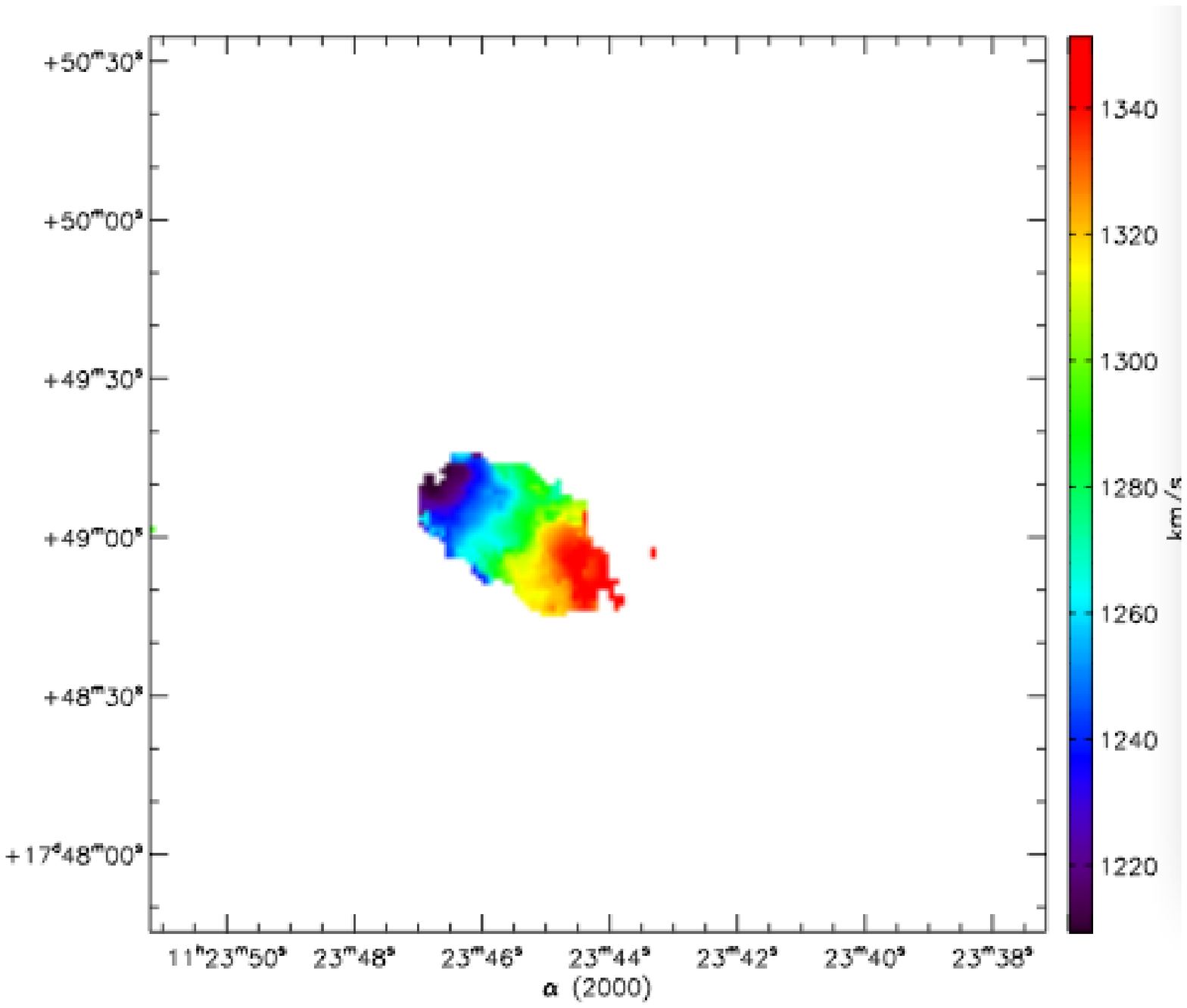} \\
  \includegraphics[width=6.8cm]{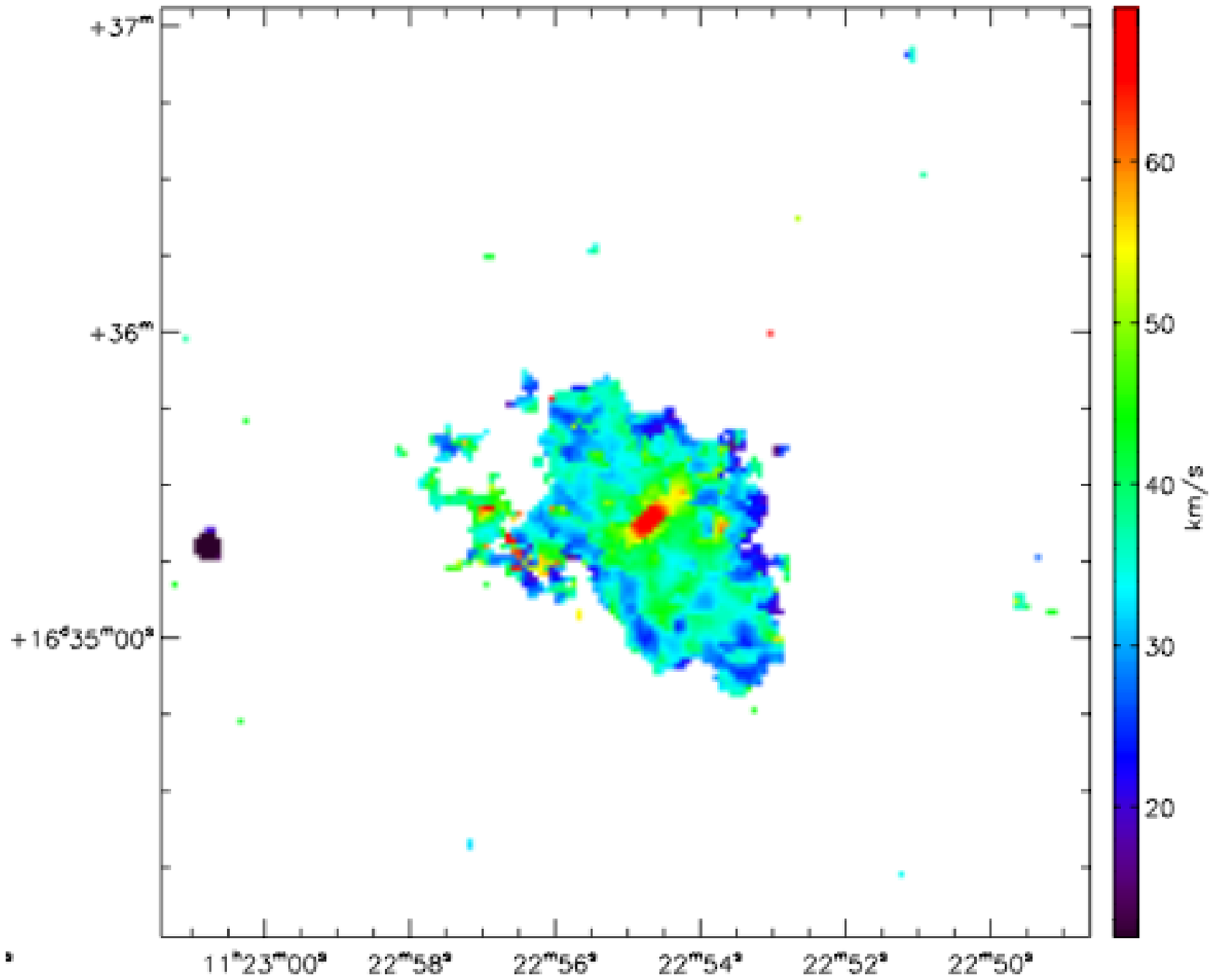} & \includegraphics[width=6.8cm]{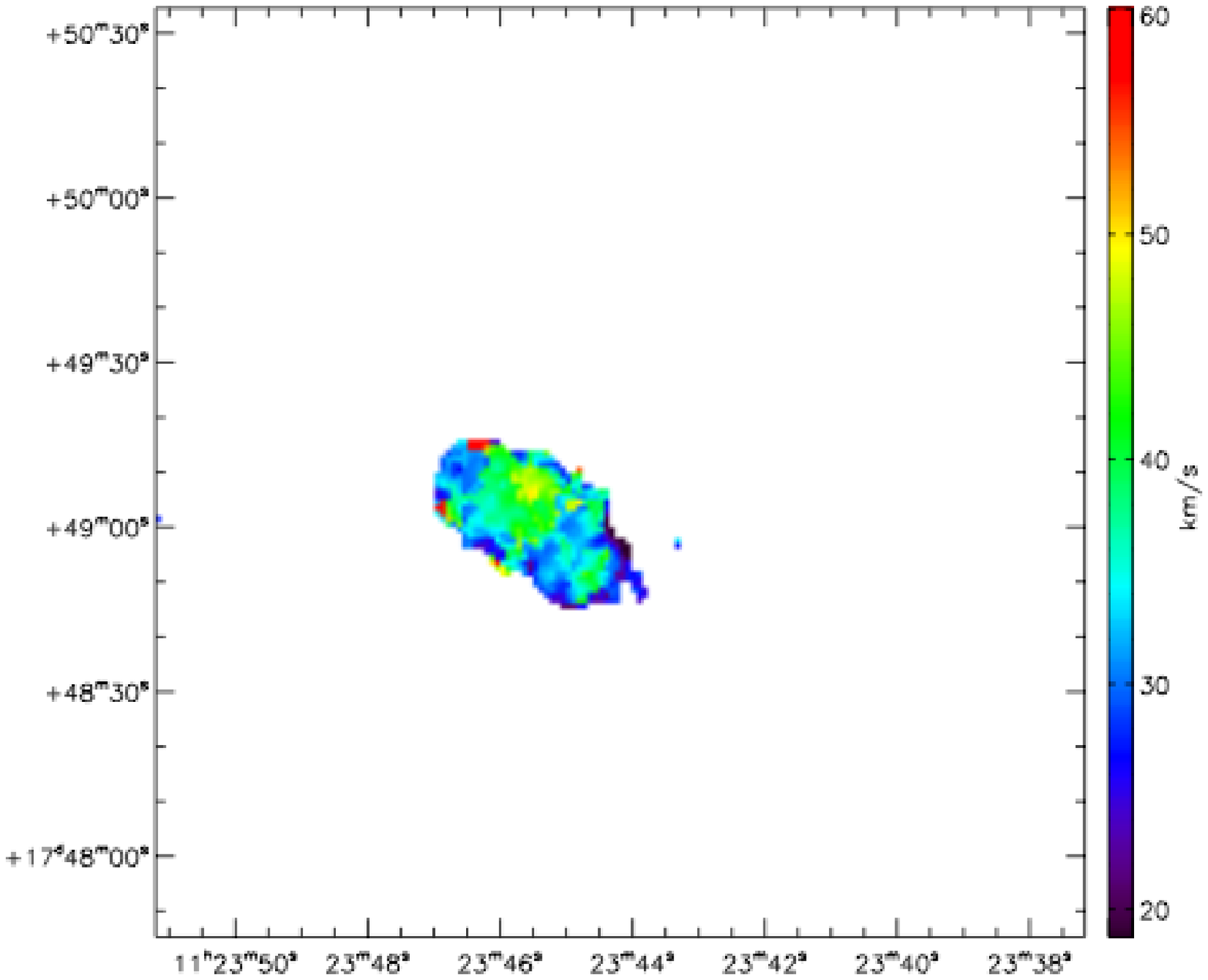} \\
  \end{tabular}
 \caption{As in Figure \ref{maps1} for NGC 3655 (left panels) and NGC 3659 (right panels) in U376. } 
 \label{maps2}
 \end{figure*}
 
\begin{figure*} 
  \begin{tabular}{ll}
 \includegraphics[width=6.5cm]{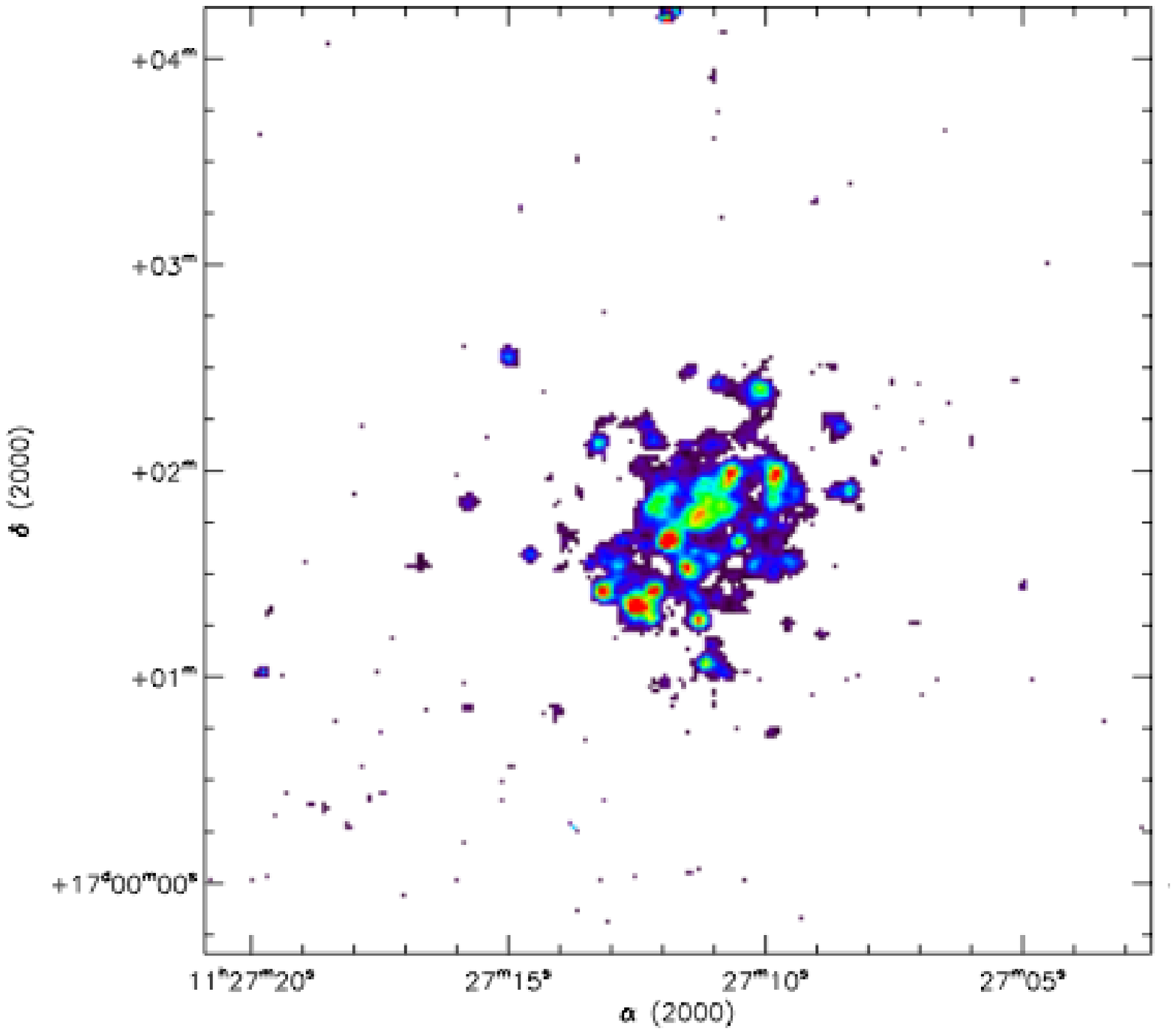}  & \includegraphics[width=6.5cm]{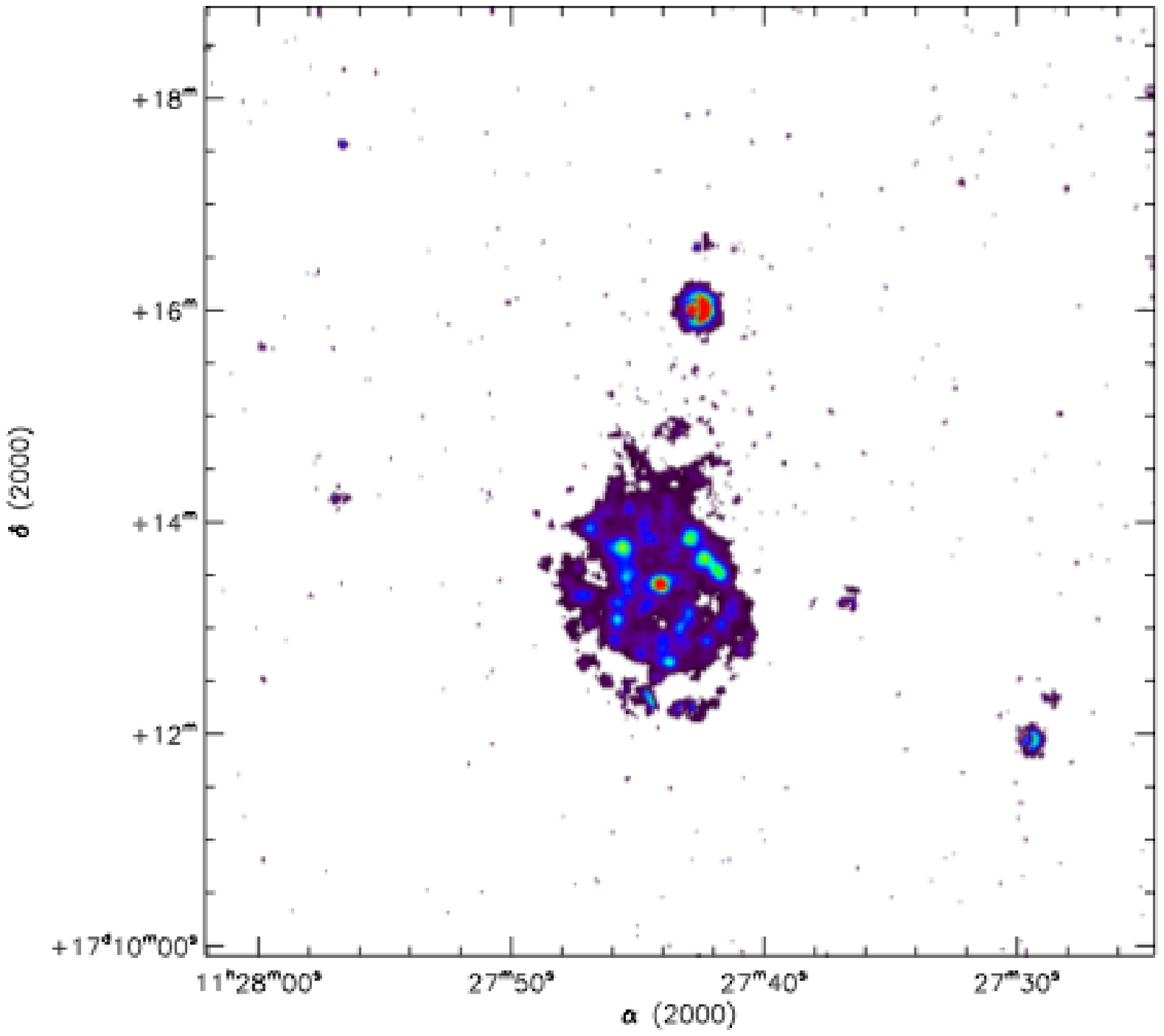} \\
 \includegraphics[width=6.8cm]{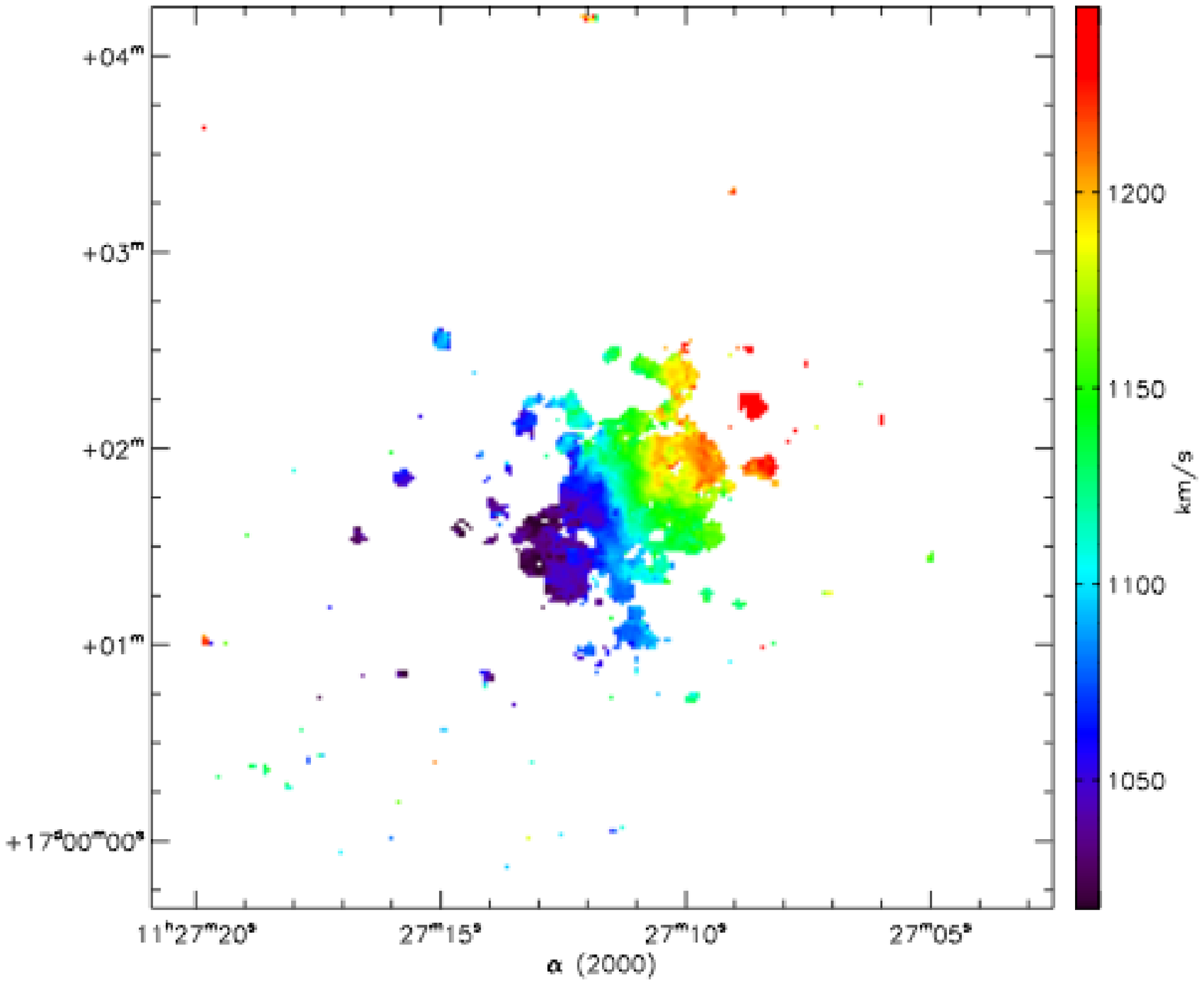}& \includegraphics[width=6.8cm]{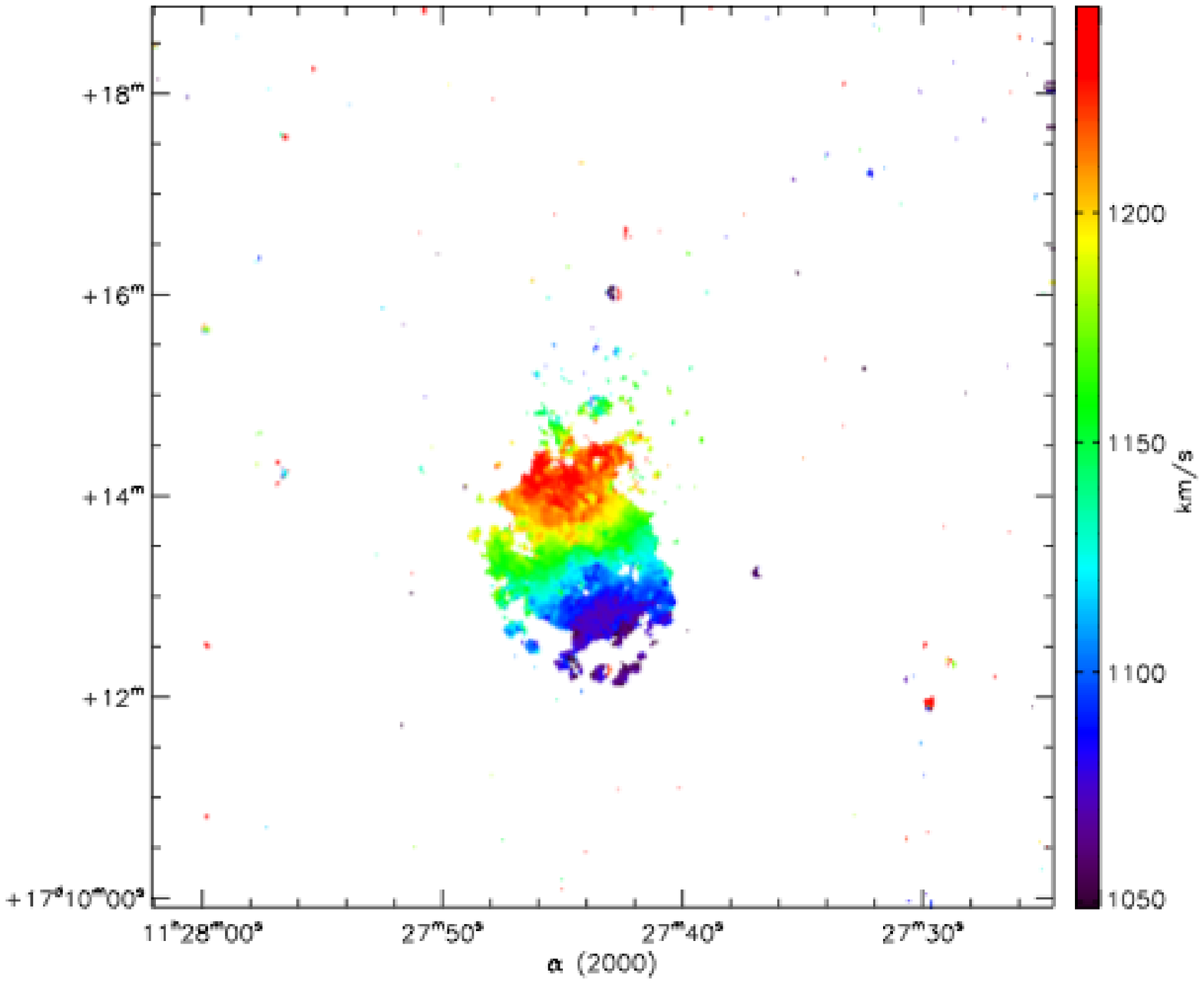} \\
  \includegraphics[width=6.8cm]{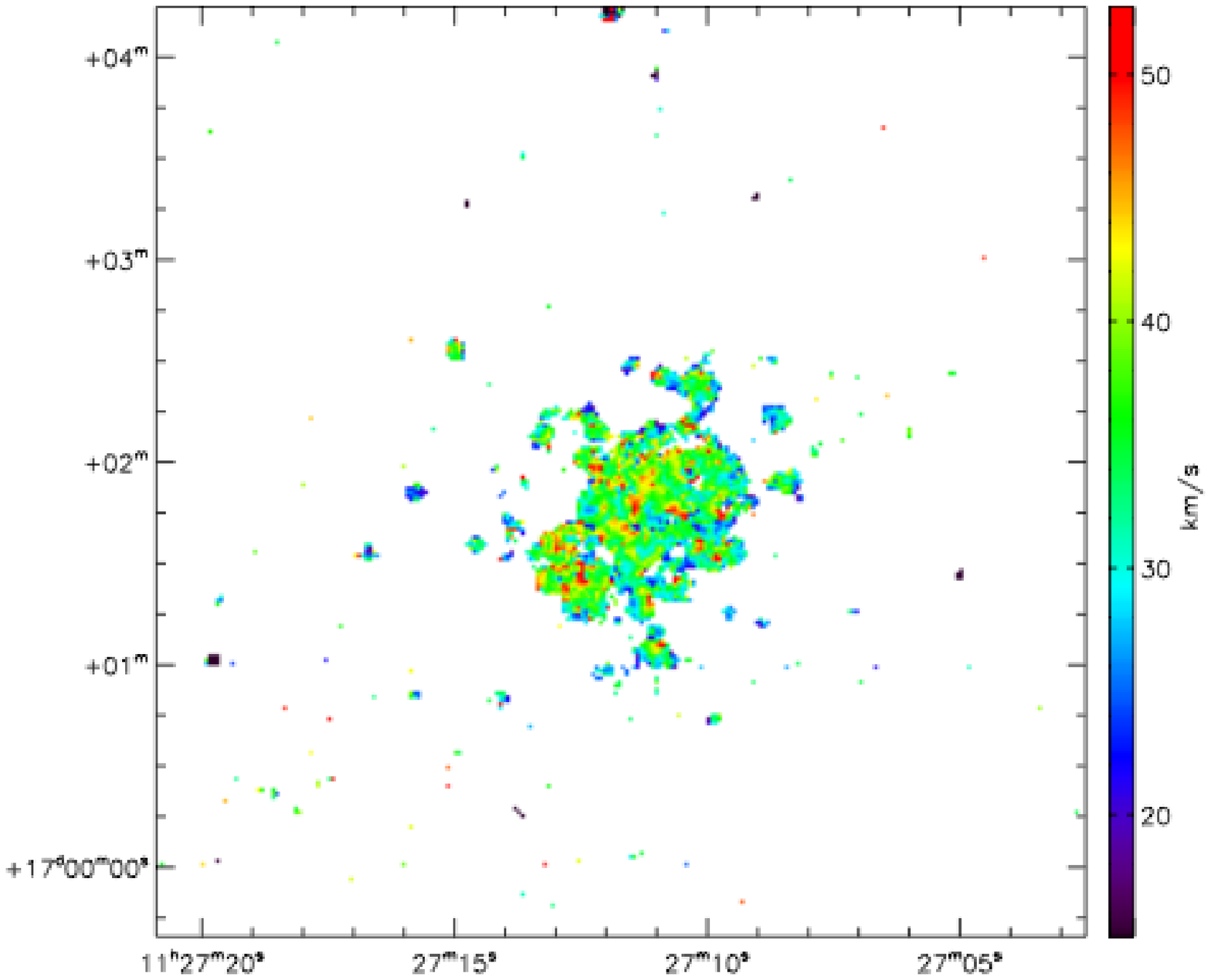} & \includegraphics[width=6.8cm]{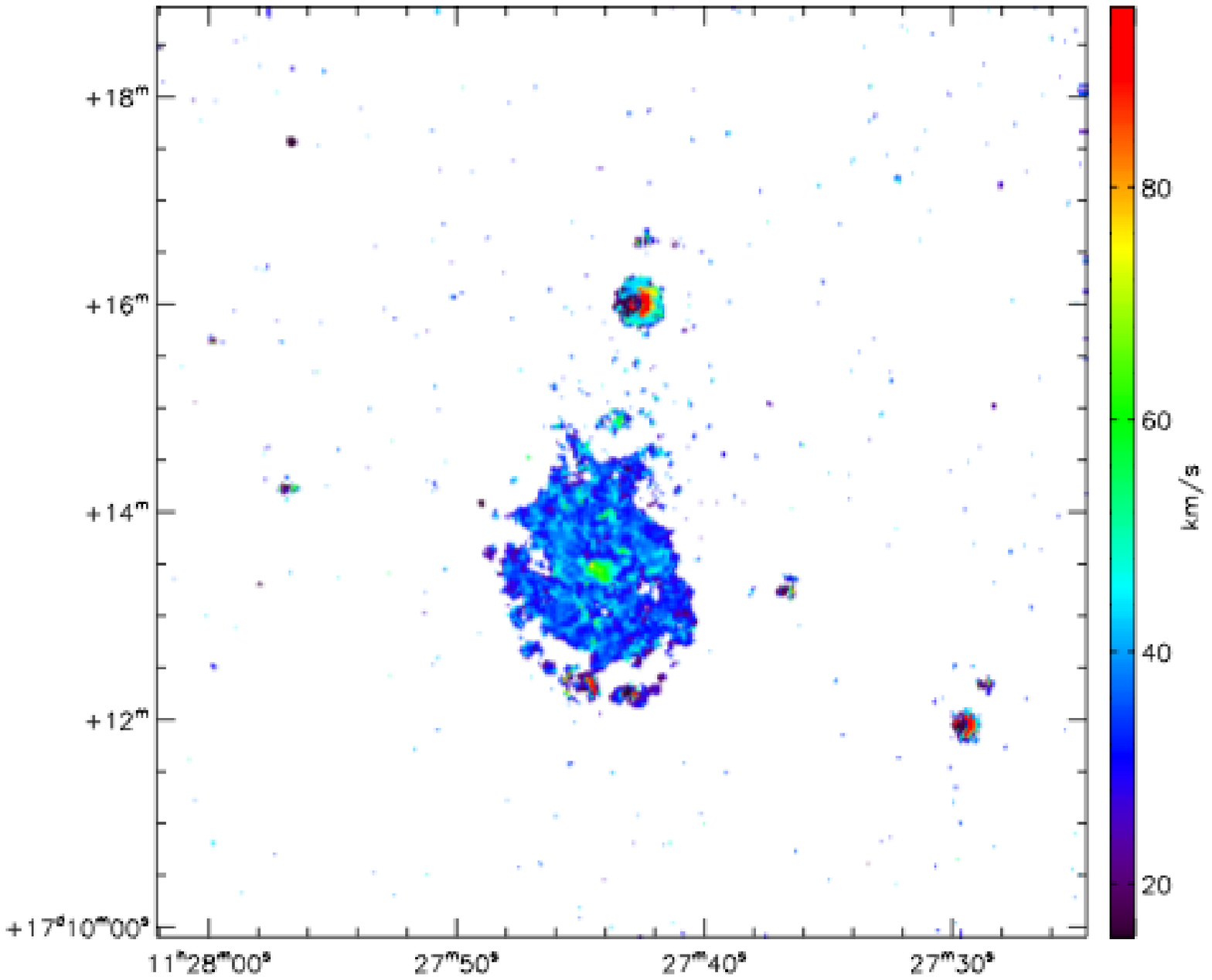} \\
  \end{tabular}
 \addtocounter{figure}{-1}
 \caption{As in Figure \ref{maps1} for  NGC 3684 (left panels) and NGC 3686 (right panels)  in U376. }
 \end{figure*}

 \begin{figure} 
  \begin{tabular}{ll}
  \includegraphics[width=6.5cm]{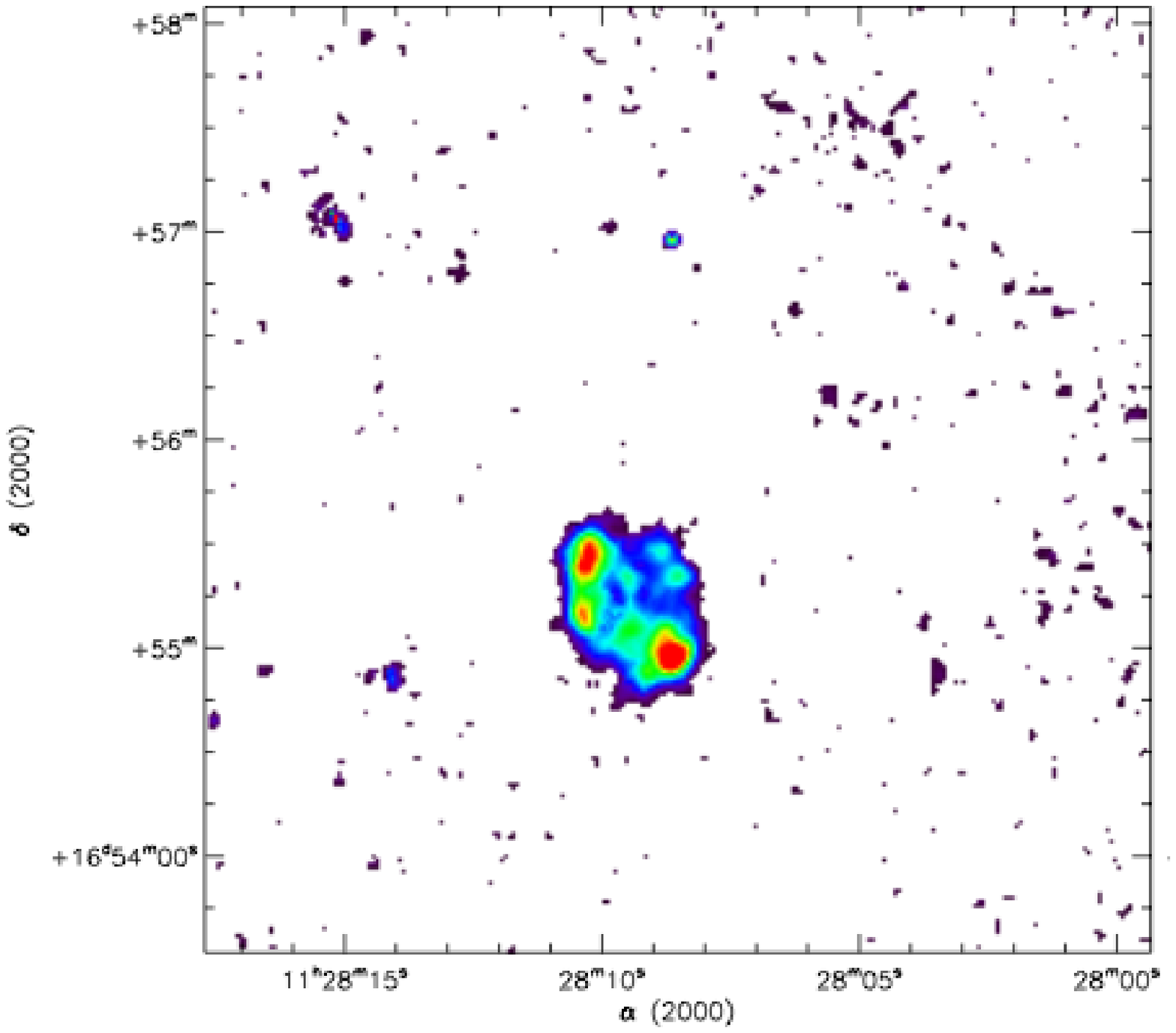}  &  \\
 \includegraphics[width=6.8cm]{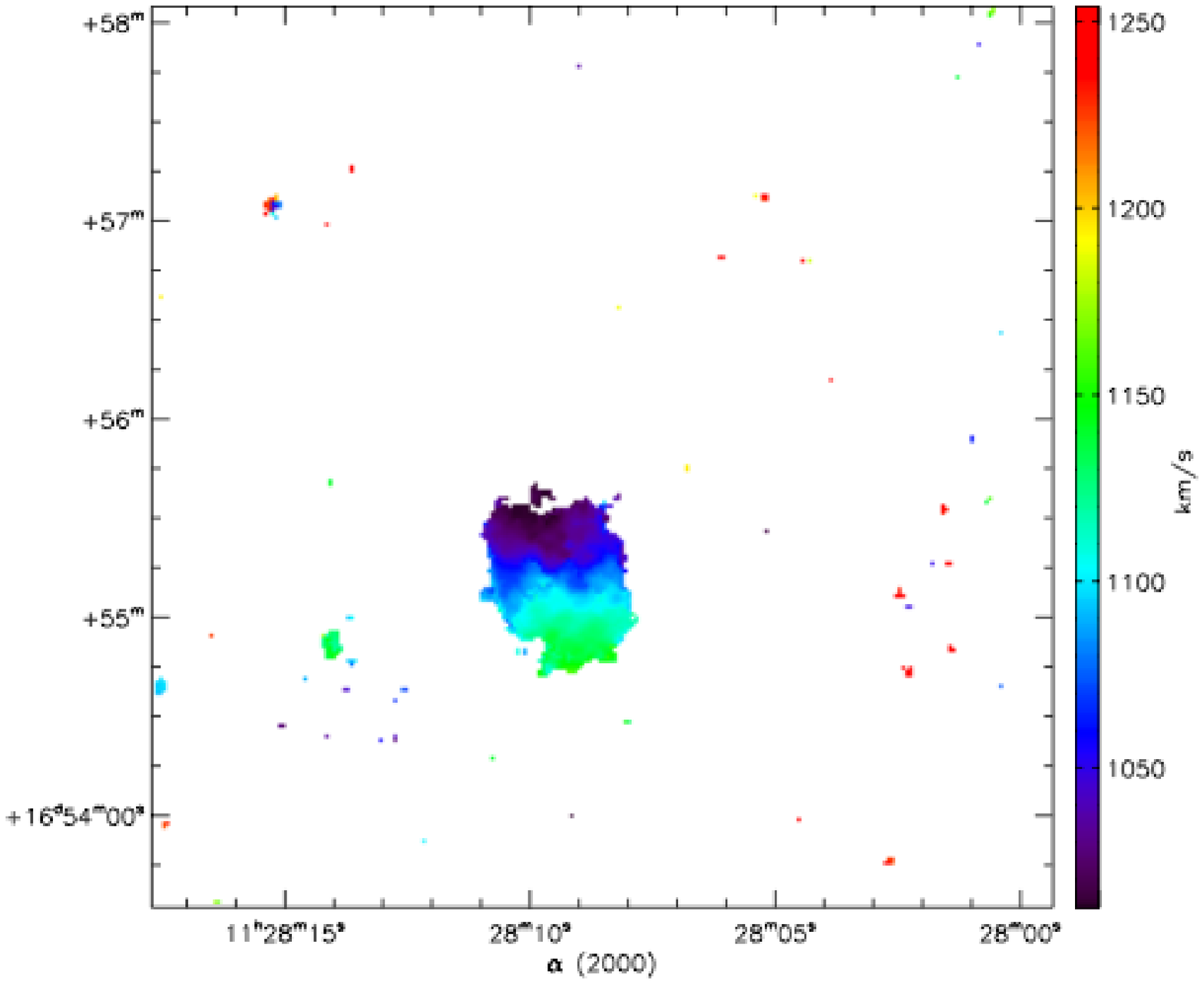}& \\
  \includegraphics[width=6.8cm]{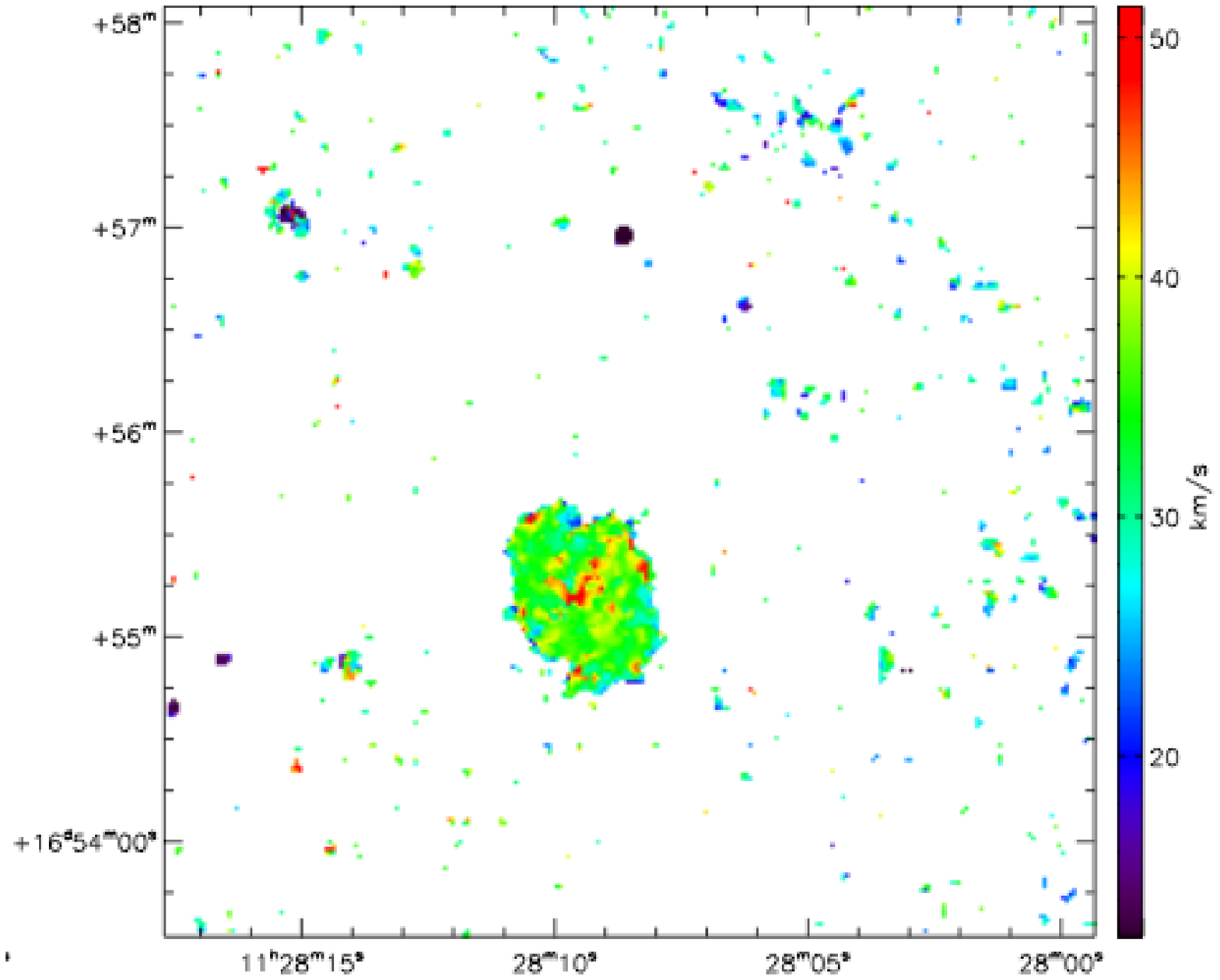} & \\
  \end{tabular}
 \addtocounter{figure}{-1}
 \caption{As in Figure \ref{maps1} for NGC 3691  in U376. }
 \end{figure}

\section{Results}

 \subsection{Photometry and surface brightness profiles} 
The  UV and optical surface photometry was carried out  using  the ELLIPSE
fitting routine in the STSDAS package of IRAF \citep{Jedrzejewski87}.
The SDSS images  (corrected frames with the soft bias 
of 1000 counts subtracted) in the five bands were registered 
to the corresponding  {\it GALEX} NUV intensity images before to evaluate brightness profiles,
using the IRAF tool {\texttt register}.
We masked the foreground stars and the background galaxies in the regions where we 
measured the surface brightness profiles. 
{\bf To secure a reliable background measure, we force the measure of 5 
   isophotes  well beyond the galaxy emission.}

From the surface brightness profiles, we derived total apparent magnitudes as follows.
For each profile, we computed the integrated 
apparent magnitude within elliptical isophotes up to the radius where 
the mean isophotal intensity is 2$\sigma$ above the background.
The background was computed around  each source, as the mean of sky value  of the outer five 
isophotes.
 Errors of the UV and optical  magnitudes where estimated by propagating the 
statistical errors on the  mean isophotal
intensity provided by ELLIPSE. In addition to the statistical error, we added 
an uncertainty to account for systematic 
uncertainties in the zero-point calibration of 0.05 and 0.03 mag in FUV and NUV 
respectively \citep{Morrissey07}.
Surface photometry was corrected for galactic extinction assuming Milky Way dust with R$_v$=3.1 \citep{Cardelli89}, A$_{FUV}$/E(B-V) = 8.376 and A$_{NUV}$/E(B-V) = 8.741, A$_{r}$/E(B-V) = 2.751.

 Figure \ref{A1} and Figure  \ref{A2} display brightness and colour profiles of 
bright members of the 2 groups.
 Table \ref{tab3} lists the measured  foreground extinction uncorrected AB
 magnitudes both in UV and optical\footnote{We  converted SDSS counts to 
 magnitudes following the recipe provided in {\tt http://www.sdss.org/df7/algorithms/fluxcal.html
 \#counts2mag}.} bands. We  used UV and optical images and luminosity profiles to obtain a more 
 robust morphological classification, we also report in Table \ref{tab3}. 
  UV and optical  magnitudes of fainter added members were
 extracted from the {\it GALEX} and SDSS pipelines. {\bf We used the FUV and NUV calibrated magnitudes  and the   Model magnitudes\footnote{http://www.sdss.org/dr5/algorithms/photometry.html}
 from the {\it GALEX} and SDSS pipelines, respectively.}
 
In the next two sections we describe the morphology, the luminosity
and the colour profiles of individual galaxies. The main purpose is to identify 
peculiarities possibly associated to interaction events.
 
\subsubsection{U268: individual morphological and photometric notes}

The group is composed of a large fraction of late-type galaxies.

\underbar{MRK 408} ~~~ The galaxy is nucleated both in UV and optical images.
Its outer isophotes appear elongated toward the SE. Neither arms,
nor flucculent structures are visible. 
The UV and optical luminosity profiles may be decomposed into a bulge+disk
component so that  the galaxy is
a S0. The galaxy is blue ($FUV-NUV\approx$0) along its extention.

\underbar{NGC 3003} ~~~ The galaxy is a nearly edge-on spiral. Its SW and   
NE parts  are asymmetric, and signatures of perturbation are likely present
in this latter part. In the southern part, the UV composite image shows distorted arm/tails
not revealed in the optical image. Luminosity profiles are typical of a spiral galaxy. The outer disk
is bluer than the nuclear part ($FUV-NUV \approx $0 vs.  $FUV-NUV \approx $ 0.7). 
The $r$ profile is truncated since the image is at the edge of the FOV and
does not allow an accurate surface photometry. 

\underbar{NGC 3011} ~~~ In this S0 galaxy, the presence of an outer ring is
visible in the UV. The $r$ luminosity profile
shows the presence of two components, the bulge and the disk (emerging at $\approx$20\arcsec).
The ring shows a blue ($FUV-NUV$= 0.3) colour  as those presented in \citet{Marino11b}.

\underbar{UGC 5287} ~~~ Galaxy is a nearly face-on spiral with open arms starting 
from a bar. The Northern and the Southern arms appear asymmetric in the optical image.
At the end of the southern arm, in the NUV image is well visible a bright region that  could be a possible dwarf companion.

\underbar{UGC 5326} ~~~ This spiral is reminiscent of the Cartwheel. The nucleus is
displaced in the NW direction with respect to the centre of the outer ring.
 All the galaxy is blue in the UV ($FUV-NUV \approx$ 0), while in the more resolved optical images, 
the ring is bluer than the nucleus  ($FUV-r$  and  $NUV-r \approx $ 1).
Deeper higher resolution images are needed to clarify  the nature
 of this galaxy.

\underbar{IC 2524} ~~~ The major axis of the UV and optical images of the galaxy, likely an S0, are
different. The P.A. of the optical isophotes is about 45$^{\circ}$ NE at odds
with UV isophotes having a P.A.$\approx$10$^{\circ}$ NE. The galaxy is quite blue for its
morphological class especially in the outskirts ($FUV-NUV$= 0). Outer isophotes 
seem boxy.  

\underbar{NGC 3067} ~~~ The UV and optical images indicate that this is a disk galaxy.  It 
is classified as a spiral in  {\tt HYPERLEDA}: our images do not show unambiguous arms.
 Blue spots ($FUV-NUV \approx$  0.5) are 
detected in the inner SE part.   
The nuclear part is red ($NUV-r \approx$ 3.9) and a dust system is likely present. 

\underbar{UGC 5393} ~~~ The UV image is more extended than the optical one. 
In this latter the arms appear to start from a bar and close forming a ring. 
Signatures  of
interaction (tail) are present in the SE. A blue ring ($FUV-NUV \approx$ 0) is visible in the both
UV luminosity colour profiles.

\underbar{UGC 5446} ~~~ The galaxy is seen edge-on. We suspect a warp in the NE side 
of the galaxy. UV and optical surface photometry do not show evidence of a nucleus.
The galaxy outskirts are bluer than the inner parts   ($FUV-r \approx 2$; $NUV-r \approx 2$, and
$\approx 0 $) in the centre.  

\underbar{NGC 3118} ~~~ The galaxy is seen edge-on showing  unambiguous warps. 
The luminosity and colour profiles show a red bulge ($FUV-r \approx 4$) and a quite
blue disk ($FUV-r$ and  $FUV-NUV \approx0$).

\subsubsection{U376: individual morphological and photometric notes}

\underbar{NGC 3592} ~~~ The galaxy is seen edge-on with a dust lane along the major axis. 
The luminosity profiles suggest that it is basically composed of a disk which is quite red ($FUV-NUV \approx 1$) compared with edge-on galaxies described in this paper. No signature of interaction.

\underbar{NGC 3599} ~~~ This S0 galaxy does not show signature of 
interaction. The FUV emission has a very limited extension. The galaxy is 
quite red ($FUV-NUV$ between 1 and 2; $FUV-r \approx 6$; $NUV-r \approx 5$). 

\underbar{NGC 3605} ~~~ We found indication  
 of outer   shells,  likely due to a past accretion event \citep{Dupraz86}
or a weak interaction \citep{Thomson91}.
The galaxy luminosity profile is typical of an elliptical. Far UV and optical colours are red
($FUV-NUV \approx 1.5$; $FUV-r \approx7$; $NUV-r \approx5$). 

\underbar{UGC 6296} ~~~ This is a disk galaxy, not an Irregular as in the classification of Table 1  from  {\tt HYPERLEDA}. Both the UV and the optical images suggest the presence of
dust absorption features. It is reminiscent of NGC~3067 in U268. In the SE part, a
blue knot is detected in the UV images ($FUV-NUV \approx 0.3$).

\underbar{NGC 3607} ~~~ This bright elliptical does not show signature of interaction. The galaxy
is red as shown by the UV and optical luminosity and colour profiles.  Average colours are:
$FUV-NUV \approx 1$; $FUV-r \approx 6$; $NUV-r \approx 5.9$.

\underbar{NGC 3608} ~~~ No signature of interaction in this red-and-dead elliptical galaxy
(Rampazzo et al. 2012 in prep.).

\underbar{CGCG 096-024} ~~~ Faint nucleated galaxy undetected in FUV, likely an elliptical as suggested 
by the optical profile. The galaxy is red  ($NUV-r \approx 5$).

\underbar{UGC 6320} ~~~ A blue ($FUV-NUV \approx 0.5$; $FUV-r \approx 2$; $NUV-r \approx 2$), 
face-on disk galaxy, apparently without a bulge. A blue polar ring is visible in the 
optical image. UV images are faint due to the low exposure time.

\underbar{UGC 6324} ~~~ Nucleated disk galaxy, likely an S0 as classified in Table~1. No
interaction signatures are visible. The luminous objects in the SW of the nucleus are
a foreground star and a background galaxy as can be seen in the SDSS composite image.

\underbar{NGC 3626} ~~~ Barred S0 galaxy with inner and outer ring structures. No UV
images are available for this galaxy.

\underbar{NGC 3655} ~~~ Classified as Sb in  {\tt HYPERLEDA}, the galaxy shows 
a red dusty bulge ($FUV-NUV \approx 1$ ; $FUV-r \approx 5$; $NUV-r \approx 4$) and a blue 
($FUV-NUV \approx 0.5$; $FUV-r \approx 3$; $NUV-r \approx 2.1$) disk. The faint outer arms may be
generated by interaction (see UV image). The galaxy in the SDSS composite image shows
a possible polar ring-like structure. 

\underbar{NGC 3659} ~~~ Barred spiral with regular arms seen at high inclination. UV
observations are lacking. 

\underbar{NGC 3681} ~~~ Face-on spiral galaxy. Arms appear wrapped up. The far-UV image
is much more extended than the optical one and arms have a flocculent aspect. No signature
of interaction. The UV image and the surface photometry reveal evidence of an inner ring
which can be guessed also in the optical. The ring may be connected with the presence
of a bar structure visible only in the optical bands. The colours of the ring are
$FUV-NUV \approx 0$; $FUV-r \approx 4$; $NUV-r \approx 3.8 $.

\underbar{NGC 3684} ~~~ Spiral galaxy. The UV image is more extended than
the optical one. No unambiguous signature of interaction.

\underbar{NGC 3686} ~~~ Spiral galaxy  seen nearly edge-on. The bar is barely visible.
No obvious signature of interaction. UV and UV-optical colours are typical of spirals  {\bf (see figures 9
and 10 in \citet{Marino10}, and figure 4 in \citet{Marino11b}}.

\underbar{NGC 3691} ~~~ Disk galaxy. The bar is barely visible
as well as arms in the blue disk. The quality of the optical images do not allow photometric measures.

\subsection{Description of the kinematics of the spiral galaxies}

\subsubsection{U268: individual kinematics notes}
We obtained kinematical data for two spiral galaxies in U268, namely UGC 05287 and UGC 05393. 

\underbar{UGC 05287} ~~~~~
The signature of the bar  is present in the velocity field which is
irregular and highly asymmetric. The blue-shifted side of the velocity field is almost 
inexistent. In the $r$-SDSS and UV images, the morphology of the
galaxy is very asymmetric. Beside the bar, the galaxy exhibits a large 
spiral arms towards South. The monochromatic H$\alpha$ map shows intense 
emission where the bar is located, and several emission regions  as in the  {\it GALEX} image. The velocity dispersion map shows 
higher dispersion within the bar region. The H$\alpha$ line profiles seems to show a double 
component on the northern extremity of the galaxy and can be followed to the 
bar region. H$\alpha$ line profiles do not 
show multiple components.

\underbar{UGC 05393} ~~~~~
The galaxy presents the most  irregular velocity map of the sample. 
It  is classified as an SBd, but the bar is not visible on the velocity
field. Emission is mainly present in the  emitting regions visible 
on the UV images.
A velocity gradient is present, but the RC (or position velocity)   cannot
be derived with  confidence.   

\begin{figure*}
   \begin{tabular}{ll}
    \vspace{0.5cm}
\includegraphics[width=8cm]{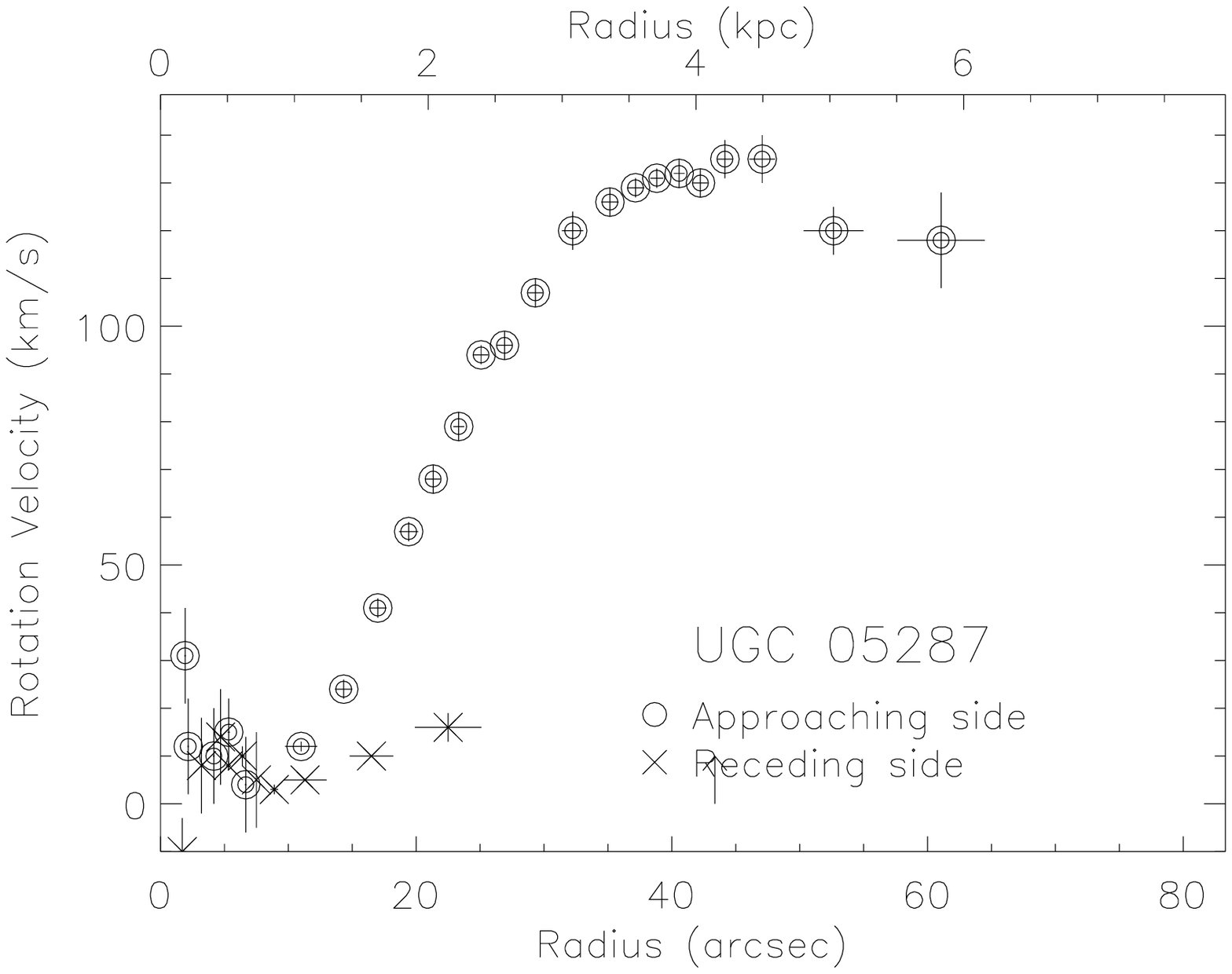}&
\includegraphics[width=8cm]{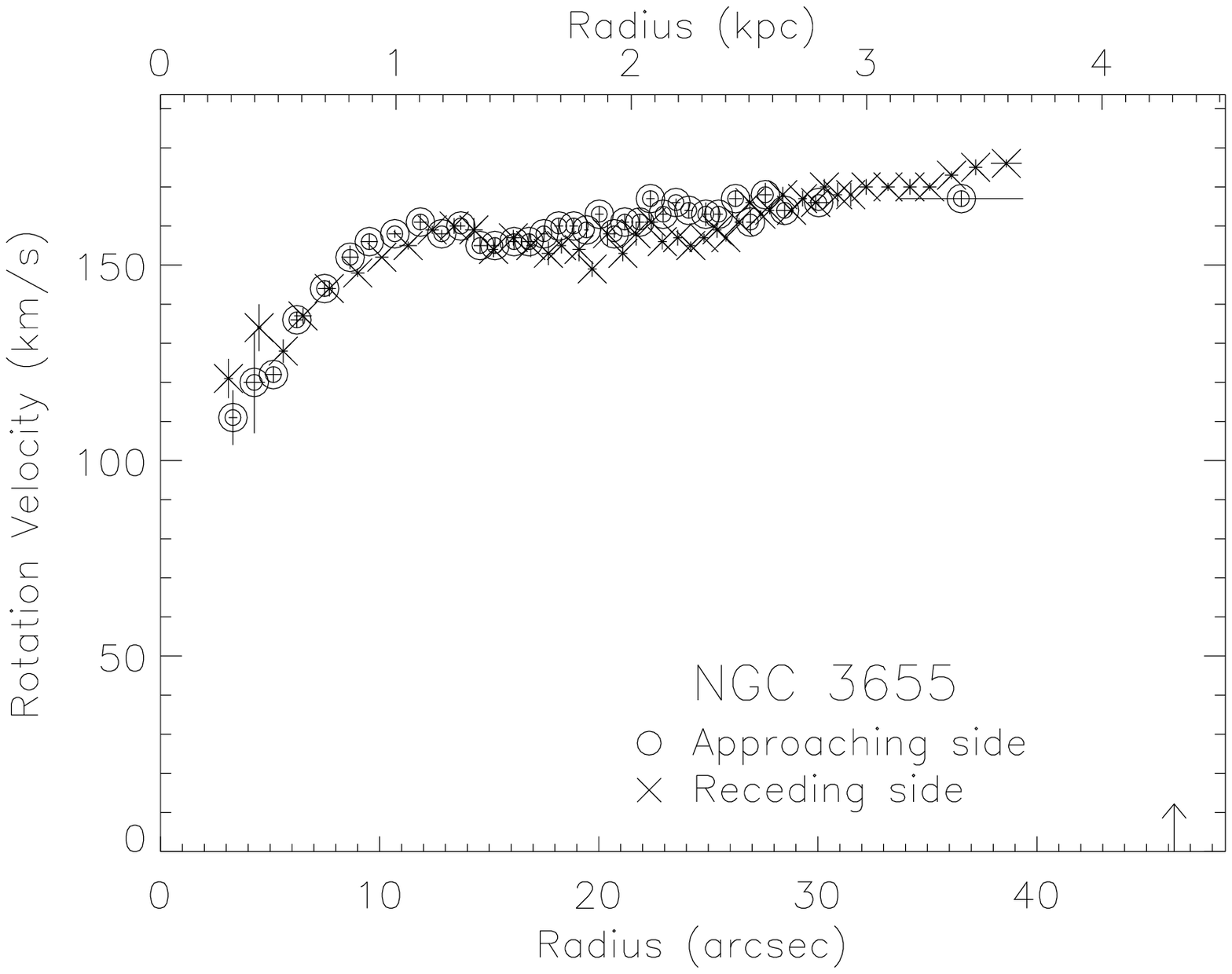}\\
 \vspace{0.5cm}
 \includegraphics[width=8cm]{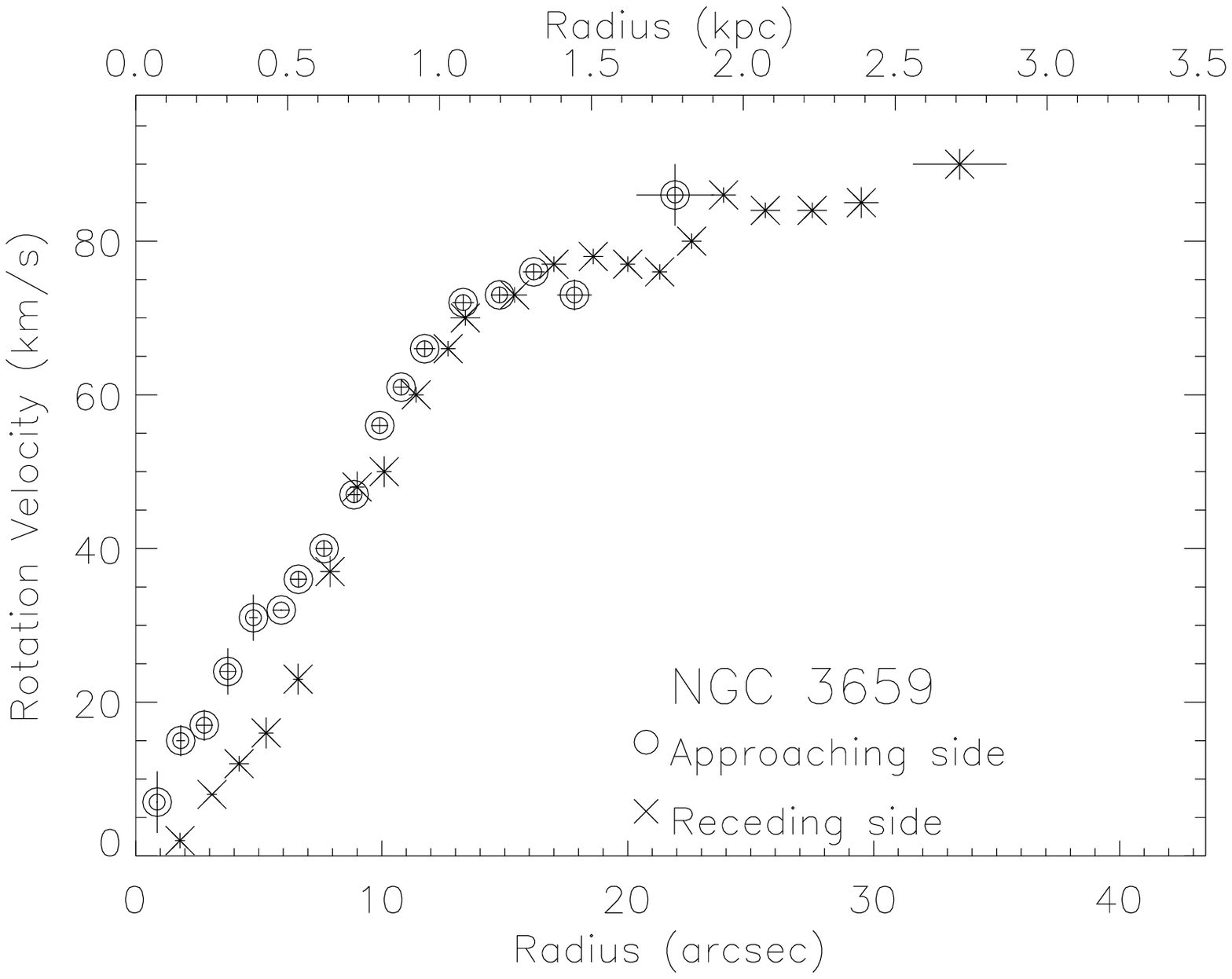}&
  \includegraphics[width=8cm]{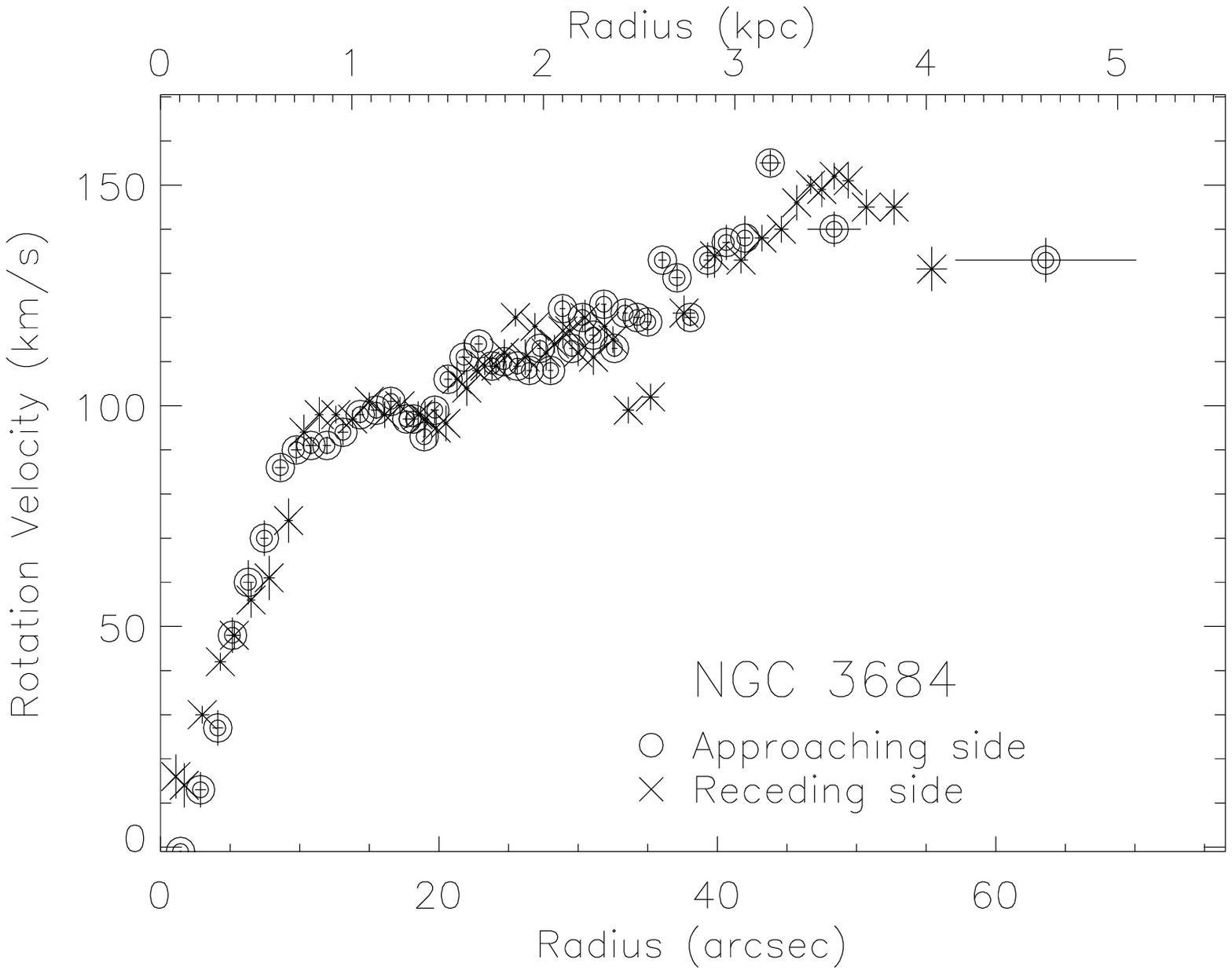}\\
  \includegraphics[width=8cm]{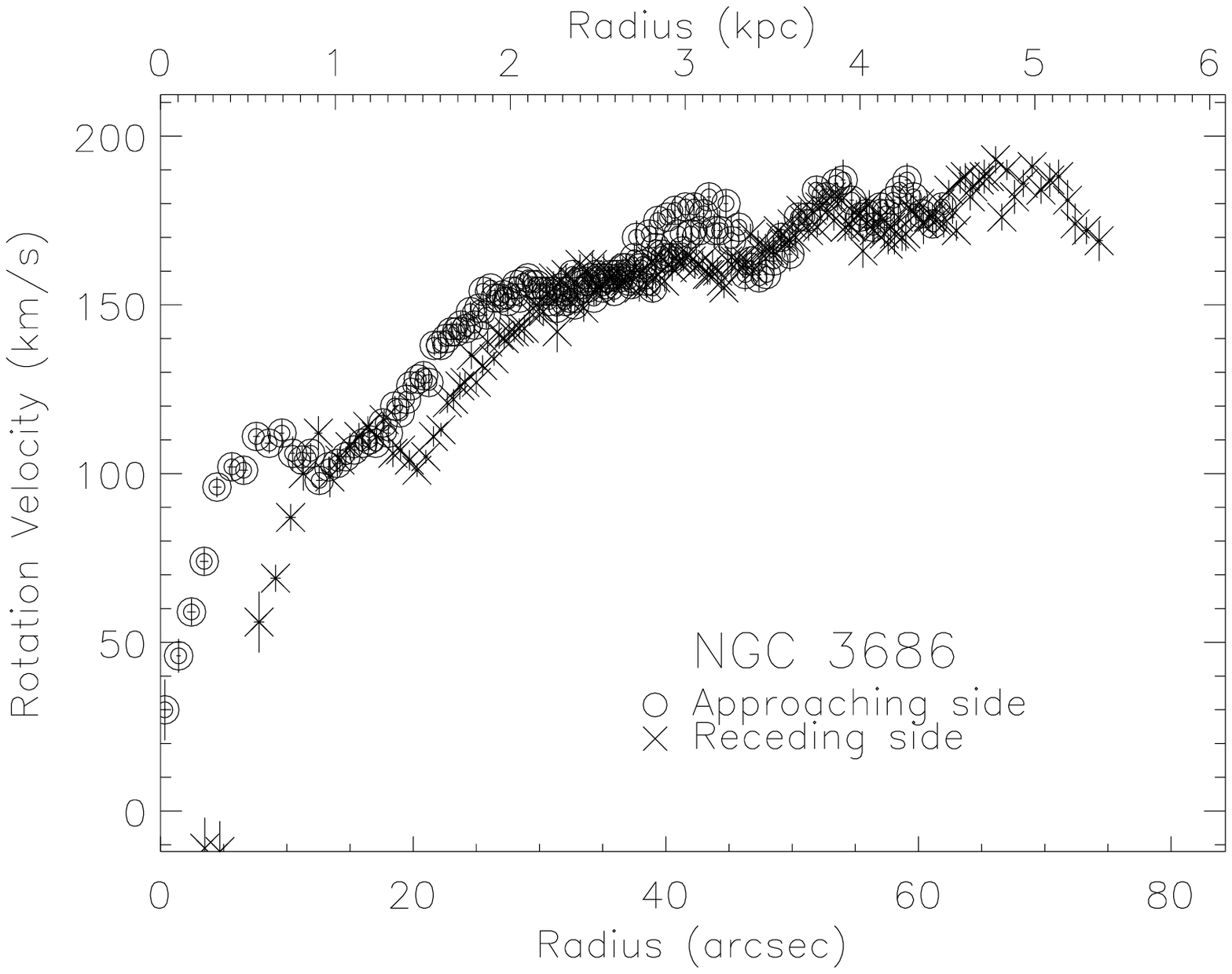}&
\includegraphics[width=8cm]{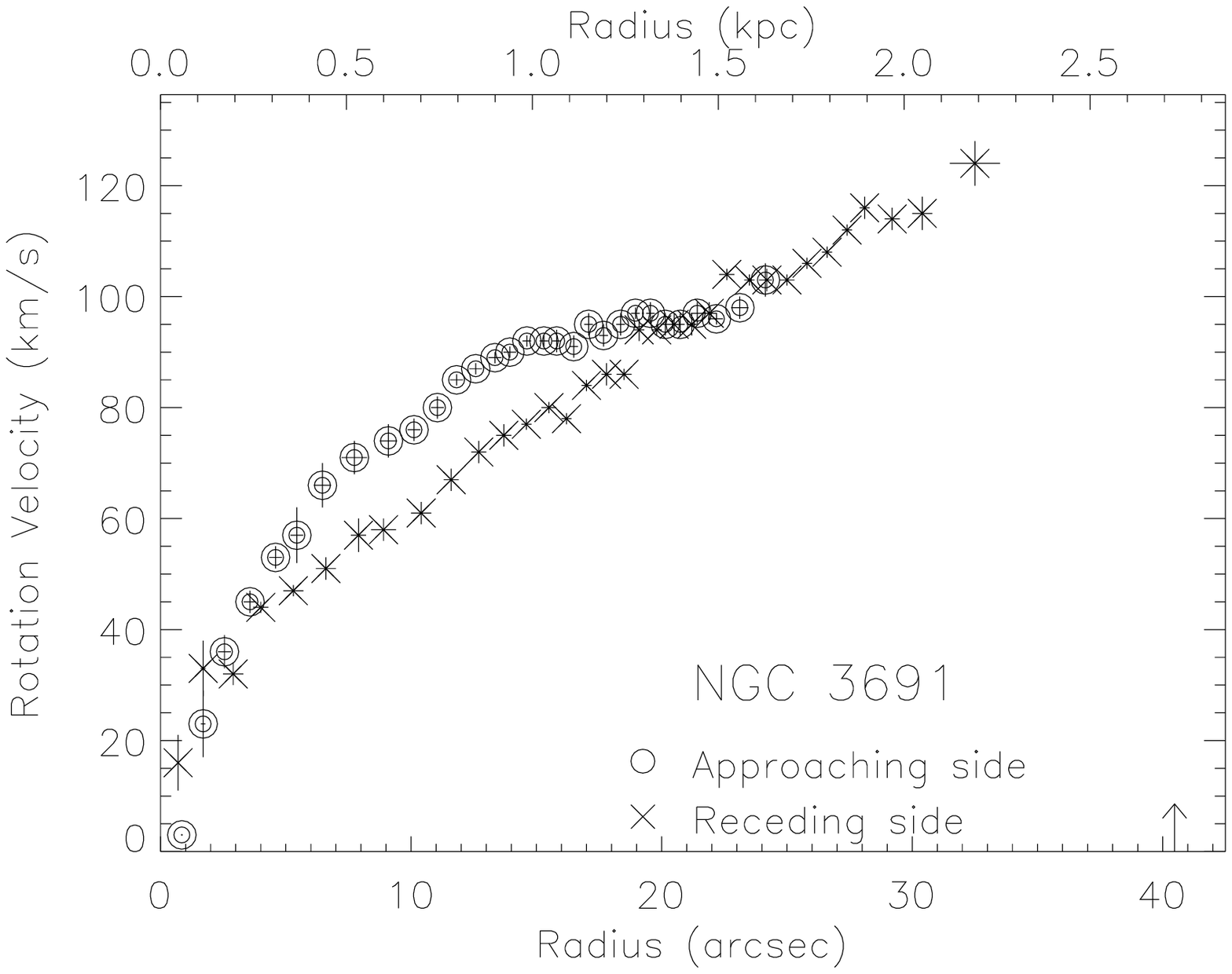}\\
\end{tabular}
 \caption{Rotation curves, extracted from 2D velocity fields, of spirals in U268 and U376.}
 \label{rc}
\end{figure*}	 

\subsubsection{U376: individual kinematical notes}

\underbar{NGC 3655} ~~~~~
The velocity field of NGC 3655 is characteristic of a rotating disk with a 
radial velocity amplitude ranging from 1280 to 1550 km~s$^{-1}$. The map 
shows a sharp rise in velocity in the centre and a plateau extended up to 
25\arcsec\ from the centre on the blue-shifted side and 38\arcsec\ on the 
red-shifted side. Even though the velocity map is very regular, it is not symmetric.
H$\alpha$ line profiles, throughout  the galaxy, are symmetric without
 signature of multiple components.
The monochromatic H$\alpha$ map shows  a non uniform emission 
for the ionized gas.  Two distinct bright regions are visible  near the centre, 
corresponding to a  polar ring-like structure.
Furthermore, a tail of ionized gas is visible, both from the monochromatic map
and from the velocity field, in the East side of the galaxy.  
The velocity dispersion map  shows a high dispersion toward the centre,
 with values between 60 and 70 km~s$^{-1}$.
The  RC in figure \ref{rc} shows a maximum velocity of 165 km~s$^{-1}$
and then a plateau extending almost up to 30\arcsec\ (3.5 kpc). Although 
the approaching and receding sides of the RC match well, the curve is not 
totally symmetric.  

\underbar{NGC 3659}~~~~~
The velocity field of NGC 3659 shows the rotation of a disk, but  the plateau 
is not as extended as in NGC 3655. The velocity map does not show the classical 
spider  pattern. Iso-velocities    
are not aligned with the major axis of the map. On the blue-shifted side, they are 
almost straight.  H$\alpha$ single component line profiles are very symmetric . 
  The monochromatic  H$\alpha$ map shows 
 strong emission regions on the NE of the 
galaxy. These large regions in that area 
 may be responsible for the behaviour of the velocity field. 
Interestingly, on the velocity dispersion map, high dispersion is present on 
the edge of the high emission regions, on the North. The lowest velocity 
dispersion corresponds to high emission regions.
The RC shows almost the same extension of  NGC 3655, but the shape is 
quite different.  The RC rises slowly
toward a maximum velocity of 85 km/s. A small plateau is present after
20\arcsec (1.6 kpc). The RC is not totally symmetric.  
  A small bump is 
visible at 22\arcsec (18kpc), the velocity raising to 85 km/s, but no 
peculiar features can be seen on the velocity map.  In the inner
9\arcsec (0.8kpc), the two sides do not agree, probably due to the effect 
of the bar.

\underbar{NGC 3684}~~~~~
The velocity field of NGC 3684 is not as typical  and regular as NGC 3655. 
Nevertheless, it shows a clear disk rotation pattern. Iso-velocity contours are 
almost straight toward the centre (as expected for a typical rotating disk). 
The position angle of the iso-velocities is not constant along the velocity 
field major axis. The pattern is particularly visible on the blue-shifted side. 
 H$\alpha$ map shows several bright emission regions, 
 limited to the central  part of the galaxy, and  
 few HII  regions around the disk.  
 The ionized gas velocity dispersion is low (between 30 to 45 km/s) compared to NGC 3659 
for example. It is not obvious, as for NGC 3659, that HII regions are located 
where the velocity dispersion is lower.  H$\alpha$ emission lines do not show 
multiple components.
Despite a peculiar velocity field,  both sides of the RC agree remarkably 
well.  The RC raises sharply starting from the inner 10\arcsec\ (0.7 kpc) to reach 
a small plateau of about 100 km~s$^{-1}$ stable for more than 10\arcsec\ (1.4 kpc). 
After that, the curve continues slowly to raise up to 150 km~s$^{-1}$ at 50\arcsec\ (3.6 kpc). 
The plateau might be due to the presence of two bright  HII regions on both side.

\underbar{NGC 3686}~~~~~
The velocity field of the galaxy is also characteristic of a rotating disk. 
The field is quite symmetric and no distortions are present 
(such as, large variation of the position angle versus radius), except 
on the centre where we can notice a distortion due to the presence of the bar.  
The H$\alpha$ monochromatic image  shows  bright HII regions along the spiral arms. 
Those regions are well defined in the UV image (see Figure 3). 
This galaxy is the only one showing a bright emission in the bulge. 
The small regions on the South are part of the lower spiral arm visible
 on the  UV image. The bar is barely visible on the H$\alpha$ image.
Contrary to the other galaxy, high velocity dispersion corresponds to high intensity
H$\alpha$  emission. Dispersion map shows a velocity dispersion of 75 km~s$^{-1}$
in the centre, while has values around 35-38 km~s$^{-1}$ in the rest of the galaxy. 
H$\alpha$ line profiles on the central area are broader than elsewhere but they are
symmetric, without signature of multiple components. The region of high velocity
dispersion is roughly coincident  with the location of the bar.
The RC shows the two side of the RC differ in the inner 12\arcsec\ (1 kpc), 
then agree almost perfectly to the end of the curve at 70\arcsec\ 
(5 kpc).  The influence of the bar certainly plays a role in the disagreement 
of both side of the RC.

\underbar{NGC 3691}~~~~~
The velocity field confirms the presence of a bar.   H$\alpha$ line profiles are symmetric
across the galaxy and there are no signatures of multiple components.  The H$\alpha$
monochromatic map shows several emission regions possibly on the spiral arms.
Contrary to NGC 3686, no emission is detected in the bulge.
Velocity dispersion map shows a higher value in the centre (around 50 km~s$^{-1}$), 
and  $\approx$35 km~s$^{-1}$ elsewhere. The region where the 
dispersion is higher corresponds to the bar.
The two sides of the {\bf folded RC  differ between $\approx$4\arcsec\ (0.5 kpc) and $\approx$20\arcsec\ (1.4 kpc)}.\\

\input{tab3a.tex}

\begin{figure}
\includegraphics[width=8cm]{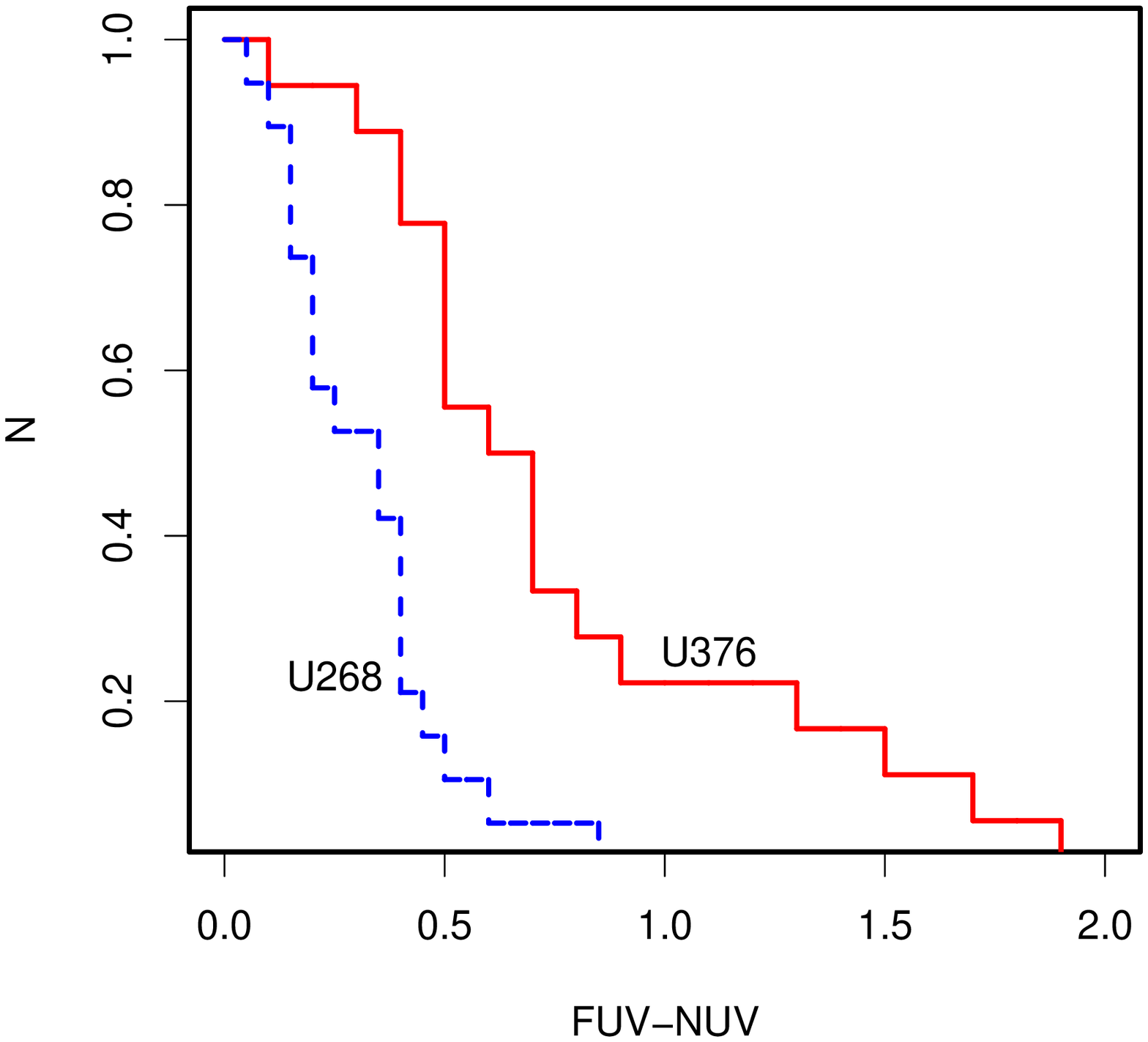}
\caption{Cumulative distributions of FUV - NUV colours of galaxies in  U268, U376.
According to a Kolmogorov Smirnov test, the null hypothesis that the two distribution are 
drawn from the same parent distribution can be rejected  at a confidence level $>99\%$. }     
\label{cum}
\end{figure}

\section{Discussion}

\subsection{U268 and U376 as evolving  groups}
U268 and U376  are dynamically different since their velocity dispersion differs by a factor $\sim$ 4 
(65 vs. 230 km$s^{-1}$).
 The morphological classification shows that U268 does not contain elliptical galaxies;
the overall population of ETGs is composed by S0s only, which are  about 24\% of the 
total population. This fraction is comparable with that found for S0s in the general
field by  \citet{Calvi12}. However,  \citet{Calvi12} found that Ellipticals  
are at least 20\% of the galaxy population even in the less massive environments. 

ETGs in U376 represent about 38\% of the galaxy population, lower than the average
(46\%) found for ETGs in field  \citep{Calvi12}. Our groups are then likely representative 
of very loose galaxy associations. Notice from Figure~1 (bottom left panels), that, at least in projection,
 U268 has a dispersed configuration at odds with U376 which is more compact than
 many other groups in the region and which, not surprisingly, show a larger fraction 
 of ETGs. 

Figure \ref{cum} shows the cumulative distributions of FUV-NUV colours of galaxies in  U268 and U376.
 The distribution gives the fraction of galaxies in each group having a colour   
greater (redder) than a given value of FUV-NUV.  E.g. $\sim$10\% of galaxies in U268 has FUV-NUV $>$ 0.5 AB mag    while in U376 they are nearly 80\%.
Using a Kolmogorov-Smirnov two-sample test, we tested the null 
hypothesis that the two distributions come from the same population, finding that they are different
(confidence level $>$ 99\%),   
galaxies in U268 being bluer than those in U376.\\

\begin{figure*}
 \includegraphics[width=8.5cm]{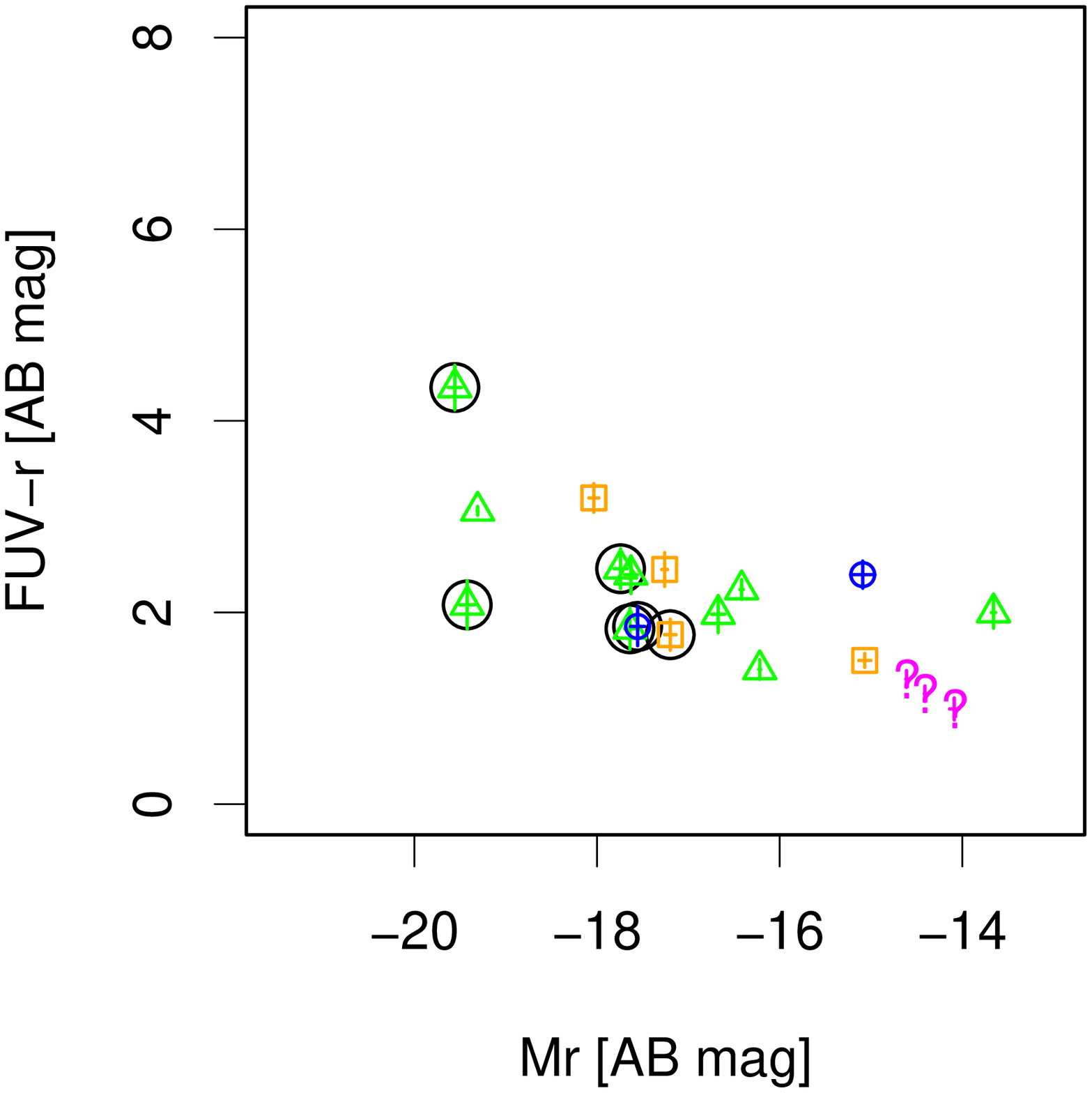}
\vspace{-1.8cm}
\includegraphics[width=8.5cm]{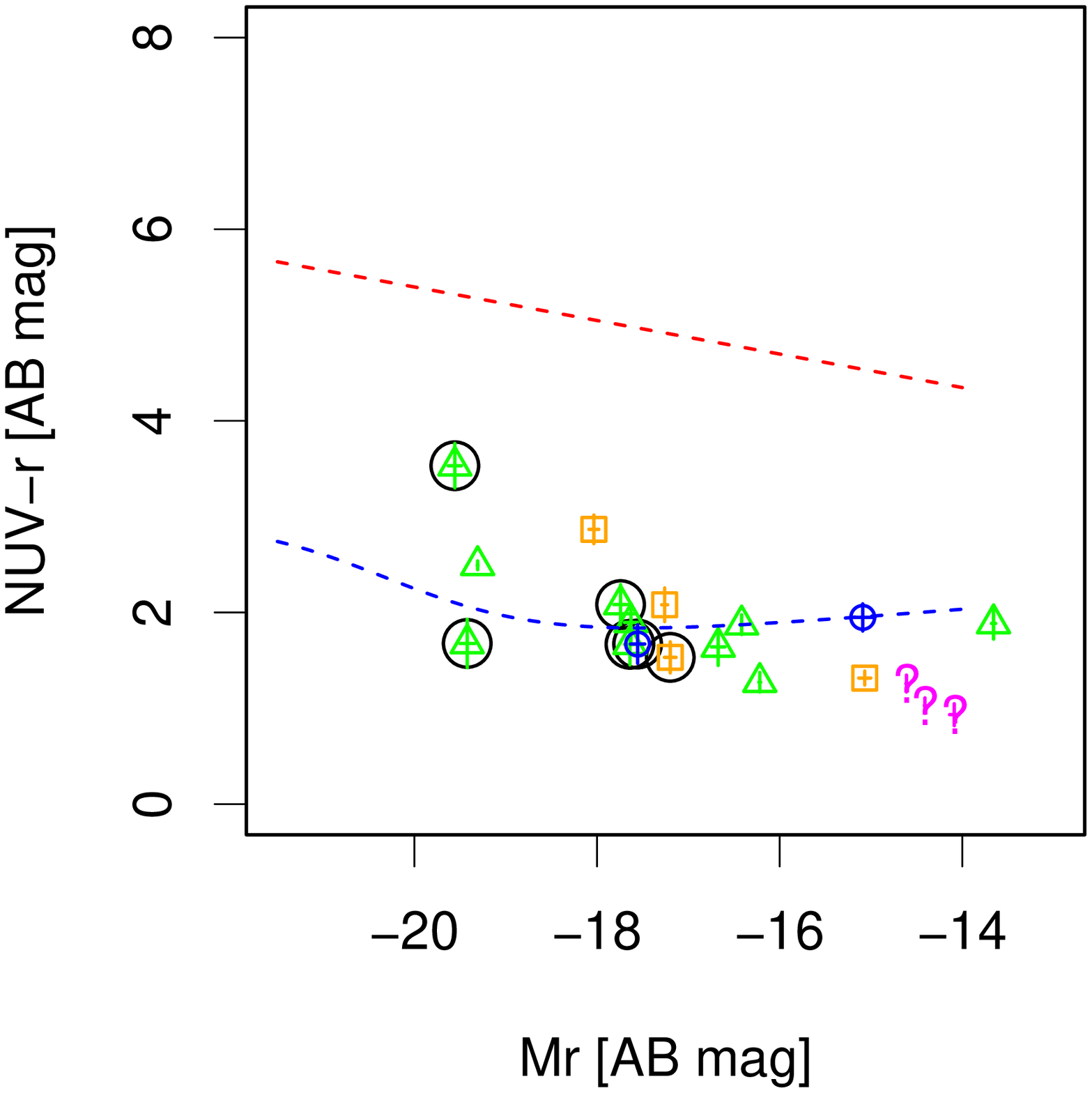}
 \includegraphics[width=8.5cm]{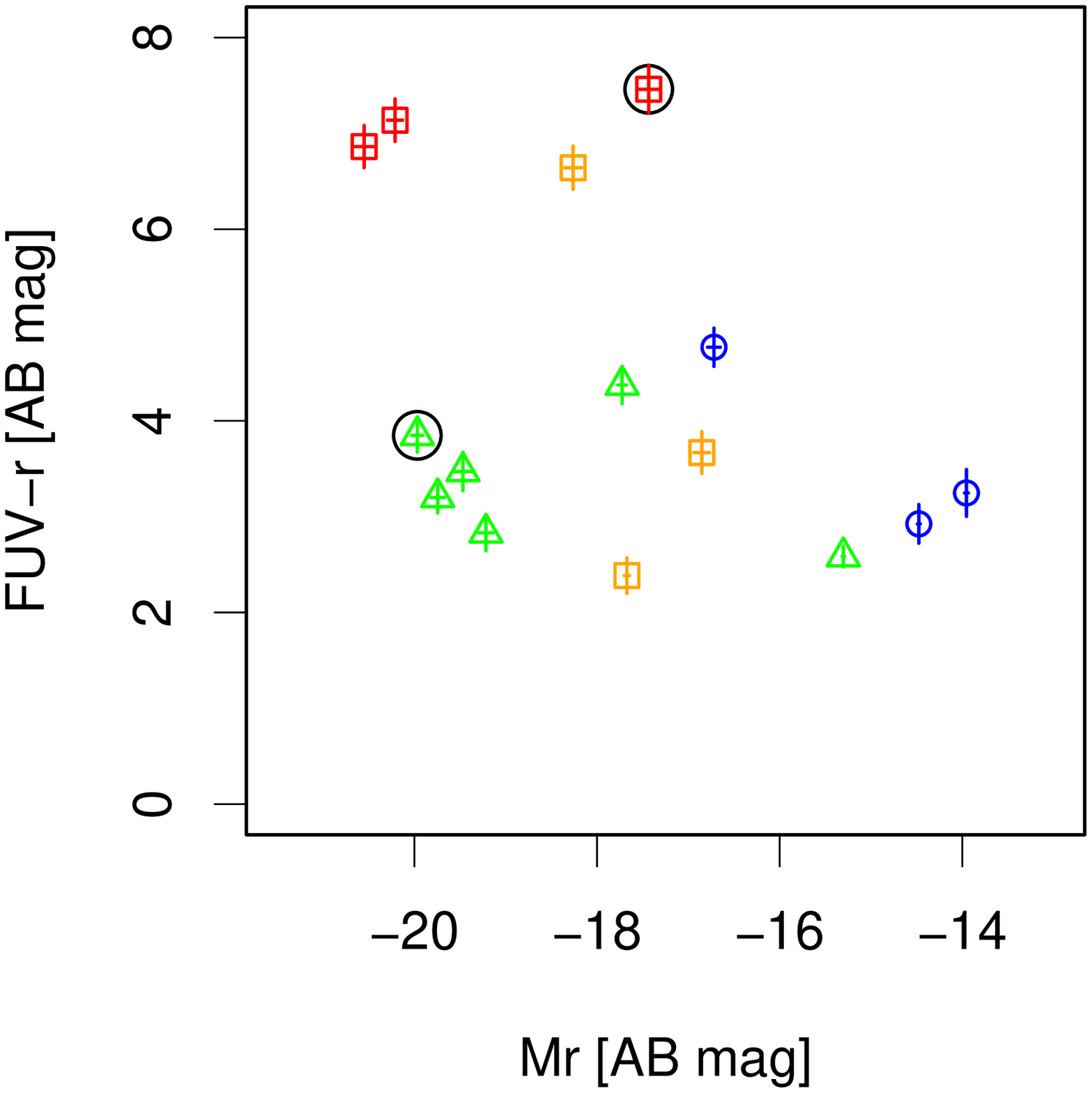}
 \includegraphics[width=8.5cm]{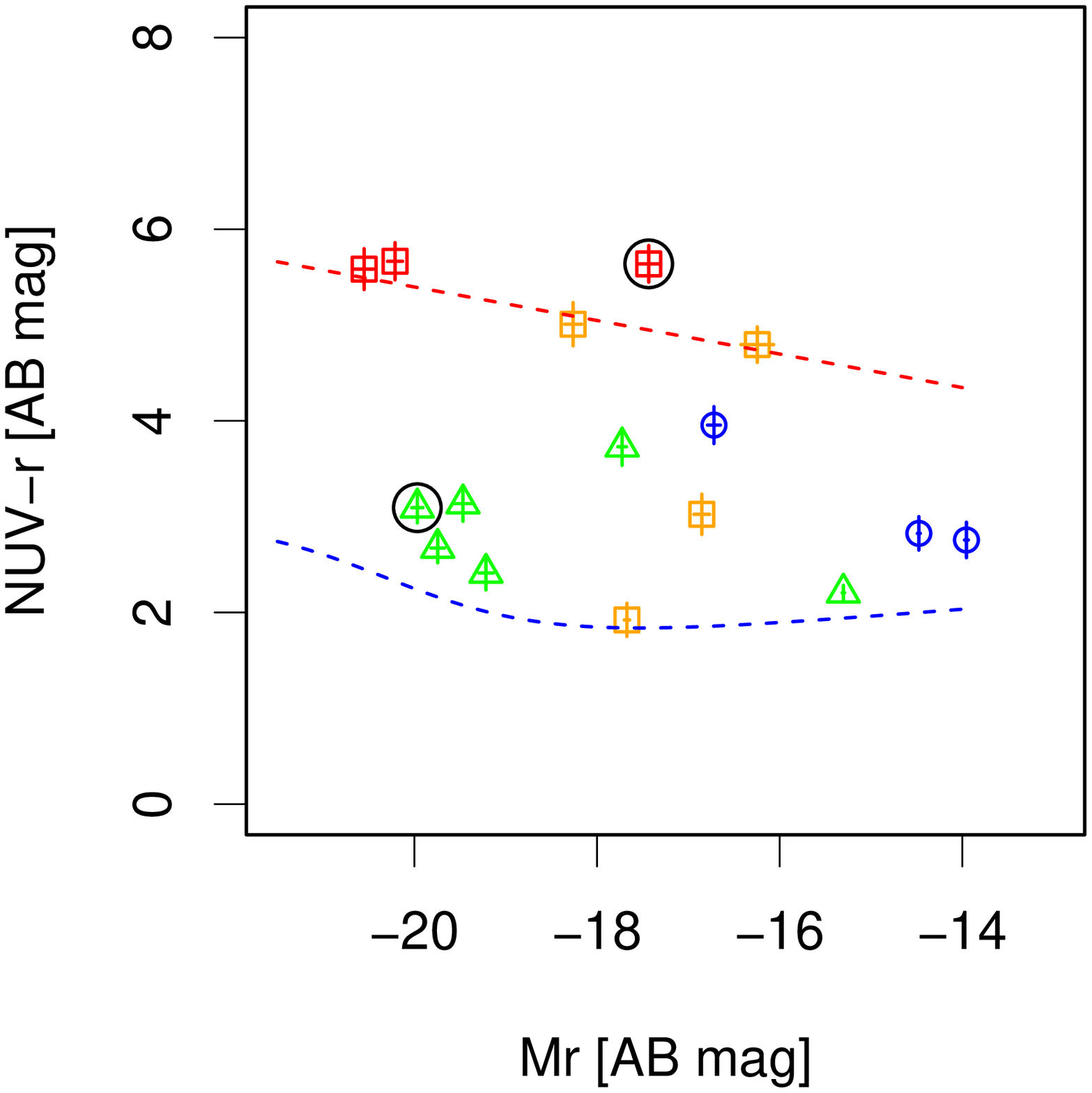}
 \caption{({\it Top}) UV - optical CMDs of the members in U268.  (M$_r$ vs. FUV -r ({\it left}), M$_r$ vs. NUV -r 
 ({\it right}). ({\it Bottom}): as top panels for U376 members.  
 In the M$r$ vs. NUV-r CMD, we overplot the \citet{Wyder07}
  fits to the red and blue galaxy sequences. Triangles indicate spirals, squares early-type galaxies 
  (red Ellipticals and yellow S0s),  circles irregulars. `?'  indicate galaxies without morphological classification (see our revised classification in  
  Table 3). Encircled symbols indicate galaxies with interaction signatures. Magnitudes of all galaxies but 
  SDSSJ095430.02+320342.0 were corrected by galactic  extinction.}
    \label{CM}
\end{figure*}

In Figure \ref{CM} we show the UV - optical CMDs  of members of the two groups.

In U268, the `red sequence', where passively evolving galaxies are located, is completely unpopulated.
 Even S0s (yellow squares), like MRK 408, populate either the `blue sequence' or 
the `green valley'. Most of the galaxies lie
in the `blue sequence'. Many bright spiral galaxies show signature of 
interaction  or morphological distortions (encircled symbols in  Figure \ref{CM}). 
The spiral NGC~3067, lying in the `green valley', is the reddest of the bright members.

On the contrary, U376 shows a well defined `red sequence' populated by both
ellipticals and S0s.  Also the shell galaxy NGC 3605   
is  located in  the `red sequence'. Since an accretion event 
is believed to produce shells \citep{Dupraz87},  the latter 
 have been likely produced by a dry accretion that did not ignite  star formation or 
 the galaxy already come back to the `red sequence' if shells are long lasting \citep[e.g.][]{Longhetti00}.
  Line strength indices of NGC 3607 
\citep{Rampazzo05, Annibali06} have been used to compute the
luminosity weighted age, metallicity and [$\alpha$/Fe] enhancement of the
galaxy. \citet{Annibali07} report a luminosity weighted  age of 3.1$\pm$0.5 Gyr
for NGC 3607,  high metallity ($Z_\odot$, nearly two times solar 
0.047$\pm$0.012) and  [$\alpha$/Fe]=0.24$\pm$0.03. The  age  suggests that  the galaxy may have 
had a recent star formation episode.
The galaxy shows emission lines with LINER characteristics \citep{Annibali10}.
Notwithstanding, NGC 3607 lies in the `red sequence' 
\citep[see also][for the photometry of the inner regions r$_e$/8]{Marino11b}.  
This fact is not unexpected: far-UV colours are very 
sensitive to recent ($<1$Gyr) star formation episodes  \citep[e.g.][]{Rampazzo07, Bianchi11}, 
 line-strength indices may trace better older episodes.
 Late-type galaxies are not found in the `blue sequence' but in the `green valley'.
Only NGC~3655 shows signature of on-going interaction but it shares with
the other spirals the `green valley' region.
 
\begin{figure}
 \includegraphics[width=8.5cm]{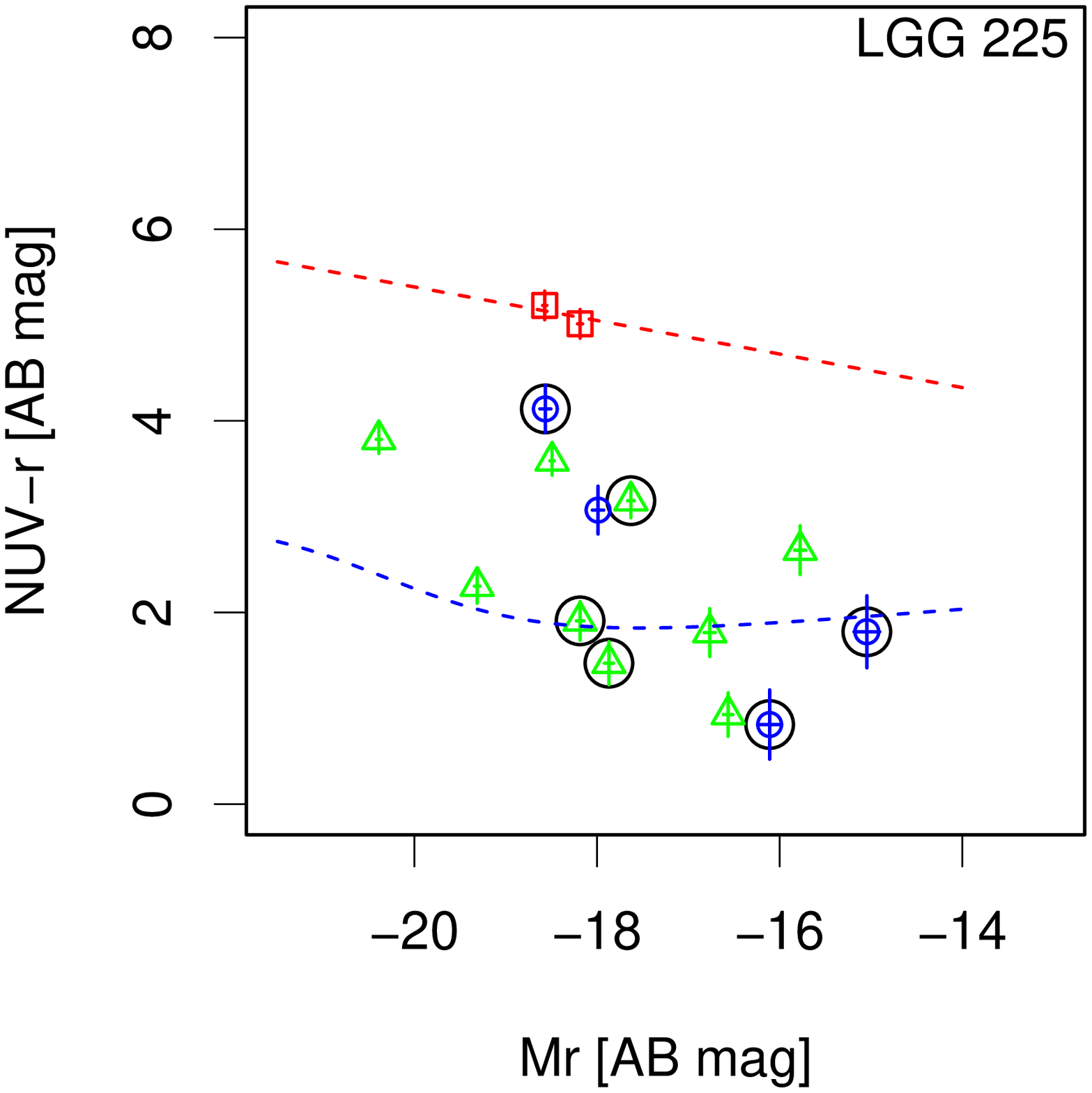}
 \caption{  M$_r$ vs. NUV -r for the LGG 225 association \citep[see][]{Marino10} which is
 also part of the Leo could as shown in Figure~1 mid right panel.  
Symbols are as in Figure~\ref{CM}.}
    \label{CM225}
\end{figure}

 Figure~\ref{CM225} shows  the  galaxies  of  the group LGG~225 \citep{Garcia93}, also
belonging to the Leo cloud  (Figure 1 right bottom panel)  in the (M$_r$ vs. NUV-r) plane, using the NUV and SDSS-r photometry  published in Paper I. 
Some LGG 225 members display interacting and  distorted morphologies  particularly evident in the UV images. 
UV colors suggest very recent episodes of star formation triggered by recent and on-going interaction.
 In the (M$_r$ vs. NUV-r) plane, the phase of LGG 225 appears 
intermediate between U268 and
 U376.  The two Ellipticals, located in the `red sequence', are  not the most massive and luminous galaxies in the group. 
No S0s are present.  The  `blue sequence' is populated by star forming and interacting galaxies. There is
 a large fraction ($\simeq$ 40\%) of galaxies in the `green valley'. 
 
 Summarising we suggest that in different regions of the Leo cloud the majority of
 galaxies, irrespective of their morphological type, are experiencing a
 transformation.  In particular,  in U376 this   
  transformation  brings the galaxies to the  `green valley'.

\subsection{Transforming  mechanisms}

In U268 the galaxies that show signature of interaction are all
Spirals (NGC~3003, UGC~5287, UGC~5326, NGC~3067,  PGC028169 
and UGC~5093). All these galaxies but NGC~3067,
that inhabits the `green valley', are located in the `blue sequence'. 

The kinematics of NGC 3003 and NGC 3067 have been studied by 
\citet{Epinat08b}. They found that the H$\alpha$ emission in NGC 3003
 is rather faint and have an asymmetric distribution as if the galaxy was disturbed 
 by a dwarf companion. Also the velocity field is rather asymmetric. 
 \citet{Epinat08b} found the signature of a bar in the velocity field
 of NGC~3607. This  galaxy is seen nearly edge-on with  dust
 patchy structure in its centre \citep{Carollo98}. Interaction signatures
 are visible in the 2MASS images \citep{Jarrett00}: a tidal arm/tail is detected
 in the SE direction.  This arm/tail appears closed as a ring in the
 SDSS image of Figure~2, while the nucleus is displaced to the NW
 with respect to the centre of the ring.
   
 The 2D kinematics of two additional spiral members 
 (UGC 5287 and UGC~ 5393) analysed here, shows that 
 both galaxies are barred, although the bar signatures are not visible in the
 velocity field of UGC~5393. In UGC~5287 we detect a double component
 H$\alpha$ profile from the northern galaxy outskirts to the bar region, likely a
 signature of an infalling gas component. \citet{Rampazzo06} interpreted a 
 similar feature in the interacting 
spiral NGC~470 as signature of infalling gas.  
If we consider the southern bright region as a possible companion, the rotation curve can be
 separated in two parts, the northern one, typical of a disk galaxy, and the southern one, 
 very irregular indicating that the companion has a disturbed motion.
 
 We find that NGC~3011 and UGC~5326 both have a ring structure.
 \citet{Paz03} found that  NGC~3011   
 has an inner and an outer ring.   They suggest  that the inner disk could  be produced by  a 
starburst-driven shock interacting with the surrounding
 interstellar medium. \citet{Comeron10} classify the galaxy as a S0a, and did
 not find a signature of  bar, which is generally connected to the 
 presence of rings.   UGC~5326 may be suffer   a head-on collision.

 Also S0s lie in the `blue sequence': they are  blue compact dwarfs, like
 MRK 408 \citep{Hunter04} with an intense nuclear star
 formation and even  ``Wolf-Rayet (WR) galaxies'' as IC 2524 (MRK 411), i.e. galaxies
 with signature of very recent star formation (WR stars appear $2\times 10^6$ 
 years after a star formation episode and disappear within some $5\times  10^6$
\citep{Brinchmann08}. Dwarf early-type galaxies may have acquired
 gas by interaction with a gas rich donor as mentioned above in the case of the
velocity field of NGC~3003 \citep[see e.g.][]{Rampazzo95, Domingue03}.
 
Summarising, the main evolutionary mechanism at work in U268 is
 the interaction/collision between galaxies which could lead to the generation
 of  bars and rings, possibly igniting a star formation episode 
 \citep{Noguchi90}. Kinematical studies 
 confirm such  picture showing that both asymmetries in the
  velocity field and  bar are common features in the  Spirals of the group.

U376, which appears  more concentrated than U268, 
shows a larger number of ETGs. 
We refer to a forthcoming paper for the analysis of the SEDs of ETGs, that 
will give us further insight into the processes driving their evolution.

 NGC~3626 presents an extended gas counterrotation \citep{Ciri95},
with ionized and atomic hydrogen rotating in opposite direction to the stars.
\citet{Em04} shows that the velocity field of NGC~3608 displays the
 presence of a counter-rotating core \citep{Jedrzejewski88,  
  Halliday01} in the central  region (inner 8$\arcsec$).  
 Counterrotations  
 are widely interpreted as generated by accretion events \citep{Galletta96, Corsini98}.
Neither in NGC~3605 nor in NGC~3608 CO has been detected  
 \citep{Sage07}. Mid infrared Spitzer-IRS
spectra of NGC~3608  show that the galaxy is passively evolving, 
without PAH emission (Rampazzo et al. 2012, in prep.). 
NGC 3608 has a low X-ray luminosity 2.7$\times$10$^{39}$ erg~s$^{-1}$ 
\citep{David06} and does not show an AGN activity although there has been a past 
``activity'' witnessed by the presence of cavities in the X-ray emission
 \citep{Cavagnolo10}.  
 
A tentative picture of the ETGs  evolution in U376 may be connected to 
accretion episodes that leave their morphological, kinematical,  and photometric 
(young age of the stellar population) imprinted on the ETGs.
Moreover, signatures of secular evolution are detected in a large fraction of Spirals.
 There is a larger presence of bars  with respect to U268 (29\% vs. 10\%).
The bar signature is detected also in the  velocity
field together with several kinds of asymmetries  as described in Section 4.2.
Only  NGC 3655 shows recent/on-going interactions. What bring them to the `green valley'? Can be the
higher concentration of the group the answer?
The study of 
\citet{Sengupta06} about the HI content of galaxies in 
loose groups give us further insight about this point. These authors compute the HI deficiency of each galaxy in the group ,  as
  def$_{HI}$ = log$\frac{(M_{HI}}{D^2}\mid_{pred}$ - log$\frac{(M_{HI}}{D^2}\mid_{obs}$, with $M_{HI}$ in solar masses and D, the RC3 galaxy diameter, in kpc.  The 
average of the predicted \citep{Sengupta06} minus the observed HI surface densities of all the galaxies of a group is used as a measure 
of the HI deficiency of the group.
 
They found that galaxies in groups with diffuse 
X-ray emission, on average, are HI deficient, and have 
lost more gas compared to those in groups without X-ray emission.  
The group U376  is included in the  \citet{Sengupta06}  study 
 (NGC 3686 in their Table 3)  among groups with diffuse X-ray emission.     Their 21 members   correspond, with three exceptions, to our 
 independent list reported in Table~1. 
A large fraction of spirals/Irr in the list  (UGC~6296, UGC~6341, NGC~3592,
 PGC 035087, PGC~35096, NGC~3686 and NGC~ 3691) are HI deficient. 
 Conversely, UGC~6320, NGC~3659, NGC~3655, NGC~3681,
NGC~3684 show an excess of HI.    
\citet{Sengupta06}  suggest that tidal aided ram-pressure stripping 
 and evaporation  are the possible mechanisms leading to the excess gas loss 
 found in groups with diffuse X-ray.  U376  as a whole is only barely 
 HI deficient (0.07$\pm$0.062)
  but single galaxies  are likely  undergoing a 
 transformation by mechanisms sketched above.
 
We estimated the HI deficiency of U268,  using the HI  flux of the galaxy members given in NED
{\bf \citep{Springob05, Huchtmeier89, Schneider90}}.
 We find that the spirals 
 UGC~5446 and NGC~3118, $\sim$ 1.6 Mpc apart from the center of the group  are the HI richest members , while NGC~3067
 and UGC~05282 are HI deficients.
   NGC~3003, UGC~05287, UGC~5326 have def$_{HI}$ of about   -0.05, 0.05 and 0.01 respectively. The def$_{HI}$ of the group U268 
is $\sim$ -0.01.

\section {Summary} 
 We have presented the photometric and kinematic analysis of  U268 and U376, two galaxy groups  
  in the Leo cloud.
Starting from the membership of the two groups 
defined in  \citet{Ramella02}, we enriched  
 the original group  with
members sharing the same spatial and velocity extent given  in the {\tt HYPERLEDA} database.
The velocity dispersion of the two groups differs by a factor $\sim$ 4 (67 vs. 240 kms$^{-1}$ in U268 and U376 respectively).
U268 does not contain Ellipticals, the overall population of ETGs is composed by S0s only, which are a small (24\%) fraction of the total, 
and comparable with  that found in the field \citep{Calvi12}.
ETGs in U376 are 38\% of the total whose 8\% of Ellipticals and 30\% of S0s.   
We investigated the presence of substructures and the environment of  each group.
We find substructure to be evident from the Dressler-Shectman test  in both groups.
Galaxy members of U376 have a more compact configuration than U268 with 
the surrounding of U376 richer of galaxies and crowded by tight associations.

We obtained the FUV and NUV  ({\it GALEX}) and optical (SDSS) surface photometry of
bright members and integrated photometry of all members. 
We have also obtained  2D kinematics of a set of bright spirals in the two
groups looking for distortions in the velocity field resulting from galaxy-galaxy interaction.
A large fraction of galaxies in U268 shows interaction signatures (60\% vs. 13\%) both from their photometry and kinematics.
 
(FUV - NUV) colours of galaxies in U268 are bluer 
than those in U376. 
No galaxies are  populating the `red sequence' of U268,  all  lying in the `blue sequence' and
in the `green valley', including S0s. At odds, the `blue sequence' of U376 is 
un-populated with respect to U268,
and  a large set of galaxies populates the `green valley'. At the same time, very few galaxies
in U376 show a distorted morphology due to on-going interactions.  
  \citet{Sengupta06}  show
that U376 members are, at least partly, depleted of gas.  We estimate that  
the members of U268 are less deficient in HI  than U376. 
Galaxy-galaxy interaction and gas removal seem the basic mechanisms leading
the transformation of late-type galaxies in our groups in Leo. ETGs in U376 show 
signatures of past accretion events, e.g. shells, stellar population rejuvenation, counter
rotating core.   
{\bf  The presence of  rings, likely connected to the internal secular evolution   (e.g. NGC 3011 in U268, NGC 3681 in U376),
 polar rings produced by minor/major accretion
events  and collisional rings recognizable from
the off-centered nucleus (e.g UGC 5326 in U268, UGC 6320 in U376), have been found.  

We also compared the CMDs of the two groups analysed here
with LGG225, a group for which comparable measurements are available in  Paper I}.
 U268 is likely in a early evolutionary stage in comparison to U376,
 with LGG225 in between. 

In a forthcoming paper we will use the multi-wavelength SEDs  
and the kinematics data presented here
to obtain (1) the  star formation history and the stellar masses of ETGs  
using chemo-photometric SPH models \citep[see][]{Bettoni11}, (2)
the stellar and kinematic masses of spiral members, 
(3) a luminosity-weighted kinematical and dynamical analysis of each group.
We will also expand the study to other 
11 groups, with  a wide range of characteristics (in richness of members, velocity dispersions, morphological types)
for which we have UV and  optical  images.

\section*{Acknowledgments} 	    
 
 A.M. and L.B. acknowledge support from NASA grant
NNX10AM36G.   
A.M., R.R., D.B., L.B., and P.M. acknowledge financial contribution from the
agreement ASI-INAF I/009/10/0.
This work has been partially supported by the Padova University funds 2011/12 (ex 60\%).
M. R.  acknowledges support from grants P-82389 from CONACYT and IN102309 from DGAPA-UNAM.	    
 {\it GALEX} is a NASA Small Explorer, launched in April
2003.			   	    
This work is based on  {\it GALEX} GI data from program GI6 - 6017 and {\it GALEX} archival data taken from MAST (http://galex.stsci.edu). 
 STScI is operated by the Association of Universities for Research in Astronomy,
Inc., under NASA contract NAS5-26555. Support for MAST for non-HST data is
provided by the NASA Office of Space Science via grant NAG5-7584 and by other
grants and contracts.	   	    
We acknowledge the usage of the {\tt HYPERLEDA} (http://leda.univ-lyon1.fr) 
and NED databases.
{\bf We thank the anonymous referee for useful comments which helped to improve the paper.} 			   	    
			   	    
{\it Facilities:}  {\it GALEX}, Sloan
 
 \bibliographystyle{mn2e}   	    
 \bibliography{amarino}	   	    

\appendix

 \section{UV and optical  images of the added members of U268 and U376}

\begin{figure*} 
 \includegraphics[width=7cm]{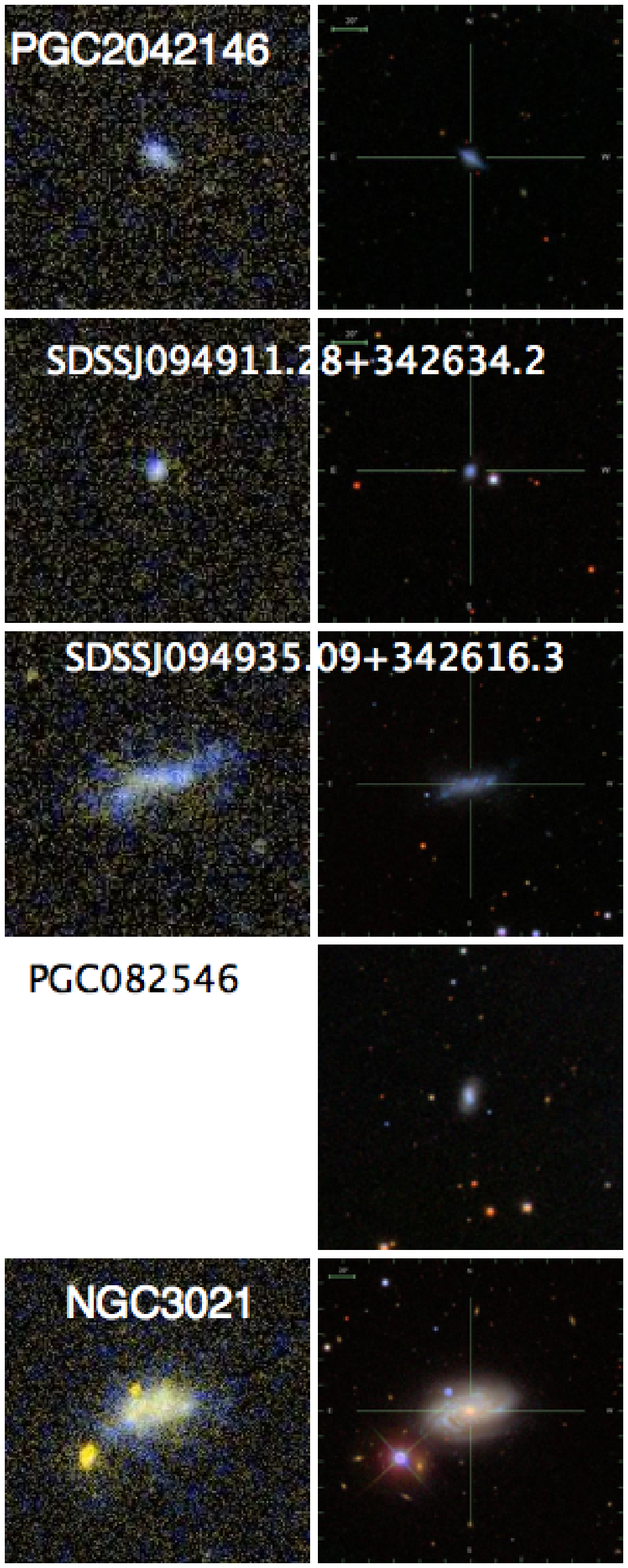}
\hspace{0.8cm}
\includegraphics[width=7cm]{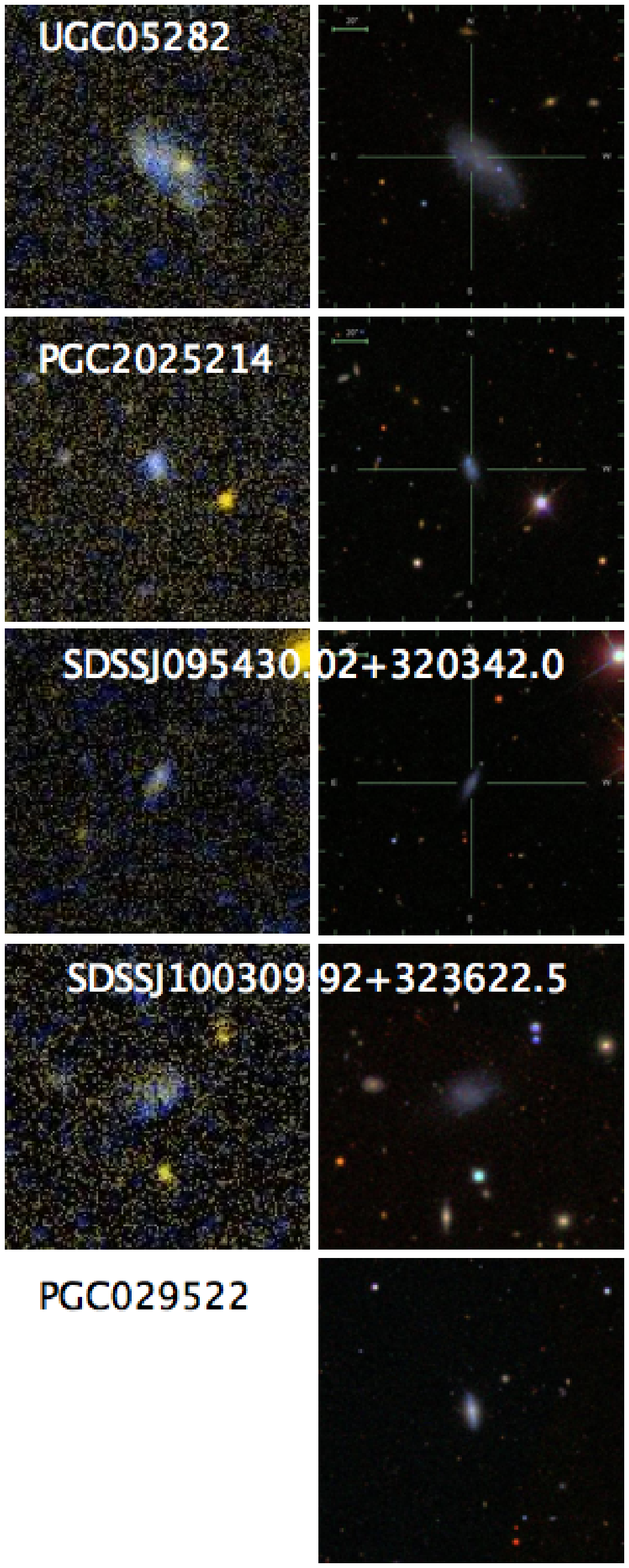}
 \caption{Color composite UV (FUV blue, NUV yellow, first and third panel from left) and optical
(SDSS, g blue, r green, i red, second and fourth panel)  images  of the new member galaxies  in U268. }
\label{N1}
 \end{figure*}

\begin{figure*} 	
 \includegraphics[width=7cm]{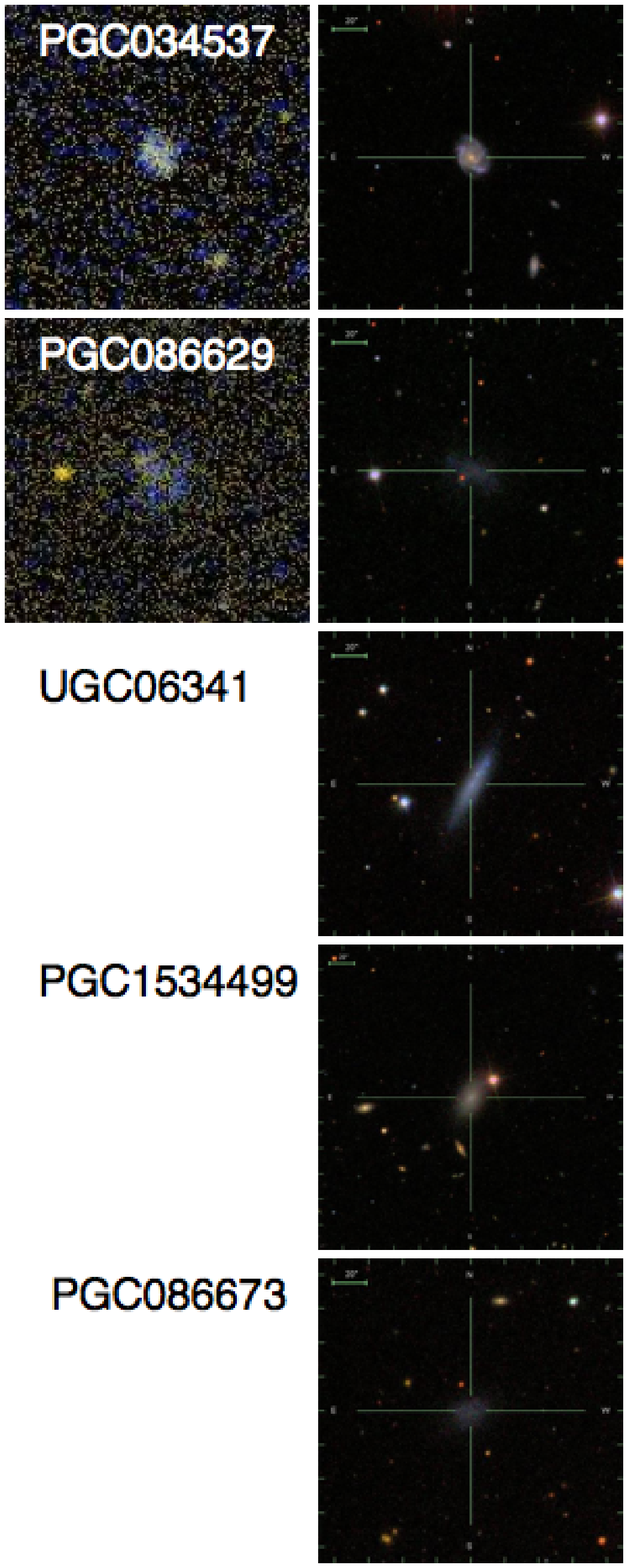}
\hspace{0.8cm}
\includegraphics[width=7cm]{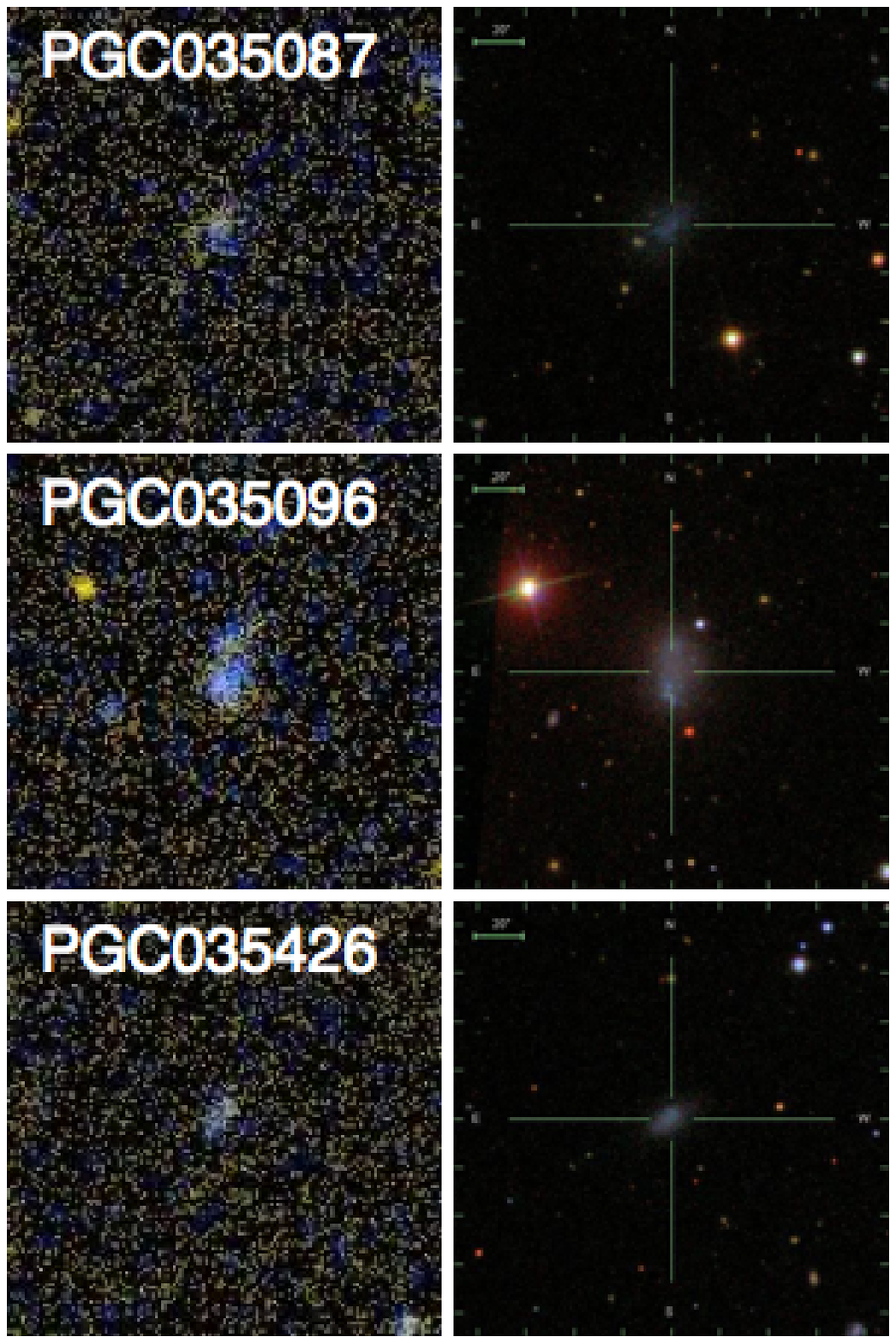}
 \caption{Color composite UV (FUV blue, NUV yellow, first and third panel from left) and optical (SDSS, g blue, r green, i red, second and fourth panel)  images  of the new member galaxies of U376. }
\label{N2}
 \end{figure*}

Figures \ref{N1} and \ref{N2} show the UV and optical colour composite images of the added members of
U268 and U376 respectively. The image of  PGC2016633 in U268 is shown in Figure 2.

 \section{Environment of U268 and U376}
 Tables \ref{a1} and \ref{a2} compile the main characteristics of the galaxies in a box of 4 Mpc $\times$ 4 Mpc centred on the B brightest galaxies of  U268 and U376 respectively and having a heliocentric radial velocity within $\pm$ 3$\sigma$ of the group mean velocity in the catalog of \citet{Ramella02}.   
  \input{tabA1.tex}

 \input{tabA2.tex}

\end{document}

%% file: tab1.tex
\begin{table*}
	\caption{Characteristics of the galaxy members$^a$.}	 
	 \scriptsize
	\begin{tabular}{llllllllllll}
	\hline\hline
	   Group  & RA & Dec.&  Morph.& RC3 & E(B-V)$^b$& Incl. & logD$_{25}$ &logr$_{25}$ & P.A.& Mean Hel.& B$_T$ \\ 
	   Galaxies & (J2000) & (J2000) & type & type & &  &   &&  &  Vel  & \\
		   & [h:m:s] & [d:m:s]   &  &  &[mag]  & [deg] &[log(0.1 arcmin)] & &[deg] &  [km/s] & [AB mag]\\
				  
	\hline
	U268   & & & & & & & & & &  \\
	\hline
 MRK 0408   &  09 48 04.8 &+32 52 58 & S0$^f$ 	& -2.0	&0.017 & 40.5  &	0.84& 0.11 &   163.9  &1470$\pm$40$^c$ & 14.84$\pm$0.48\\  
 NGC 3003   &  09 48 35.7 &+33 25 17 & Sbc	& 4.3	&0.013 & 90.0  &	1.68& 0.65 &	78.3  &1498$\pm$36$^c$ & 12.09$\pm$0.10 \\  
 NGC 3011   &  09 49 41.2 &+32 13 16 & S0	&-1.8	&0.016 & 38.6  &	0.92& 0.08 &	59.5  &1548$\pm$36$^c$ & 14.38$\pm$0.44\\  
 UGC 05287  &  09 51 28.1 &+32 56 35 & Sc	& 5.9	&0.014 & 41.2  &	0.98& 0.12 & 15$^e$   &1482$\pm$42$^c$ & 14.68$\pm$0.42 \\  
 UGC 05326  &  09 55 24.5 &+33 15 47 & IB	& 9.8	&0.016 & 16.9  &	0.97& 0.02 &	0$^d$ &1415$\pm$45$^c$ & 14.46$\pm$0.40 \\  
 IC 2524    &  09 57 33.0 &+33 37 11 & S0-a	&-1.0	&0.013 & 63.1  &	0.87& 0.23 &	61.8  &1487$\pm$10$^c$ & 14.84$\pm$0.49 \\  
 NGC 3067   &  09 58 21.3 &+32 22 11 & SABa	& 2.1	&0.015 & 81.8  &	1.31& 0.49 &   104.3  &1491$\pm$36$^c$ & 12.57$\pm$0.12 \\  
 UGC 05393  &  10 01 42.1 &+33 08 12 & SBd	& 8.0	&0.013 & 68.8  &	1.14& 0.28 &   117.0  &1448$\pm$25$^c$ & 14.79$\pm$0.38 \\  
 UGC 05446  &  10 06 30.9 &+32 56 49 & Sc	& 5.9	&0.014 & 80.5  &	1.12& 0.61 &	49.8  &1383$\pm$42$^c$ & 15.22$\pm$0.40 \\  
 NGC 3118   &  10 07 11.6 &+33 01 40 & Sbc	& 4.1	&0.018 & 90.0  &	1.32& 0.67 &	38.6  &1315$\pm$37$^c$ & 14.18$\pm$0.35 \\  
\hline
PGC2016633\#     &09 48 02.68 &+32 54 01.7              &    &     & 0.016   &   53.1 &     0.63 &     0.21 &  165.0 &     1552$\pm$47 &  17.09$\pm$0.50 \\
 SDSSJ094838.45+332529.1$^*$\# & 09 48 38.45 &+33 25 29.1    &	    &	    &      &	     &       &	   &   86.8 &	  1419$\pm$31	 &  	\\
PGC2042146\#     & 09 49 03.1 &+33 59 28.95	      &     &	    &0.010 &	44.5 &     0.50 &     0.14 &   53.8 &	  1494$\pm$2	 &  17.69$\pm$0.50 \\
 SDSSJ094911.28+342634.2\# &  09 49 11.28 &+34 26 34.2  &     &       &0.012 &     &       &  &164.1 &                     1489$\pm$2 &           \\
 SDSSJ094935.09+342616.3\# & 09 49 35.09 &+34 26 16.3   &       &     &0.010&     &      &  & 98.3&                            1488$\pm$2 &   \\
PGC082546\#      & 09 50 20.9 &+33 35 02                & Sc  &	4.6 &0.015 &	46.3 &     0.56 &     0.15 &  171.0 &	  1586$\pm$29	 &  16.89$\pm$0.36 \\
NGC3021\#        & 09 50 57.1 &+33 33 13                & Sbc &	4.0 &0.014 &	55.7 &     1.13 &     0.23 &  108.3 &	  1540$\pm$3	 &  12.38$\pm$ 0.27 \\
SDSSJ095058.02+333319.3$^*$\# & 09 50 58.02 &+33 33 19.3    &	    &	    &0.014 &         &	        &   &  125.7 &	  1596$\pm$3	 &    \\
UGC05282\#	&09 51 10.4 & +33 07 53	              & Sm  &	8.8 &0.014 &    67.6 &     0.97 &     0.27 &   46.0 &	  1557$\pm$10	 &    	   \\
PGC2025214\#	&09 53 45.2 & +33 09 52.0	      &     &	    &0.013 &	57.2 &     0.55 &     0.25 &   18.5 &	  1520$\pm$35	 &   17.92$\pm$ 0.50 \\
SDSSJ095430.02+320342.0\# & 09 54 30.02 &+32 03 42.0    &     &	    &      &         &          &	   &  152.9 &     1421$\pm$2     & 	  \\
SDSSJ100309.92+323622.5\# & 10 03 09.92 &+32 36 22.5    &     &	    & 0.015&         &	        &          &  123.3 &     1562$\pm$96    & 	  \\
SDSSJ100714.59+330221.5$^*$\# & 10 07 14.59 &+33 02 21.5    &     &	    &      &         &	        &  &   32.5 &     1252$\pm$4     & 	 \\
PGC029522\#  &	10 08 58.0 & +32 00 38 	              & Sc  & 4.7   &0.018  &   57.7 &	 0.66   &     0.25 &   11.8 &	  1468$\pm$34    &   16.75$\pm$ 0.44 \\

\hline \hline	
 											 	      		       
	U376    & & & & & & & & & &  \\
\hline
NGC 3592	&11 14 27.5 &+17 15 34&  Sc   &     5.3  &0.015 &  79.4 &	  1.33&       0.57  &  117.3 &1298$\pm$20$^c$  &  14.31$\pm$0.30 \\
NGC 3599	&11 15 27.0 &+18 06 37&  S0   &    -2.0  &0.021 &  28.3 &	  1.38&       0.04  &  0$^d$ &812$\pm$31$^c$  &  12.72$\pm$0.08 \\
NGC 3605	&11 16 46.6 &+18 01 01&  E    &    -4.5  &0.021 &  90.0 &	  1.10&       0.24  &	19.1 &695$\pm$39$^c$  &  13.00$\pm$0.28 \\
UGC 06296	&11 16 51.0 &+17 47 55$^b$&  I&    10.0  &0.019 &  90.0 &	  1.11&       0.48  &  168.0 &909$\pm$26$^c$  &  14.02$\pm$0.41 \\
NGC 3607	&11 16 54.7 &+18 03 06&  E-SO &    -3.1  &0.021 &  34.9 &	  1.66&       0.06  &  120.0 &955$\pm$41$^c$  &  10.77$\pm$0.17 \\
NGC 3608	&11 16 59.0 &+18 08 55&  E    &    -4.8  &0.021 &  47.0 &	  1.50&       0.08  &	80.0 &1199$\pm$42$^c$ &  11.41$\pm$0.26 \\
CGCG 096-024	&11 17 58.0 &+17 26 29&       &          &0.020 &  28.9 &	  0.94&       0.06  &  157.3 &807$\pm$66$^c$  &  14.94$\pm$0.29 \\
UGC 06320	&11 18 17.5 &+18 50 48&       &          &0.023 &  20.3 &	  1.13&       0.03  & 0$^d$  &1125$\pm$5$^c$  &  13.67$\pm$0.41 \\
UGC 06324	&11 18 22.1 &+18 44 18&  S0   &    -1.9  &0.022 &  90.0 &	  1.14&       0.33  &  166.8 &1068$\pm$38$^c$ &  14.61$\pm$0.32 \\
NGC 3626	&11 20 03.9 &+18 21 24&  S0-a &    -1.0  &0.020 &  56.1 &	  1.47&       0.18  &  156.3 &1570$\pm$37$^c$ &  11.64$\pm$0.23 \\
NGC 3655	&11 22 54.7 &+16 35 24&  Sc   &     5.0  &0.025 &  47.1 &	  1.18&       0.16  &	31.8 &1490$\pm$39$^c$ &  12.16$\pm$0.07 \\
NGC 3659	&11 23 45.3 &+17 49 05&  SBd  &     7.8  &0.019 &  68.8 &	  1.29&       0.28  &	59.2 &1282$\pm$39$^c$ &  12.76$\pm$0.48 \\
NGC 3681	&11 26 29.8 &+16 51 48&  Sbc  &     4.0  &0.026 &  15.2 &	  1.24&       0.01  &  0$^d$ &1236$\pm$53$^c$ &  12.26$\pm$0.16 \\
NGC 3684	&11 27 11.2 &+17 01 48&  Sbc  &     4.0  &0.026 &  50.8 &	  1.36&       0.18  &  126.2 &1131$\pm$39$^c$ &  12.15$\pm$0.11 \\
NGC 3686	&11 27 44.1 &+17 13 26&  SBbc &     4.1  &0.024 &  42.0 &	  1.46&       0.12  &	22.7 &1096$\pm$36$^c$ &  11.84$\pm$0.07 \\
NGC 3691	&11 28 09.4 &+16 55 14&  SBb  &     3.0  &0.026 &  47.4 &	  1.17&       0.15  &	22.0 &1081$\pm$39$^c$ &  12.48$\pm$0.17 \\
\hline
PGC 034537\# &  11 18 06.0  &+18 47 53       & Sc    &   4.9 &0.023 & 23.3 &0.73 & 0.04 &	 & 1140$\pm$7 & 16.19$\pm$0.38 \\
PGC 086629\# &  11 18  21.4  &+17 41 51       & I     &   9.8&0.021 & 0.0  &0.89 & 0.00 &	 & 1055$\pm$5 & 	 \\
UGC 06341\#  &  11 20 00.7  &+18 15 38       & Sd    &   7.9 &0.019 & 90.0 &1.00 & 0.44 &  152.1 & 1611$\pm$42& 15.96$\pm$0.39 \\
PGC1534499\# &  11 21 25.41 &+17 30 35.16    & Sm    &   9.0 &0.024 & 65.5 &0.92 & 0.25 &  145.0 & 979$\pm$57 & 16.30$\pm$0.50 \\
PGC 086673\# &  11 22 59.3  &+17 28 27       & I     &  10.0 &0.021 & 0.0  &0.79 & 0.00 &	 & 1383$\pm$4 &        \\
PGC 035087\# &  11 25 01.8  &+17 05 09       & Sm    &	9.0  &0.025 &	   &	 &	&	 & 1208$\pm$5 &        \\
PGC035096\#  &  11 25 10.8   & +16 53 04     & I       & 9.5 &0.026&  51.0 & 0.76 &  0.15 &  147.0 & 1021$\pm$3 &  15.96$\pm$1.17\\
PGC 035426\# &  11 29 54.4  &+16 25 46       &Im$^b$ &       &0.027 &	   &	 &	&	 & 1067$\pm$5 &       \\

 \hline
\end{tabular}
 										   
$^a$data from \tt HYPERLEDA http://leda.univ-lyon1.fr \citep{Paturel03}. 					   
$^b$ Taken from NED. 												   
$^c$ Taken from \citet{Ramella02}. 
$^d$ no value in HYPERLEDA and RC3. 
$^e$ Taken from RC3. $^f$ Taken from \citet{Petrosian07}. 
\#{\bf Added members  from  Hyperleda}.
$^*$ SDSS galaxy misidentification.
 \label{tab1}
\end{table*}

%% file: tab2.tex
\begin{table}
	\caption{Journal of the {\it GALEX} observations.    }	 
	 \scriptsize
	\begin{tabular}{llll }
	\hline\hline
	   Group  & FUV & NUV  \\ 
	   Galaxies & Exp. Time & Exp. Time \\ 
		   & [sec]) & [sec]  \\ 
				  
	\hline
	U268   & & \\ 
	\hline
 MRK 0408   & 186& 186 \\  
 NGC 3003   & 186&2579  \\  
 NGC 3011   & 1543.05&1688\\  
 UGC 05287  & 206&1533  \\  
 UGC 05326  & 278& 278  \\  
 IC 2524    & 278& 1626 \\  
 NGC 3067   & 200& 1884 \\  
 UGC 05393  & 109& 1643 \\  
 UGC 05446  & 112& 1661  \\  
 NGC 3118   & 112& 1884  \\  
\hline
PGC 2016633 & 186 &186 \\ 
PGC 2042146 & 192 & 192 \\
SDSS J094911.28+342634.2 & 192 &192 \\
SDSS J094935.09+342616.3 & 192 & 192 \\
NGC 3021 &206 & 206 \\
UGC 05282 &  206 & 1533 \\
PGC 2025214 & 206 & 1533\\
SDSS J095430.02+320342.0 & 199 & 420 \\
SDSS J100309.92+323622.5 & 109 & 109\\
\hline\hline														 	      
	U376    & & \\ 
\hline
NGC 3592	&3715&3715\\
NGC 3599	&2468.05&2468.05\\
NGC 3605	&2468.05& 2468.05\\
UGC 06296	&2468.05& 2468.05\\
NGC 3607	&2468.05& 2468.05\\
NGC 3608	&2468.05& 2468.05\\
CGCG 096-024	&135&1658\\
UGC 06320	&90&90\\
UGC 06324	&90&90\\
NGC 3626$^1$	& &\\
NGC 3655	&106.1&106.1\\
NGC 3659$^1$	&&\\
NGC 3681	&106.1&106.1\\
NGC 3684	&107.05&107.05\\
NGC 3686	&105&105\\
NGC 3691	&107.05&107.05\\
\hline
PGC 034537 &  93 & 93\\
PGC 086629 & 135 & 135\\
PGC 035087 & 112 & 112\\
PGC 035096 & 112& 112\\
PGC 035426 & 110 & 110\\

\hline														 	      

\end{tabular}

$^1$ No {\it GALEX} observations for these galaxies. {\bf Unfortunately the {\it GALEX} observations of the two galaxies, awarded as part of the GI6-6017 proposal, have not been completed. } 

\label{tab2}
\end{table}

%% file: tab2a.tex
\begin{table*}
 \centering
 \begin{minipage}{140mm}
  \caption{Journal of H$_\alpha$ observations}
  \begin{tabular}{lcc}
  \hline
  \hline
                 &   NGC3659 / NGC3684 / NGC3691/ UGC05287    &  NGC3655 / NGC3686 / UGC05393     \\
{\bf Observations}     &             &            \\
Telescope             &   OAN 2.1m    &  OAN2.1m  \\
Date                      &   Feb. 2$^{rd}$  2011   &  Feb. 4$^{th}$ 2011    \\
Instrument            &   PUMA                       &  PUMA          \\
                           &                                  &                      \\
{\bf Interference Filter}  &                              &                                                       \\
Central Wavelenght  (\AA) &           6607         &                                6607                 \\
FWHM (\AA)                      &            89                           &      89                    \\
Transmission (\%) @ V$_{sys_{FP}}$  & 65 / 66 / 67 / 63   &      67 / 66 / 63                 \\
                                            &                                  &                      \\
{\bf Calibration}                &                                  &                      \\
Neon Calib. Lamp (\AA)  &    6598.95                   &   6598.95       \\
                                             &                                  &                      \\
{\bf Perot-Fabry}               &                                  &                      \\
Interference Order           &  330@H$\alpha$         &       330@H$\alpha$               \\
Free Spectral Range      &  910km.s$^{-1}$           &         910km.s$^{-1}$             \\
Finesse@H$\alpha$      &   24                             &        24           \\
Spec. Resolution@H$\alpha$   &    8000                         &   8000                \\
                                                         &                                  &                      \\
{\bf Sampling}                                &                                   &                     \\
N. of Channels                              &    48                            &      48              \\
Sampling step (\AA)                     &   0.43                          &      0.43               \\
equivalent (km.s$^{-1}$)             &   19                             &        19            \\
Pixel Size                                       &    1.07"                           &      1.07"            \\
                                                         &                                  &                      \\
{\bf Detector}                                 &      Thomson 2k      &   Thomson 2k                     \\
                                                         &                                  &                      \\
{\bf Exposure Time}                      &                                   &                     \\
Total exp. (h)                                 &    1.6                          &         3.2             \\
Eleme. exp./chan (s)                    &   120                          &        120       \\

 \hline
\hline
\end{tabular}
\end{minipage}
\end{table*}

%% file: tab3.tex
\begin{table*}
 \caption{UV and optical photometry. Column 2 also reports our morphological type classification.}
	\label{tab3}       
        \scriptsize
	\begin{tabular}{lllllllll}
	\hline\hline\noalign{\smallskip}
	Group & Type & FUV & NUV & u & g & r & i &z  \\
	galaxy & & [AB mag] &[AB mag] & [AB mag] &[AB mag] &[AB mag] &[AB mag]& [AB mag] \\
	 \hline
 
	U268   & & & &  & & & &  \\
	\hline
 MRK 0408  & S0       &16.126$\pm$0.062 &15.896$\pm$0.047& 15.368$\pm$0.041& 14.602$\pm$0.024& 14.263$\pm$0.024 &14.123$\pm$0.035& 14.044$\pm$0.049 \\  
 NGC 3003  & Sbc      &14.233$\pm$0.079 &13.835$\pm$0.040& 13.371$\pm$0.083& 12.654$\pm$0.076& 12.081$\pm$0.062 &11.933$\pm$0.055& 11.666$\pm$0.063  \\  
 NGC 3011  & S0(R)    &16.824$\pm$0.064 &16.502$\pm$0.046& 15.433$\pm$0.071& 14.118$\pm$0.021& 13.539$\pm$0.020 &13.229$\pm$0.029& 13.040$\pm$0.036 \\  
 UGC 05287 & SBc      &16.272$\pm$0.092 &15.903$\pm$0.045& 15.199$\pm$0.158& 14.051$\pm$0.042& 13.737$\pm$0.041 &13.541$\pm$0.067& 13.433$\pm$0.103  \\  
 UGC 05326 & Sc(R)pec &15.767$\pm$0.115 &15.589$\pm$0.058& 14.926$\pm$0.121& 14.099$\pm$0.031& 13.823$\pm$0.037 &13.647$\pm$0.055& 13.646$\pm$0.082  \\  
 IC 2524   & S0       &16.748$\pm$0.075 &16.385$\pm$0.043& 15.636$\pm$0.070& 14.599$\pm$0.025& 14.227$\pm$0.029 &14.063$\pm$0.033& 13.963$\pm$0.053  \\  
 NGC 3067  & Sapec    &16.367$\pm$0.095 &15.556$\pm$0.050& 13.838$\pm$0.066& 12.597$\pm$0.054& 11.935$\pm$0.051 &11.563$\pm$0.052& 11.330$\pm$0.045  \\  
 UGC 05393 & SBb      &15.691$\pm$0.093 &15.534$\pm$0.050& 14.780$\pm$0.149& 14.015$\pm$0.048& 13.789$\pm$0.043 &13.709$\pm$0.057& 13.789$\pm$0.110  \\  
 UGC 05446 & Sc       &16.717$\pm$0.090 &16.380$\pm$0.056& 15.777$\pm$0.102& 14.966$\pm$0.053& 14.656$\pm$0.034 &14.482$\pm$0.043& 14.522$\pm$0.103  \\  
 NGC 3118  & Sbc      &16.086$\pm$0.102 &15.594$\pm$0.054& 14.766$\pm$0.094& 13.867$\pm$0.042& 13.591$\pm$0.036 &13.343$\pm$0.041& 13.370$\pm$0.089  \\  
\hline
PGC2016633	        & S0?  & 18.100$\pm$0.060 &17.922$\pm$0.035& 17.728$\pm$0.020& 16.824$\pm$0.004& 16.511$\pm$0.004 &16.399$\pm$0.004 & 16.311$\pm$0.016\\	   
PGC2042146	        & ?    & 18.247$\pm$0.059 &18.203$\pm$0.041& 17.767$\pm$0.023& 17.054$\pm$0.005& 16.885$\pm$0.006 &16.868$\pm$0.006 & 16.826$\pm$0.033\\
SDSSJ094911.28+342634.2 & ?    & 18.300$\pm$0.057 &18.180$\pm$0.038& 17.792$\pm$0.015& 17.186$\pm$0.005& 17.080$\pm$0.005 &17.056$\pm$0.005 & 17.006$\pm$0.020\\
SDSSJ094935.09+342616.3 & Sd   & 16.731$\pm$0.028 &16.602$\pm$0.019& 15.510$\pm$0.020& 15.650$\pm$0.010& 15.270$\pm$0.010 &15.340$\pm$0.010 & 15.290$\pm$0.020\\
PGC082546	        & Sc   &                  &  	      & 17.348$\pm$0.013& 16.474$\pm$0.004& 16.142$\pm$0.004 &16.022$\pm$0.004 & 15.890$\pm$0.013 \\
NGC3021	        	& Sbc  & 15.394$\pm$0.017 &14.831$\pm$0.008& 14.120$\pm$0.003& 12.838$\pm$0.002& 12.255$\pm$0.002 &11.820$\pm$0.002 & 11.556$\pm$0.002 \\  
UGC05282	        & Sm   & 17.488$\pm$0.046 &17.126$\pm$0.024& 16.630$\pm$0.030& 15.550$\pm$0.010& 15.170$\pm$0.010 &15.370$\pm$0.010 & 14.940$\pm$0.020 \\	
PGC2025214	        & ?    & 18.512$\pm$0.066 &18.457$\pm$0.045& 18.413$\pm$0.047& 17.740$\pm$0.010& 17.446$\pm$0.011 &17.410$\pm$0.011 & 17.161$\pm$0.044\\
SDSSJ095430.02+320342.0 & Sd   & 19.729$\pm$0.112 &19.615$\pm$0.059& 19.101$\pm$0.089& 18.103$\pm$0.024& 17.730$\pm$0.023 &17.675$\pm$0.023 & 17.431$\pm$0.065\\
SDSSJ100309.92+323622.5 & Irr  & 18.982$\pm$0.116 &18.541$\pm$0.071& 17.935$\pm$0.059& 17.228$\pm$0.035& 16.503$\pm$0.015 &16.372$\pm$0.015 & 16.470$\pm$0.071\\		
PGC029522	        & S0? &                  &  	      & 17.302$\pm$0.016& 16.383$\pm$0.004& 16.013$\pm$0.004 &15.849$\pm$0.004 & 15.664$\pm$0.012\\

\hline\hline														 	      
	U376    & & & & & & & & \\
\hline
NGC 3592       & Sd &17.924$\pm$0.064 &17.285$\pm$0.061 &15.586$\pm$0.102& 14.090$\pm$0.037 &13.466$\pm$0.034 &13.122$\pm$0.044& 12.959$\pm$0.064\\
NGC 3599       & S0 &18.670$\pm$0.092 &17.045$\pm$0.089 &14.314$\pm$0.099& 12.585$\pm$0.049 &11.911$\pm$0.043 &11.569$\pm$0.042& 11.408$\pm$0.042\\
NGC 3605       & E &19.979$\pm$0.179 &18.165$\pm$0.081 &15.003$\pm$0.065& 13.092$\pm$0.029 &12.401$\pm$0.030 &11.993$\pm$0.029& 11.810$\pm$0.037\\
UGC 06296      & Sc &18.573$\pm$0.084 &17.767$\pm$0.064 &15.786$\pm$0.118& 14.358$\pm$0.035 &13.698$\pm$0.033 &13.300$\pm$0.032& 13.042$\pm$0.050\\
NGC 3607       & S0&16.954$\pm$0.087 &15.683$\pm$0.067 &12.723$\pm$0.062& 10.742$\pm$0.044 &9.974$\pm$0.041 &9.534$\pm$0.042& 9.301$\pm$0.037\\
NGC 3608       & E &18.063$\pm$0.132 &16.598$\pm$0.076 &13.505$\pm$0.063& 11.539$\pm$0.034 &10.807$\pm$0.032 &10.405$\pm$0.033& 10.149$\pm$0.037\\
CGCG 096-024   & S0 &  		   & 18.829$\pm$0.097 &16.144$\pm$0.083 &14.481$\pm$0.046 &13.914$\pm$0.025 &13.516$\pm$0.041 &13.466$\pm$0.071\\
UGC 06320      & S0? &15.722$\pm$0.087& 15.268$\pm$0.054& 14.388$\pm$0.050& 13.607$\pm$0.029& 13.208$\pm$0.027& 13.026$\pm$0.032& 12.954$\pm$0.066\\
UGC 06324      & S0 &17.707$\pm$0.098& 17.071$\pm$0.078& 15.831$\pm$0.075& 14.391$\pm$0.032& 13.915$\pm$0.038& 13.688$\pm$0.050& 13.574$\pm$0.063\\
NGC 3626       & SB0r(R) &   &    &  & & & &  \\
NGC 3655       & Sb& 15.512$\pm$0.087 &14.766$\pm$0.047 &13.336$\pm$0.039& 12.113$\pm$0.024& 11.523$\pm$0.023 &11.252$\pm$0.023 &11.058$\pm$0.036 \\
NGC 3659       & SBd&       & 	  & 	 & &&&	 \\
NGC 3681       & SAB(r)bc&15.235$\pm$0.119& 14.909$\pm$0.092& 13.592$\pm$0.097& 12.274$\pm$0.024& 11.617$\pm$0.026 &11.260$\pm$0.028& 11.112$\pm$0.052\\
NGC 3684       & SAB(r)bc&14.653$\pm$0.089& 14.244$\pm$0.067& 13.245$\pm$0.082& 12.141$\pm$0.028& 11.676$\pm$0.027 &11.640$\pm$0.039& 11.261$\pm$0.060\\
NGC 3686       & SB(s)bc&14.414$\pm$0.073& 13.895$\pm$0.052& 12.847$\pm$0.059& 11.614$\pm$0.022& 11.079$\pm$0.021 &10.784$\pm$0.020& 10.633$\pm$0.050\\
NGC 3691       & SBb&15.608$\pm$0.051 & 15.165$\pm$0.030      & 	 & &  & &	 \\
\hline
PGC034537 & SB(r)bc&18.319$\pm$0.095 &17.950$\pm$0.054&17.614$\pm$0.028 &16.225$\pm$0.004 &15.607$\pm$0.003 &15.235$\pm$0.003 &15.000$\pm$0.007 \\	 
PGC086629 & Irr &19.351$\pm$0.128 &18.667$\pm$0.087&21.915$\pm$0.376 &21.409$\pm$0.099 &20.746$\pm$0.085 &20.256$\pm$0.085 &21.476$\pm$0.793\\
UGC06341  & Sd&                 &                &16.915$\pm$0.020& 15.709$\pm$0.005& 15.391$\pm$0.004 &15.235$\pm$0.004 &15.192$\pm$0.016  \\
PGC1534499& S0?&                 &                &17.564$\pm$0.041& 16.108$\pm$0.005& 15.475$\pm$0.005 &15.111$\pm$0.005 &15.184$\pm$0.032 \\
PGC086673 &Irr &                 &                &18.271$\pm$0.082& 17.637$\pm$0.055& 16.822$\pm$0.027 &16.643$\pm$0.027 &16.549$\pm$0.067 \\
PGC035087 & Irr? &19.624$\pm$0.167 &19.534$\pm$0.132&17.946$\pm$0.086 &16.961$\pm$0.012 &16.559$\pm$0.013 &16.465$\pm$0.013 &16.318$\pm$0.081\\
PGC035096 & Irr &18.254$\pm$0.081 &17.972$\pm$0.056&21.977$\pm$0.230 &22.937$\pm$0.196 &22.871$\pm$0.266 &23.533$\pm$0.266 &22.083$\pm$0.659\\
PGC035426 & Im &20.210$\pm$0.205 &19.729$\pm$0.127&18.598$\pm$0.108 &17.306$\pm$0.012 &16.811$\pm$0.018 &16.665$\pm$0.018 &16.603$\pm$0.058 \\

\hline														 	      

\end{tabular}													  
 \end{table*}

%% file: tab3a.tex
\begin{table}
 \centering
 \caption{Kinematical parameters}
  \begin{tabular}{lccc}
  \hline
  \hline

Group	&  V$_{sys_FP}$   & 	PA$_{Kin} $	 	&   i$_{Kin}$    \\
Galaxies          		&	kms$^{-1}$  &	 	[deg]			& 	[deg]		\\
\hline
U268 & & & \\
\hline			
UGC 5287   	& 1517$\pm$5		&	-	 		&	-	\\
UGC 5393   	& 1963$\pm$3	 	&	153$\pm$3	 		&	12$\pm$15 	\\
\hline	
U376 & & & \\	
\hline		
NGC 3655   	&  1441$\pm$6   	&	27$\pm$4 	 	         &	51$\pm$1	\\
NGC 3659   	& 1274$\pm$2    	&	58$\pm$2		 		&	62$\pm$3	\\
NGC 3684   	& 1117$\pm$1    	&	117$\pm$1	 		&	51$\pm$3	\\
NGC 3686   	& 1146$\pm$5    	&	17$\pm$4		 		&	27$\pm$2	\\
NGC 3691   	& 1073$\pm$1    	&	9 $\pm$1	 		&	37$\pm$11	\\
\hline

\end{tabular}
\label{kp}
\end{table}

%% file: tabA1.tex
\begin{table*}
	\caption{Galaxies in a box of 4 Mpc $\times$ 4 Mpc centred on NGC 3003 in U268 with heliocentric radial velocity   within  $\pm$ 3$\sigma$
	 of the group mean velocity given in the catalog of \citet{Ramella02}.}
	   \label{a1}	 
	 \scriptsize
	\begin{tabular}{llllllllllll}
	\hline\hline
	   Galaxies  & RA & Dec.&  Morph.& Mean Hel.& logD$_{25}$ & logr$_{25}$ & P.A. &B$_T$ \\ 
	                   & (J2000) & (J2000) & type &  Vel.& &  &   &  \\
		          & (hours) & (degrees )   & & [km/s]  &[arcmin] & &[deg]  & mag \\
				  
	\hline
	U268   & & & & & & & &   \\ 
\hline
UGC05015                 & 9.42999 & 34.2771 & 7.8 & 1648$\pm$5 & 1.21 & 0.02 &      & 15.60$\pm$0.32 \\
UGC05020                 & 9.43374 & 34.65339 & 5.8& 1619$\pm$7 & 1.31 & 0.59 & 77.6 & 15.16$\pm$0.27 \\
PGC027157                & 9.56196 & 33.60034 & -1 & 1601$\pm$23 &0.88 & 0.14 & 127.2& 15.57$\pm$0.31 \\
SDSSJ093441.75+323203.5  & 9.57826 & 32.53437 &    & 1277$\pm$54 &   &   & 119.6 &   &   \\
UGC05105                 & 9.58888 & 35.91365 & 9.9& 1588$\pm$7 & 0.99 & 0.15 & 147.2 & 16.11$\pm$0.50 \\
PGC027311                & 9.60209 & 29.11222 & 3.5& 1652$\pm$2 & 0.87 & 0.42 & 8.5   & 15.62$\pm$0.45 \\
SDSSJ094029.36+365443.5  & 9.67482 & 36.91214 & 10 & 1629$\pm$63 & 0.58 & 0.07 & 7.1  & 17.76$\pm$0.50 \\
PGC027626                & 9.6794  & 32.47254 &    & 1288$\pm$3 & 0.31 & 0.06 & 82.4  & 18.03$\pm$0.50 \\
NGC2964                  & 9.71512 & 31.84703 & 4.1& 1318$\pm$3 & 1.47 & 0.13 & 97    & 12.04$\pm$0.09 \\
SDSSJ094254.50+315103.5  & 9.71515 & 31.85104 &    & 1321$\pm$36 &   &   &   &   &   \\
NGC2968                  & 9.72    & 31.92873 & 1.2& 1556$\pm$13 & 1.39 & 0.18 & 54 & 12.74$\pm$0.09 \\
SDSSJ094316.02+313725.1  & 9.72111 & 31.62357 &    & 1541$\pm$91 &   &   & 73.4 &      &   \\
PGC2832070               & 9.72196 & 31.82409 &    & 1468$\pm$68 &   &   & 90.4 &      &   \\
NGC2970                  & 9.7253  & 31.9769 & -4.6& 1644$\pm$15 & 0.93 & 0.06 & 64.8 & 14.41$\pm$0.11 \\
PGC1973445               & 9.72571 & 31.96755 &    & 1416$\pm$90 & 0.63 & 0.23 & 71.4 & 17.89$\pm$0.50 \\
SDSSJ094338.13+351207.8  & 9.72726 & 35.2022 &     & 1552$\pm$41 & 0.53 & 0.2 & 93.8  & 17.77$\pm$0.50 \\
PGC027992                & 9.7637  & 30.53404 & 4.8& 1495$\pm$34 & 0.48 & 0.26 & 39    &18.10$\pm$0.50 \\
PGC028086                & 9.78139 & 31.79583 &    & 1396$\pm$42 & 0.4 & 0.1 &        & 18.39$\pm$0.50 \\
PGC028153                & 9.79657 & 39.08419 & 3.9& 1592$\pm$9 & 0.68 & 0.04 & 129.8 & 15.20$\pm$0.15 \\
PGC2800953               & 9.79722 & 39.1427 &     & 1553$\pm$25 &   &   &   &   &   \\
SDSSJ094758.45+390510.1  & 9.79957 & 39.0862 & 10  & 1501$\pm$60 & 0.6 & 0.28 & 112.1 & 18.09$\pm$0.35 \\
PGC2016633               & 9.80076 & 32.90034 &    & 1552$\pm$47 & 0.63 & 0.21 & 165  & 17.25$\pm$0.50 \\
PGC028169                & 9.8013  & 32.88255 & 3.2& 1528$\pm$7 & 0.83 & 0.11 & 163.9 & 14.76$\pm$0.28 \\
PGC1916762               & 9.80591 & 30.75928 &    & 1468$\pm$5 & 0.43 & 0.08 & 26.6  & 18.48$\pm$0.50 \\
SDSSJ094829.55+332459.3  & 9.8082  & 33.41643 &    & 1619$\pm$338 &   &   & 82 &   &   \\
NGC3003                  & 9.80992 & 33.42136 & 4.3& 1480$\pm$3 & 1.68 & 0.65 & 78.8 & 12.25$\pm$0.10 \\
SDSSJ094838.45+332529.1  & 9.8107  & 33.42466 &    & 1419$\pm$31 &   &   & 86.8 & &   \\
SDSSJ094847.46+275225.4  & 9.81319 & 27.8737 &     & 1311$\pm$2 & &   & 38.7 &   &   \\
PGC2042146               & 9.81754 & 33.99134 &    & 1494$\pm$2 & 0.5 & 0.14 & 53.8 & 17.85$\pm$0.5 \\
SDSSJ094911.28+342634.2  & 9.81979 & 34.44292 &    & 1489$\pm$2 &   &	& 164.1 &   &	\\
SDSSJ094935.09+342616.3  & 9.82641 & 34.43793 &    & 1488$\pm$2 &   &	& 98.3 &  &   \\
NGC3011                  & 9.82811 & 32.22109 &-1.7& 1539$\pm$5 & 0.92 & 0.08 & 59.5 & 14.54$\pm$0.44 \\
PGC028305                & 9.83635 & 28.01298 &-4.8& 1448$\pm$6 & 0.73 & 0.09 & 63.5 & 15.81$\pm$0.40 \\
SDSSJ095011.18+280044.8  & 9.83645 & 28.0125 &     & 1458$\pm$32 & 0.81 & 0.22 & 49.8& 16.05$\pm$0.35 \\s
PGC082546                & 9.83917 & 33.58391 & 4.6& 1586$\pm$29 & 0.56 & 0.15 & 171 & 17.05$\pm$0.36 \\
NGC3026                  & 9.84871 & 28.55119 & 9.7& 1488$\pm$4 & 1.35 & 0.57 & 82.7 & 13.85$\pm$0.43 \\
NGC3021                  & 9.84921 & 33.55341 & 4  & 1540$\pm$3 & 1.13 & 0.23 & 108.3& 12.54$\pm$0.27 \\
SDSSJ095058.02+333319.3  & 9.84946 & 33.55535 &    & 1596$\pm$3 &   &	& 125.7 &   &	\\
UGC05282                 & 9.85284 & 33.13083 & 8.8& 1557$\pm$10 & 0.97 & 0.27 & 46 &	&   \\
UGC05287                 & 9.8578  & 32.94288 & 5.9& 1470$\pm$6 & 0.98 & 0.12 &   & 14.84$\pm$0.42 \\
SDSSJ095141.68+384207.1  & 9.86158 & 38.70185 & 10 & 1426$\pm$3 & 0.63 & 0.02 &   & 17.41$\pm$0.35 \\
NGC3032                  & 9.86895 & 29.23653 &-1.9& 1538$\pm$6 & 1.15 & 0.04 &   & 13.07$\pm$0.20 \\
PGC028474                & 9.88261 & 29.31088 & 7.9& 1638$\pm$6 & 0.53 & 0.26 & 9.3 & 17.79$\pm$0.86 \\
PGC2025214               & 9.8959  & 33.16437 &    & 1520$\pm$35 & 0.55 & 0.25 & 18.5 & 18.08$\pm$0.50 \\
SDSSJ095430.02+320342.0  & 9.90835 & 32.06166 &    & 1421$\pm$2 &   &	& 152.9 &   &	\\
UGC05326                 & 9.92349 & 33.26252 & 9.8& 1413$\pm$3 & 0.97 & 0.02 &   & 14.62$\pm$0.40 \\
IC2524                   & 9.95912 & 33.61961 & -1 & 1462$\pm$13 & 0.87 & 0.23 & 61.8 & 15$\pm$0.49 \\
UGC05349                 & 9.96841 & 37.2928 & 7.8 & 1381$\pm$12 & 1.35 & 0.57 & 37.5 & 14.45$\pm$0.36 \\
NGC3067                  & 9.97253 & 32.36993 & 2.1& 1473$\pm$4 & 1.31 & 0.49 & 104.3 & 12.73$\pm$0.12 \\
PGC028991                & 10.01862 & 36.92261 &4.8& 1446$\pm$3 & 0.77 & 0.45 & 13.5 & 17.06$\pm$0.50 \\
PGC029004                & 10.02065 & 37.07109 &4.8& 1463$\pm$2 & 0.7 & 0.18 & 174.8 & 16.80$\pm$0.46 \\
SDSSJ100140.18+371538.3  & 10.02784 & 37.26055 &   & 1530$\pm$440 & 0.76 & 0.04 &   & 16.66$\pm$0.50 \\
UGC05391                 & 10.02805 & 37.2477 & 8.9& 1568$\pm$10 & 1.3 & 0.56 & 167.7 & 15.24$\pm$0.52 \\
UGC05393                 & 10.02834 & 33.13635 & 8 & 1444$\pm$12 & 1.14 & 0.28 & 117 & 14.95$\pm$0.38 \\
SDSSJ100142.11+371443.2  & 10.02837 & 37.24526 &   & 1567$\pm$369 & 1.2 & 0.55 & 163.6 & 16.79$\pm$0.50 \\
UGC05394                 & 10.02998 & 36.49881 &6.2& 1431$\pm$13 & 0.93 & 0.49 & 56.3 & 16.36$\pm$0.43 \\
PGC2125269               & 10.05056 & 38.37724 &   & 1353$\pm$40 & 0.7 & 0.31 & 67.7 & 17.11$\pm$0.40 \\
SDSSJ100309.92+323622.5  & 10.05276 & 32.60629 &   & 1562$\pm$96 &   &   & 123.3 &   &   \\
SDSSJ100429.42+365652.6  & 10.07484 & 36.94803 &   & 1602$\pm$61 & 0.59 & 0.17 & 59 & 18.15$\pm$0.50 \\
PGC029347                & 10.10504 & 28.94493&-2.4& 1362$\pm$7 & 0.82 & 0.03 &   & 14.54$\pm$0.09 \\
UGC05446                 & 10.10859 & 32.94659 &5.9& 1361$\pm$3 & 1.12 & 0.61 & 49.8 & 15.38$\pm$0.40 \\
NGC3118                  & 10.11988 & 33.02738 &4.1& 1342$\pm$4 & 1.32 & 0.67 & 38.6 & 14.34$\pm$0.35 \\
PGC029522                & 10.14943 & 32.01043 &4.7& 1468$\pm$34 & 0.66 & 0.25 & 11.8 & 16.91$\pm$0.44 \\
UGC05478                 & 10.1588  s& 30.1503 &9.9& 1377$\pm$4 & 1.19 & 0.03 &   & 14.75$\pm$0.06 \\
\hline
\end{tabular}

\end{table*}

%% file: tabA2.tex
\begin{table*}
	\caption{Galaxies in a box of 4 Mpc $\times$ 4 Mpc centred on NGC 3607 in U376 with heliocentric radial velocity  within  $\pm$ 3$\sigma$ of the group mean velocity given in the catalog of \citet{Ramella02}.}
	   \label{a2}	 
	 \scriptsize
	\begin{tabular}{llllllllllll}
	\hline\hline
	   Galaxies  & RA & Dec.&  Morph.& Mean Hel.& logD$_{25}$ & logr$_{25}$ & P.A. &B$_T$ \\ 
	                   & (J2000) & (J2000) & type &  Vel.& &  &   &  \\
		          & (hours) & (degrees )   & & [km/s]  &[arcmin] & &[deg]  & mag \\
				  
	\hline
	U376   & & & & & & & &   \\ 
\hline
PGC031819 & 10.68599 & 21.36189 & & 932$\pm$60 & 0.57 & 0.15 & 37.9 & 16.25$\pm$0.68 \\
PGC1441580 & 10.69473 & 13.82499 & & 1271$\pm$27 & 0.46 & 0.15 & 168.5 & 17.46$\pm$0.41 \\
PGC031877 & 10.70009 & 12.33506 & 0.0 & 774$\pm$5 & 0.81 & 0.31 & 33.1 & 16.01$\pm$0.51 \\
NGC3338 & 10.70210 & 13.74687 & 5.1 & 1299$\pm$2 & 1.09 & 0.24 & 97.0 & 11.44$\pm$0.21 \\
SDSSJ104226.51+135725.7 & 10.70736 & 13.95720 & & 1280$\pm$18 & 0.86 & 0.39 & 86.5 & 17.34$\pm$0.50 \\
UGC05832 & 10.71348 & 13.45988 & 4.2 & 1216$\pm$3 & 1.04 & 0.16 & 86.3 & 14.48$\pm$0.70 \\
PGC1439163 & 10.71455 & 13.74117 & & 1145$\pm$5 & 0.40 & 0.20 & 48.0 & 18.40$\pm$0.50 \\
PGC031937 & 10.71819 & 13.51109 & & 1258$\pm$6 & & & 75.2 & 15.50$\pm$0.41 \\
UGC05833 & 10.71835 & 20.42220 & -2.0 & 1333$\pm$5 & 1.08 & 0.45 & 149.0 & 14.45$\pm$0.15 \\
NGC3344 & 10.72531 & 24.92215 & 4.0 & 583$\pm$3 & 1.83 & 0.02 & & 10.50$\pm$0.08 \\
NGC3346 & 10.72747 & 14.87178 & 5.9 & 1257$\pm$9 & 1.42 & 0.08 & 118.6 & 12.45$\pm$0.34 \\
NGC3351 & 10.73269 & 11.70355 & 3.1 & 778$\pm$2 & 1.86 & 0.21 & 9.9 & 10.58$\pm$0.12 \\
SDSSJ104435.27+135622.7 & 10.74313 & 13.93964 & & 633$\pm$31 & & & 116.5 &     \\
PGC2806914 & 10.74903 & 20.86190 & 9.0 & 1219$\pm$8 & & & &     \\
PGC083326 & 10.74931 & 11.91611 & 10.0 & 871$\pm$4 & 0.60 & 0.36 & & 17.36$\pm$0.36 \\
PGC3090611 & 10.75273 & 15.44989 & & 1219$\pm$3 & 0.68 & 0.14 & 119.2 & 17.23$\pm$0.45 \\
PGC1733581 & 10.76385 & 25.35310 & 3.0 & 1418$\pm$41 & 0.68 & 0.54 & 33.4 & 17.87$\pm$0.37 \\
AGC205287 & 10.77675 & 12.62639 & & 957$\pm$8 & & & &     \\
AGC205289 & 10.77686 & 12.43167 & & 1006$\pm$8 & & & &     \\
PGC083336 & 10.77813 & 12.32714 & 10.0 & 1030$\pm$5 & 0.82 & 0.13 & 10.8 & 16.91$\pm$0.42 \\
AGC205290 & 10.77853 & 12.77972 & & 915$\pm$8 & & & &    \\
NGC3368 & 10.77936 & 11.81981 & 2.2 & 892$\pm$2 & 1.92 & 0.18 & 177.0 & 10.08$\pm$0.10 \\
PGC4689210 & 10.78144 & 12.74444 & 10.0 & 640$\pm$5 & 0.90 & 0.12 & & 18.45$\pm$0.57 \\
PGC083338 & 10.78186 & 12.78795 & 10.0 & 928$\pm$8 & 0.70 & 0.15 & & 18.20$\pm$0.29 \\
PGC083339 & 10.78246 & 12.99925 & 10.0 & 832$\pm$4 & 1.08 & 0.30 & & 17.63$\pm$0.34 \\
PGC4689200 & 10.78356 & 12.95980 & 10.0 & 843$\pm$7 & 0.60 & 0.12 & & 18.60$\pm$0.50 \\
AGC205291 & 10.78414 & 12.22417 & & 1018$\pm$8 & & & &     \\
NGC3370 & 10.78445 & 17.27369 & 5.1 & 1281$\pm$4 & 1.39 & 0.24 & 143.5 & 12.29$\pm$0.30 \\
AGC201972 & 10.78589 & 13.00472 & & 834$\pm$8 & & & &     \\
AGC205292 & 10.78594 & 13.05028 & & 824$\pm$8 & & & &     \\
AGC205293 & 10.78872 & 13.15583 & & 806$\pm$8 & & & &     \\
SDSSJ104720.05+122314.9 & 10.78891 & 12.38745 & & 1146$\pm$45 & & & 30.8 &     \\
NGC3377A & 10.78954 & 14.06997 & 8.9 & 573$\pm$2 & 1.16 & 0.02 & & 14.40$\pm$0.11 \\
AGC205294 & 10.79425 & 11.92861 & & 971$\pm$5 & & & &     \\
NGC3377 & 10.79510 & 13.98564 & -4.8 & 684$\pm$11 & 1.59 & 0.32 & 37.2 & 11.13$\pm$0.17 \\
AGC205295 & 10.79669 & 12.21639 & & 978$\pm$8 & & & &     \\
NGC3379 & 10.79711 & 12.58161 & -4.8 & 908$\pm$6 & 1.69 & 0.06 & 71.0 & 10.23$\pm$0.17 \\
SDSSJ104749.63+123432.6 & 10.79711 & 12.57587 & & 922$\pm$25 & & & 178.3 &     \\
AGC205296 & 10.79722 & 13.12694 & & 787$\pm$8 & & & &     \\
AGC205297 & 10.80128 & 13.18750 & & 794$\pm$8 & & & &     \\
AGC205301 & 10.80344 & 12.06833 & & 927$\pm$8 & & & &     \\
AGC205302 & 10.80386 & 12.14167 & & 917$\pm$8 & & & &     \\
AGC205303 & 10.80442 & 12.29806 & & 910$\pm$8 & & & &     \\
NGC3384 & 10.80469 & 12.62890 & -2.7 & 733$\pm$20 & 1.72 & 0.34 & 53.0 & 10.89$\pm$0.14 \\
SDSSJ104819.74+123825.6 & 10.80550 & 12.64040 & & 568$\pm$22 & & & 47.1 &     \\
NGC3389 & 10.80776 & 12.53310 & 5.3 & 1303$\pm$2 & 1.43 & 0.35 & 102.9 & 12.51$\pm$0.29 \\
AGC205304 & 10.80789 & 12.42889 & & 854$\pm$8 & & & &     \\
AGC205305 & 10.80850 & 12.62611 & & 648$\pm$8 & & & &     \\
AGC205306 & 10.80919 & 12.49944 & & 794$\pm$8 & & & &     \\
AGC205307 & 10.81008 & 12.04667 & & 924$\pm$8 & & & &     \\
AGC205308 & 10.81200 & 13.26528 & & 785$\pm$8 & & & &     \\
PGC032348 & 10.81492 & 14.12450 & -1.0 & 541$\pm$22 & 0.92 & 0.16 & 65.0 & 15.66$\pm$0.41 \\
PGC032346 & 10.81579 & 12.19512 & 4.2 & 1323$\pm$3 & 0.87 & 0.22 & 14.4 & 15.70$\pm$0.63 \\
PGC083348 & 10.81801 & 12.40737 & 5.4 & 1365$\pm$7 & 0.49 & 0.10 & 171.9 & 18.38$\pm$0.31 \\
AGC205309 & 10.81883 & 12.37389 & & 1342$\pm$8 & & & &     \\
AGC205310 & 10.81994 & 12.49194 & & 1379$\pm$8 & & & &     \\
AGC205311 & 10.82017 & 12.19500 & & 869$\pm$8 & & & &      \\
PGC032371 & 10.82144 & 12.42196 & 9.0 & 1382$\pm$5 & 0.95 & 0.27 & 157.6 & 15.72$\pm$0.72 \\
PGC032376 & 10.82206 & 12.37440 & 9.0 & 1346$\pm$6 & 0.89 & 0.00 & & 17.00$\pm$0.35 \\
PGC083352 & 10.82408 & 12.25776 & -5.0 & 1321$\pm$3 & 0.60 & 0.15 & 29.6 & 17.26$\pm$0.69 \\
AGC205197 & 10.82856 & 13.82806 & & 1332$\pm$8 & & & &     \\
AGC205313 & 10.83106 & 12.61139 & & 774$\pm$8 & & & &     \\
PGC1424345 & 10.83117 & 13.16179 & 10.0 & 764$\pm$6 & 0.60 & 0.09 & 81.0 & 18.98$\pm$1.08 \\
AGC205314 & 10.83119 & 13.28667 & & 787$\pm$8 & & & &     \\
AGC205315 & 10.83131 & 12.53694 & & 779$\pm$8 & & & &     \\
AGC205316 & 10.83250 & 12.67028 & & 776$\pm$8 & & & &     \\
SDSSJ105001.79+134705.1 & 10.83384 & 13.78474 & & 1311$\pm$1 & & & 129.8 &     \\
AGC205321 & 10.83414 & 13.10583 & & 788$\pm$8 & & & &      \\
AGC205322 & 10.83597 & 13.00583 & & 797$\pm$8 & & & &      \\
UGC05944 & 10.83861 & 13.27256 & 9.7 & 1073$\pm$27 & 0.97 & 0.02 & & 15.27$\pm$0.44 \\
UGC05945 & 10.84043 & 17.56435 & 9.8 & 1132$\pm$10 & 0.92 & 0.26 & 96.0 & 15.59$\pm$0.60 \\
UGC05947 & 10.84177 & 19.64430 & 10.0 & 1253$\pm$12 & 1.09 & 0.24 & 25.0 & 14.92$\pm$0.08 \\
UGC05948 & 10.84394 & 15.76333 & 10.0 & 1118$\pm$6 & 1.04 & 0.53 & 35.0 &     \\
NGC3412 & 10.84814 & 13.41218 & -2.0 & 843$\pm$15 & 1.60 & 0.26 & 155.0 & 11.44$\pm$0.14 \\

\hline 
\end{tabular}
 \end{table*}
\begin{table*}
 \addtocounter{table}{-1}
	\caption{Continue.}
	   \label{a2}	 
	 \scriptsize
	\begin{tabular}{llllllllllll}
	\hline\hline
	   Galaxies  & RA & Dec.&  Morph.& Mean Hel.& logD$_{25}$ & logr$_{25}$ & P.A. &B$_T$ \\ 
	                   & (J2000) & (J2000) & type &  Vel.& &  &   &  \\
		          & (hours) & (degrees )   & & [km/s]  &[arcmin] & &[deg]  & mag \\
				  
	\hline
	U376   & & & & & & & &   \\ 
\hline
SDSSJ105053.31+132523.9 & 10.84815 & 13.42333 & & 774$\pm$25 & & & 156.5 &     \\
SDSSJ105131.34+140653.1 & 10.85871 & 14.11477 & & 832$\pm$53 & & & 30.2 &     \\
SDSSJ105204.79+150149.6 & 10.86798 & 15.03040 & & 828$\pm$38 & & & 5.8 &     \\
PGC1388083 & 10.87208 & 11.04314 & & 824$\pm$25 & 0.87 & 0.13 & 100.7 & 16.49$\pm$0.39 \\
PGC032630 & 10.87261 & 17.93527 & -0.4 & 1526$\pm$90 & 1.03 & 0.23 & 18.7 & 15.17$\pm$0.33 \\
UGC05989 & 10.87551 & 19.79220 & 9.9 & 1126$\pm$5 & 1.16 & 0.49 & 125.1 & 14.25$\pm$0.33 \\
SDSSJ105234.93+170841.8 & 10.87636 & 17.14495 & & 1054$\pm$51 & 0.49 & 0.12 & 127.1 & 18.23$\pm$0.50 \\
NGC3437 & 10.87657 & 22.93413 & 5.3 & 1276$\pm$11 & 1.37 & 0.46 & 118.8 & 12.59$\pm$0.38 \\
AGC749324 & 10.88317 & 26.47278 & & 1507$\pm$8 & & & &     \\
NGC3443 & 10.88336 & 17.57361 & 6.6 & 1132$\pm$8 & 1.25 & 0.30 & 143.0 & 14.79$\pm$0.44 \\
NGC3447 & 10.89000 & 16.77244 & 8.8 & 1069$\pm$3 & 1.53 & 0.24 & 25.6 & 14.54$\pm$0.68 \\
SDSSJ105324.34+164735.4 & 10.89010 & 16.79327 & & 1092$\pm$3 & 0.61 & 0.26 & 174.0 & 18.04$\pm$0.50 \\
SDSSJ105329.58+164359.6 & 10.89155 & 16.73324 & & 1017$\pm$65 & 0.62 & 0.09 & 160.0 & 17.84$\pm$0.50 \\
NGC3447A & 10.89158 & 16.78601 & 9.9 & 1094$\pm$7 & 1.17 & 0.28 & 105.8 & 16.39$\pm$0.50 \\
UGC06014 & 10.89519 & 9.72750 & 8.0 & 1133$\pm$4 & 1.03 & 0.27 & 70.0 & 15.44$\pm$0.52 \\
UGC06018 & 10.90159 & 20.64460 & 9.9 & 1291$\pm$4 & 1.02 & 0.03 & &     \\
UGC06022 & 10.90427 & 17.80950 & 9.9 & 973$\pm$6 & 1.03 & 0.24 & 10.0 &     \\
NGC3454 & 10.90817 & 17.34407 & 5.5 & 1114$\pm$10 & 1.38 & 0.66 & 116.1 & 13.70$\pm$0.09 \\
NGC3455 & 10.90863 & 17.28470 & 3.1 & 1106$\pm$5 & 1.36 & 0.20 & 68.5 & 14.32$\pm$0.24 \\
NGC3457 & 10.91349 & 17.62111 & -5.0 & 1159$\pm$5 & 0.99 & 0.02 & & 12.97$\pm$0.24 \\
UGC06035 & 10.92473 & 17.14170 & 9.9 & 1073$\pm$4 & 0.99 & 0.19 & 177.0 & 15.44$\pm$0.50 \\
PGC032843 & 10.92901 & 17.00500 & 4.8 & 1143$\pm$29 & 0.80 & 0.04 & & 15.24$\pm$0.38 \\
PGC2806987 & 10.93228 & 12.33913 & 9.5 & 847$\pm$7 & 0.90 & 0.30 & & 18.40$\pm$0.50 \\
SDSSJ105603.00+234848.5 & 10.93417 & 23.81347 & & 1278$\pm$50 & 0.54 & 0.26 & 175.9 & 18.50$\pm$0.50 \\
PGC087256 & 10.93719 & 12.01109 & 10.0 & 990$\pm$4 & 0.84 & 0.20 & 153.5 & 17.18$\pm$0.59 \\
SDSSJ105619.93+170505.9 & 10.93888 & 17.08503 & & 954$\pm$3 & 0.45 & 0.04 & & 18.52$\pm$0.50 \\
SDSSJ105638.66+172301.2 & 10.94406 & 17.38381 & & 941$\pm$3 & 0.58 & 0.29 & 156.2 & 18.19$\pm$0.50 \\
PGC087257 & 10.96061 & 13.97881 & 10.0 & 1227$\pm$8 & & & 61.8 &     \\
PGC087170 & 10.98113 & 14.12973 & 5.0 & 684$\pm$12 & & & 105.3 &     \\
NGC3485 & 11.00067 & 14.84132 & 3.1 & 1434$\pm$5 & 1.34 & 0.03 & & 12.67$\pm$0.27 \\
NGC3489 & 11.00516 & 13.90119 & -1.2 & 704$\pm$7 & 1.54 & 0.23 & 70.0 & 11.06$\pm$0.10 \\
UGC06083 & 11.00660 & 16.69230 & 4.1 & 937$\pm$2 & 1.13 & 0.92 & 142.4 & 15.23$\pm$0.37 \\
SDSSJ110047.14+165255.5 & 11.01311 & 16.88218 & & 1146$\pm$39 & 0.55 & 0.16 & 103.4 & 18.13$\pm$0.50 \\
UGC06095 & 11.01791 & 19.10020 & 10.0 & 1115$\pm$10 & 1.04 & 0.43 & 5.0 &     \\
PGC033270 & 11.03085 & 22.34454 & 3.1 & 1334$\pm$24 & 0.96 & 0.33 & 147.8 & 15.25$\pm$0.35 \\
UGC06112 & 11.04311 & 16.73495 & 7.4 & 1038$\pm$6 & 1.28 & 0.52 & 120.1 & 14.79$\pm$0.52 \\
NGC3501 & 11.04647 & 17.98957 & 5.9 & 1130$\pm$8 & 1.63 & 0.82 & 28.0 & 13.58$\pm$0.21 \\
NGC3507 & 11.05705 & 18.13599 & 3.1 & 975$\pm$9 & 1.47 & 0.07 & & 12.07$\pm$0.50 \\
PGC087171 & 11.05733 & 16.01637 & 5.0 & 1228$\pm$6 & 1.01 & 0.43 & 15.9 & 16.85$\pm$0.50 \\
PGC033447 & 11.07400 & 11.75599 & 10.0 & 778$\pm$8 & 1.11 & 0.49 & 144.1 & 15.91$\pm$0.39 \\
SDSSJ110456.82+173830.5 & 11.08245 & 17.64173 & & 918$\pm$40 & 0.80 & 0.33 & 114.1 & 17.44$\pm$0.50 \\
PGC1724397 & 11.08688 & 25.06459 & & 1482$\pm$289 & 0.45 & 0.19 & 129.8 & 17.75$\pm$0.43 \\
PGC033523 & 11.09237 & 17.63963 & -1.0 & 983$\pm$56 & 1.19 & 0.44 & 24.3 & 14.86$\pm$0.34 \\
UGC06151 & 11.09896 & 19.82534 & 8.8 & 1331$\pm$3 & 1.18 & 0.02 & & 14.90$\pm$0.09 \\
NGC3524 & 11.10891 & 11.38537 & 0.1 & 1366$\pm$20 & 1.17 & 0.53 & 13.7 & 13.34$\pm$0.33 \\
AGC215262 & 11.10981 & 12.23000 & & 1606$\pm$8 & & & &     \\
NGC3522 & 11.11124 & 20.08566 & -4.9 & 1220$\pm$8 & 1.07 & 0.28 & 114.1 & 14.08$\pm$0.28 \\
SDSSJ110651.09+173002.8 & 11.11418 & 17.50090 & & 949$\pm$71 & 0.48 & 0.12 & 153.8 & 18.52$\pm$0.50 \\
AGC749189 & 11.11589 & 26.90833 & & 1159$\pm$8 & & & &     \\
UGC06169 & 11.11761 & 12.06000 & 3.0 & 1553$\pm$5 & 1.22 & 0.75 & 1.4 & 14.54$\pm$0.34 \\
UGC06171 & 11.11943 & 18.56943 & 9.9 & 1199$\pm$7 & 1.24 & 0.53 & 66.0 & 15.13$\pm$0.39 \\
UGC06181 & 11.12962 & 19.54953 & 9.7 & 1169$\pm$2 & 1.02 & 0.18 & 48.0 & 15.54$\pm$0.36 \\
PGC033816 & 11.15646 & 10.83412 & & 1554$\pm$9 & 0.92 & 0.03 & & 15.24$\pm$0.38 \\
NGC3547 & 11.16553 & 10.72052 & 3.1 & 1580$\pm$2 & 1.24 & 0.35 & 6.2 & 13.20$\pm$0.12 \\
PGC033905 & 11.17365 & 10.12589 & 9.0 & 1322$\pm$11 & 0.95 & 0.23 & 117.9 & 15.92$\pm$0.59 \\
PGC033981 & 11.18180 & 9.62186 & 3.1 & 1577$\pm$10 & 0.75 & 0.15 & 107.0 & 15.49$\pm$0.50 \\
PGC087172 & 11.20437 & 16.75397 & 10.0 & 1205$\pm$9 & 0.95 & 0.35 & & 17.40$\pm$0.50 \\
AGC219368 & 11.20606 & 24.07750 & & 761$\pm$8 & & & &     \\
$[$RGK2003$]$J111223.17+134248.9 & 11.20643 & 13.71370 & & 630$\pm$99 & & & &     \\
PGC087259 & 11.20881 & 16.28964 & -5.0 & 1250$\pm$71 & 0.91 & 0.31 & 147.3 & 16.89$\pm$0.50 \\
UGC06248 & 11.21439 & 10.20003 & 9.9 & 1286$\pm$3 & 1.05 & 0.10 & 53.1 & 23.43$\pm$0.50 \\
AGC215280 & 11.22119 & 15.40778 & & 1479$\pm$8 & & & &     \\
UGC06258 & 11.23050 & 21.52043 & 9.9 & 1454$\pm$3 & 0.92 & 0.51 & 172.7 & 15.34$\pm$0.40 \\
SDSSJ111350.75+095739.4 & 11.23078 & 9.96087 & & 1610$\pm$5 & 0.35 & 0.06 & 74.3 & 18.55$\pm$0.50 \\
AGC215387 & 11.23742 & 12.77944 & & 578$\pm$8 & & & &     \\
PGC1487160 & 11.24033 & 15.53389 & & 861$\pm$30 & 0.75 & 0.22 & 92.7 & 17.05$\pm$0.50 \\
NGC3592 & 11.24093 & 17.25999 & 5.3 & 1302$\pm$4 & 1.33 & 0.57 & 117.7 & 14.47$\pm$0.30 \\
NGC3593 & 11.24361 & 12.81765 & -0.4 & 630$\pm$3 & 1.67 & 0.40 & 91.2 & 11.85$\pm$0.12 \\
SDSSJ111445.01+123851.7 & 11.24584 & 12.64759 & & 628$\pm$8 & 0.72 & 0.24 & 95.7 & 17.51$\pm$0.50 \\
NGC3596 & 11.25173 & 14.78694 & 5.2 & 1192$\pm$2 & 1.55 & 0.02 & & 11.79$\pm$0.08 \\
SDSSJ111506.95+144708.3 & 11.25192 & 14.78565 & & 1203$\pm$35 & & & 179.9 &     \\
SDSSJ111516.15+144154.9 & 11.25450 & 14.69867 & & 1092$\pm$45 & & & 17.5 &     \\
NGC3599 & 11.25750 & 18.11033 & -2.0 & 837$\pm$20 & 1.38 & 0.04 & & 12.88$\pm$0.08 \\
SDSSJ111532.38+143438.4 & 11.25899 & 14.57736 & & 1133$\pm$8 & 0.55 & 0.09 & 145.4 & 17.84$\pm$0.50 \\

\hline 
\end{tabular}
 \end{table*}
\begin{table*}
 \addtocounter{table}{-1}
	\caption{Continue.}
	   \label{a2}	 
	 \scriptsize
	\begin{tabular}{llllllllllll}
	\hline\hline
	   Galaxies  & RA & Dec.&  Morph.& Mean Hel.& logD$_{25}$ & logr$_{25}$ & P.A. &B$_T$ \\ 
	                   & (J2000) & (J2000) & type &  Vel.& &  &   &  \\
		          & (hours) & (degrees )   & & [km/s]  &[arcmin] & &[deg]  & mag \\
				  
	\hline
	U376   & & & & & & & &   \\ 
\hline
NGC3605 & 11.27963 & 18.01717 & -4.5 & 675$\pm$15 & 1.10 & 0.24 & 19.1 & 13.16$\pm$0.28 \\
UGC06296 & 11.28086 & 17.79847 & 8.0 & 976$\pm$2 & 1.11 & 0.48 & 168.7 & 14.18$\pm$0.41 \\
NGC3607 & 11.28182 & 18.05190 & -3.2 & 930$\pm$3 & 1.66 & 0.06 & 120.0 & 10.93$\pm$0.17 \\
NGC3608 & 11.28304 & 18.14854 & -4.8 & 1211$\pm$20 & 1.50 & 0.08 & 80.0 & 11.57$\pm$0.26 \\
IC2684 & 11.28363 & 13.09921 & & 591$\pm$11 & 0.88 & 0.13 & 171.2 & 15.78$\pm$0.61 \\
SDSSJ111702.67+100836.1 & 11.28408 & 10.14330 & & 1765$\pm$5 & 0.75 & 0.43 & 31.6 & 17.72$\pm$0.50 \\
AGC215386 & 11.29750 & 13.98222 & & 871$\pm$8 & & & &     \\
PGC034522 & 11.29945 & 17.44132 & -5.0 & 766$\pm$2 & 0.95 & 0.06 & 157.3 & 15.10$\pm$0.29 \\
SDSSJ111803.87+101439.9 & 11.30107 & 10.24441 & & 957$\pm$65 & 0.54 & 0.18 & 118.5 & 18.33$\pm$0.50 \\
PGC034537 & 11.30167 & 18.79814 & 4.9 & 1140$\pm$7 & 0.73 & 0.04 & & 16.35$\pm$0.38 \\
UGC06320 & 11.30481 & 18.84708 & 8.0 & 1122$\pm$4 & 1.13 & 0.03 & & 13.83$\pm$0.41 \\
PGC086629 & 11.30593 & 17.69740 & 9.8 & 1055$\pm$5 & 0.89 & 0.00 & &     \\
UGC06324 & 11.30614 & 18.73836 & -1.9 & 1083$\pm$56 & 1.14 & 0.33 & 166.8 & 14.77$\pm$0.32 \\
AGC215389 & 11.30797 & 14.30056 & & 917$\pm$8 & & & &     \\
AGC215392 & 11.30933 & 14.53083 & & 909$\pm$8 & & & &    \\
SDSSJ111842.18+123341.7 & 11.31173 & 12.56175 & & 979$\pm$58 & 0.73 & 0.15 & 152.8 & 17.47$\pm$0.50 \\
AGC215393 & 11.31467 & 13.40667 & & 862$\pm$8 & & & &     \\
AGC215396 & 11.31500 & 12.89472 & & 581$\pm$8 & & & &     \\
AGC215397 & 11.31525 & 14.21556 & & 909$\pm$8 & & & &     \\
NGC3623 & 11.31551 & 13.09294 & 1.0 & 806$\pm$5 & 1.88 & 0.59 & 173.0 & 10.14$\pm$0.13 \\
AGC215398 & 11.31819 & 12.75528 & & 753$\pm$8 & & & &     \\
AGC215400 & 11.31894 & 12.65194 & & 753$\pm$8 & & & &     \\
AGC215401 & 11.32006 & 13.59250 & & 834$\pm$8 & & & &     \\
AGC215286 & 11.32019 & 14.32778 & & 998$\pm$8 & & & &     \\
SDSSJ111914.39+115707.7 & 11.32066 & 11.95203 & & 861$\pm$4 & 0.87 & 0.30 & 94.1 & 17.01$\pm$0.50 \\
SDSSJ111915.87+141724.8 & 11.32108 & 14.29024 & & 728$\pm$36 & & & 171.1 &   \\
SDSSJ111921.39+140431.6 & 11.32262 & 14.07550 & & 485$\pm$132 & 0.65 & 0.27 & 144.7 & 18.39$\pm$0.50 \\
AGC215402 & 11.32392 & 13.23389 & & 772$\pm$8 & & & &     \\
PGC1367320 & 11.32447 & 9.59562 & & 990$\pm$6 & 0.62 & 0.31 & 117.7 & 17.81$\pm$0.55 \\
AGC215403 & 11.32511 & 12.56361 & & 716$\pm$8 & & & &     \\
AGC215405 & 11.32592 & 12.51417 & & 695$\pm$8 & & & &     \\
AGC215406 & 11.32608 & 13.85944 & & 984$\pm$8 & & & &      \\
AGC215407 & 11.32711 & 12.39333 & & 655$\pm$8 & & & &     \\
AGC215287 & 11.32919 & 15.50222 & & 1334$\pm$8 & & & &     \\
AGC215409 & 11.33175 & 12.87528 & & 678$\pm$8 & & & &     \\
AGC215410 & 11.33303 & 13.29000 & & 785$\pm$8 & & & &     \\
UGC06341 & 11.33350 & 18.26050 & 7.9 & 1611$\pm$42 & 1.00 & 0.44 & 152.1 & 16.12$\pm$0.39 \\
NGC3626 & 11.33438 & 18.35704 & -0.9 & 1480$\pm$5 & 1.47 & 0.18 & 156.3 & 11.81$\pm$0.23 \\
SDSSJ112009.38+133555.3 & 11.33594 & 13.59866 & & 650$\pm$33 & & & &     \\
NGC3627 & 11.33750 & 12.99099 & 3.1 & 721$\pm$2 & 2.01 & 0.35 & 173.0 & 9.74$\pm$0.20 \\
NGC3628 & 11.33804 & 13.58876 & 3.1 & 846$\pm$2 & 2.04 & 0.51 & 103.3 & 9.97$\pm$0.18 \\
AGC215411 & 11.34086 & 12.86778 & & 646$\pm$8 & & & &    \\
NGC3629 & 11.34216 & 26.96328 & 5.9 & 1508$\pm$4 & 1.25 & 0.19 & 34.5 & 13.12$\pm$0.49 \\
PGC1534499 & 11.35698 & 17.51034 & 9.0 & 979$\pm$57 & 0.92 & 0.25 & 145.0 & 16.46$\pm$0.50 \\
AGC215412 & 11.36328 & 13.61861 & & 908$\pm$8 & & & &    \\
$[$RGK2003$]$J112216.25+205158.7 & 11.37117 & 20.86628 & & 810$\pm$60 & & & &     \\
PGC2806917 & 11.37124 & 20.87560 & 10.0 & 810$\pm$8 & & & &     \\
IC2763 & 11.37180 & 13.06512 & 6.0 & 1569$\pm$2 & 1.16 & 0.57 & 98.2 & 15.14$\pm$0.35 \\
SDSSJ112223.18+130440.1 & 11.37311 & 13.07784 & & 1078$\pm$2 & 0.94 & 0.47 & 90.7 & 17.08$\pm$0.50 \\
AGC215413 & 11.37317 & 13.64583 & & 905$\pm$8 & & & &   \\
PGC1398872 & 11.37317 & 11.79393 & & 1571$\pm$4 & 0.53 & 0.21 & 111.1 & 17.98$\pm$0.45 \\
PGC1420152 & 11.37333 & 12.97956 & & 626$\pm$26 & 0.63 & 0.05 & 142.1 & 17.31$\pm$0.49 \\
PGC1407849 & 11.38074 & 12.34474 & & 1544$\pm$6 & 0.56 & 0.03 & & 17.25$\pm$0.35 \\
NGC3655 & 11.38186 & 16.59003 & 5.0 & 1453$\pm$13 & 1.18 & 0.16 & 31.8 & 12.32$\pm$0.07 \\
IC2782 & 11.38205 & 13.44105 & 8.0 & 860$\pm$24 & 0.97 & 0.05 & 22.6 & 15.12$\pm$0.41 \\
AGC215290 & 11.38308 & 12.46056 & & 1613$\pm$8 & & & &    \\
PGC086673 & 11.38316 & 17.47390 & 10.0 & 1383$\pm$4 & 0.79 & 0.00 & &     \\
AGC215414 & 11.38642 & 13.70556 & & 878$\pm$8 & & & &     \\
HIPASSJ1122+13 & 11.38700 & 13.71541 & 10.0 & 883$\pm$10 & & & &     \\
IC2787 & 11.38863 & 13.62967 & 6.0 & 708$\pm$27 & 0.90 & 0.04 & & 15.70$\pm$0.45 \\
SDSSJ112337.61+125344.8 & 11.39377 & 12.89597 & 9.0 & 666$\pm$6 & 0.76 & 0.20 & 158.6 & 17.15$\pm$0.50 \\
NGC3659 & 11.39594 & 17.81839 & 7.7 & 1284$\pm$11 & 1.29 & 0.28 & 59.2 & 12.92$\pm$0.48 \\
PGC091112 & 11.40367 & 26.24551 & 5.6 & 1483$\pm$11 & 0.78 & 0.73 & 61.8 & 17.31$\pm$0.50 \\
AGC729579 & 11.40481 & 25.71056 & & 1523$\pm$8 & & & &     \\
NGC3666 & 11.40726 & 11.34195 & 5.2 & 1060$\pm$2 & 1.54 & 0.54 & 97.5 & 12.69$\pm$0.11 \\
AGC215415 & 11.40950 & 12.67778 & & 1004$\pm$8 & & & &     \\
PGC1480186 & 11.41236 & 15.27553 & & 1125$\pm$6 & 0.84 & 0.44 & 5.9 & 17.45$\pm$0.50 \\
PGC035087 & 11.41717 & 17.08572 & 9.0 & 1208$\pm$5 & & & &     \\
PGC035096 & 11.41966 & 16.88431 & 9.5 & 1021$\pm$3 & 0.76 & 0.15 & 147.0 & 16.12$\pm$1.17 \\
AGC749436 & 11.42331 & 23.91889 & & 1024$\pm$8 & & & &     \\
IC0692 & 11.43153 & 9.98722 & 1.9 & 1158$\pm$4 & 0.88 & 0.16 & 117.7 & 14.18$\pm$0.35 \\
NGC3681 & 11.44161 & 16.86328 & 4.0 & 1240$\pm$3 & 1.24 & 0.01 & & 12.42$\pm$0.16 \\
AGC215296 & 11.44867 & 14.83417 & & 913$\pm$8 & & & &     \\
NGC3684 & 11.45311 & 17.03018 & 4.0 & 1159$\pm$6 & 1.36 & 0.18 & 126.2 & 12.31$\pm$0.11 \\

\hline 
\end{tabular}
 \end{table*}
\begin{table*}
 \addtocounter{table}{-1}
	\caption{Continue.}
	   \label{a2}	 
	 \scriptsize
	\begin{tabular}{llllllllllll}
	\hline\hline
	   Galaxies  & RA & Dec.&  Morph.& Mean Hel.& logD$_{25}$ & logr$_{25}$ & P.A. &B$_T$ \\ 
	                   & (J2000) & (J2000) & type &  Vel.& &  &   &  \\
		          & (hours) & (degrees )   & & [km/s]  &[arcmin] & &[deg]  & mag \\
				  
	\hline
	U376   & & & & & & & &   \\ 
\hline
NGC3686 & 11.46220 & 17.22409 & 4.1 & 1157$\pm$3 & 1.46 & 0.12 & 22.7 & 12.00$\pm$0.07 \\
NGC3691 & 11.46928 & 16.92044 & 3.0 & 1078$\pm$10 & 1.17 & 0.15 & 22.0 & 12.64$\pm$0.17 \\
NGC3692 & 11.47342 & 9.40753 & 3.1 & 1719$\pm$4 & 1.51 & 0.63 & 93.4 & 13.03$\pm$0.55 \\
PGC086634 & 11.48088 & 18.28267 & 9.5 & 1303$\pm$9 & 1.01 & 0.15 & & 17.10$\pm$0.50 \\
PGC035380 & 11.48730 & 20.58080 & 10.0 & 1412$\pm$33 & 0.31 & 0.18 & & 16.31$\pm$0.16 \\
IC0700 & 11.48759 & 20.58336 & 4.2 & 1415$\pm$6 & 0.97 & 0.31 & 70.9 & 14.09$\pm$0.23 \\
PGC035384 & 11.48788 & 20.58566 & 10.0 & 1420$\pm$35 & 0.48 & 0.04 & & 17.23$\pm$0.14 \\
PGC035385 & 11.48793 & 20.58816 & 10.0 & 1670$\pm$50 & 0.39 & 0.35 & 54.4 & 18.73$\pm$0.19 \\
SDSSJ112934.60+104835.4 & 11.49295 & 10.80987 & & 744$\pm$46 & 0.55 & 0.06 & 160.8 & 17.74$\pm$0.50 \\
PGC035426 & 11.49847 & 16.42957 & & 1067$\pm$5 & & & &     \\
NGC3705 & 11.50209 & 9.27667 & 2.4 & 1019$\pm$3 & 1.63 & 0.39 & 121.0 & 11.76$\pm$0.10 \\
PGC2807141 & 11.51478 & 14.14596 & 10.0 & 878$\pm$1 & & & &    \\
SDSSJ113108.88+133413.4 & 11.51912 & 13.57052 & & 1013$\pm$7 & & & 114.4 &    \\
PGC035565 & 11.53386 & 14.61087 & 4.8 & 1123$\pm$7 & 0.79 & 0.41 & 85.6 & 17.03$\pm$0.50 \\
PGC1468083 & 11.56391 & 14.82472 & 3.2 & 1129$\pm$6 & 0.54 & 0.21 & 8.2 & 17.75$\pm$0.50 \\
SDSSJ113350.91+140315.0 & 11.56414 & 14.05420 & & 933$\pm$2 & & & 93.1 &    \\
IC2934 & 11.57212 & 13.32139 & 4.9 & 1196$\pm$6 & 0.87 & 0.29 & 152.9 & 15.34$\pm$0.59 \\
PGC2807142 & 11.58147 & 11.02016 & 10.0 & 882$\pm$2 & & & &     \\
SDSSJ113456.50+161452.5 & 11.58237 & 16.24797 & & 1135$\pm$4 & 0.54 & 0.13 & 137.9 & 17.80$\pm$0.50 \\
SDSSJ113708.79+131504.9 & 11.61912 & 13.25120 & & 983$\pm$5 & 0.83 & 0.47 & 38.9 & 17.50$\pm$0.50 \\
PGC086675 & 11.62536 & 18.59347 & 9.0 & 958$\pm$60 & 0.84 & 0.00 & &     \\
UGC06594 & 11.62696 & 16.55621 & 6.4 & 1043$\pm$5 & 1.33 & 0.74 & 132.3 & 14.89$\pm$0.37 \\
UGC06599 & 11.62847 & 24.13256 & 9.9 & 1569$\pm$2 & 0.99 & 0.45 & 172.0 & 16.61$\pm$0.50 \\
NGC3773 & 11.63695 & 12.11216 & -2.0 & 985$\pm$4 & 1.09 & 0.10 & 160.0 & 13.51$\pm$0.53 \\
PGC1448560 & 11.68241 & 14.07445 & & 910$\pm$7 & 0.83 & 0.49 & 129.0 & 17.12$\pm$0.50 \\
NGC3810 & 11.68300 & 11.47120 & 5.2 & 993$\pm$2 & 1.52 & 0.17 & 31.6 & 11.27$\pm$0.10 \\
UGC06655 & 11.69739 & 15.97359 & -1.0 & 740$\pm$4 & 0.83 & 0.21 & 31.0 & 14.93$\pm$0.65 \\
UGC06669 & 11.70516 & 14.99519 & 9.9 & 1022$\pm$2 & 1.12 & 0.29 & 68.8 &     \\
UGC06670 & 11.70814 & 18.33374 & 9.9 & 922$\pm$1 & 1.18 & 0.47 & 154.3 & 13.39$\pm$0.28 \\
PGC036412 & 11.71957 & 14.22419 & 10.0 & 1015$\pm$1 & & & &    \\
PGC1393002 & 11.72417 & 11.39835 & & 898$\pm$7 & 0.57 & 0.06 & 164.6 & 16.82$\pm$0.48 \\
PGC2806928 & 11.80454 & 18.64242 & 10.0 & 976$\pm$6 & 0.52 & 0.00 & &     \\
SDSSJ114843.14+171053.1 & 11.81197 & 17.18155 & & 1092$\pm$337 & 0.70 & 0.32 & 96.2 & 17.91$\pm$0.50 \\
SDSSJ114850.14+102655.9 & 11.81393 & 10.44876 & & 753$\pm$102 & & & &     \\
UGC06782 & 11.81593 & 23.83769 & 9.8 & 523$\pm$16 & 1.24 & 0.00 & & 15.16$\pm$0.16 \\
PGC1528400 & 11.81822 & 17.25570 & & 623$\pm$29 & 0.38 & 0.13 & 63.9 & 18.49$\pm$0.54 \\
NGC3900 & 11.81928 & 27.02194 & -0.2 & 1800$\pm$2 & 1.41 & 0.31 & 1.5 & 12.29$\pm$0.07 \\
SDSSJ114925.75+253700.4 & 11.82382 & 25.61673 & & 1799$\pm$8 & 0.48 & 0.13 & 155.8 & 18.34$\pm$0.50 \\
SDSSJ114931.03+151539.7 & 11.82528 & 15.26094 & & 857$\pm$29 & & & 148.9 &     \\
SDSSJ114957.11+161743.6 & 11.83254 & 16.29550 & & 1188$\pm$45 & 0.52 & 0.22 & 154.7 & 18.13$\pm$0.50 \\
PGC036976 & 11.83409 & 15.02331 & 4.6 & 751$\pm$2 & 0.71 & 0.17 & 89.0 & 15.76$\pm$0.54 \\
NGC3912 & 11.83459 & 26.47946 & 3.1 & 1787$\pm$3 & 1.18 & 0.35 & 3.3 & 13.34$\pm$0.51 \\
PGC1739846 & 11.84440 & 25.52629 & 4.8 & 1819$\pm$3 & 0.91 & 0.49 & 152.3 & 16.58$\pm$0.36 \\
PGC037048 & 11.84887 & 14.59490 & & 1008$\pm$5 & 0.54 & 0.21 & 51.9 & 16.66$\pm$1.22 \\
PGC037162 & 11.86718 & 13.87873 & 4.4 & 967$\pm$4 & 0.78 & 0.10 & 41.7 & 15.28$\pm$0.41 \\
SDSSJ115220.20+152736.2 & 11.87228 & 15.46006 & & 787$\pm$6 & 0.79 & 0.48 & 52.3 & 18.35$\pm$0.50 \\

\hline
\end{tabular}
\end{table*}